\documentclass[aps,nofootinbib,prd,twocolumn,showpacs,preprintnumbers,amsmath,amssymb,superscriptaddress,groupedaddress]{revtex4}

\usepackage{psfrag}
\usepackage{bm}

\usepackage{amsmath,amssymb,bm,graphicx,epsfig,color,gensymb}

\usepackage{dcolumn}% Align table columns on decimal point

\newcommand{\order}{ {\cal O} }

\providecommand{\be}{\begin{equation}}
\providecommand{\ee}{\end{equation}}

\newcommand{\nl}{\nonumber \\ }

\def\max{\text{max}}

\def\s{\sigma}

\begin{document}

\preprint{EFI Preprint 14-35}

\title{Extraction of the proton radius from electron-proton scattering data}

\author{Gabriel Lee}
\email{leeg@physics.technion.ac.il}
\affiliation{
 Enrico Fermi Institute and Department of Physics, \\
 The University of Chicago, Chicago, Illinois, 60637, USA
}
\affiliation{
Physics Department,
Technion--Israel Institute of Technology, Haifa 32000, Israel
}

\author{John R. Arrington}
\email{johna@anl.gov}
\affiliation{
Physics Division,
Argonne National Laboratory, Argonne, Illinois, 60439, USA
}

\author{Richard J. Hill}
\email{richardhill@uchicago.edu}
\affiliation{
 Enrico Fermi Institute and Department of Physics, \\
 The University of Chicago, Chicago, Illinois, 60637, USA
}

\date{May 8, 2015}

\begin{abstract}
We perform a new analysis of electron-proton scattering data to
determine the proton electric and magnetic radii, enforcing
model-independent constraints from form factor analyticity. A
wide-ranging study of possible systematic effects is performed. An
improved analysis is developed that rebins data taken at
identical kinematic settings, and avoids a scaling assumption of
systematic errors with statistical errors.  Employing standard models
for radiative corrections, our improved analysis of the 2010 Mainz A1
Collaboration data yields a proton electric radius $r_E =
0.895(20)$~fm and magnetic radius $r_M = 0.776(38)$~fm.  A similar
analysis applied to world data (excluding Mainz data) implies $r_E =
0.916(24)$~fm and $r_M = 0.914(35)$~fm.   The Mainz and world values
of the charge radius are consistent, and a simple combination yields
a value $r_E = 0.904(15)$~fm that is $4\sigma$ larger than the CREMA
Collaboration muonic hydrogen determination.  The Mainz and world values of the
magnetic radius differ by $2.7\sigma$, and a simple average yields
$r_M= 0.851(26)$~fm.  The circumstances under which published muonic
hydrogen and electron scattering data could be reconciled are
discussed, including a possible deficiency in the standard radiative
correction model which  requires further analysis.

\end{abstract}


\maketitle{}

\tableofcontents
  
\newpage

\section{Introduction}

The electromagnetic form factors of the nucleons provide basic inputs
to precision tests of the Standard Model and to the determination of
fundamental constants~\cite{Mohr:2012tt}. These form factors are also
of critical importance for the accelerator neutrino
program~\cite{deGouvea:2013onf}. The development of muonic atom
spectroscopy~\cite{Pohl:2010zza, Antognini:1900ns} has introduced a
powerful new probe of proton and nuclear structure, challenging
existing results obtained from (electronic) hydrogen and electron
scattering~\cite{Mohr:2012tt}. Taken at face value, in the absence of
new physics explanations, the muonic hydrogen Lamb shift
measurement~\cite{Pohl:2010zza} necessitates a $\gtrsim 5\sigma$
revision of the Rydberg constant, in addition to discarding or
revising the predictions from a large body of previous results in both
electron-proton scattering and hydrogen spectroscopy. Sources of
systematic error in electron-proton scattering measurements also
impact neutrino-nucleus scattering observables and hence the
extraction of fundamental neutrino oscillation parameters at current
and future facilities~\cite{Bhattacharya:2011ah,Day:2012gb,deGouvea:2013onf}.
Resolution of the so-called proton radius puzzle thus has important
implications across the fields of high energy, nuclear, and atomic
physics~\cite{Pohl:2013yb, Carlson:2015jba}.

The muonic hydrogen measurement~\cite{Antognini:1900ns} yields $r_E =
0.84087(39)$ fm, compared to $r_E = 0.8758(77)$~fm for Lamb shift
measurements from ordinary (electronic) hydrogen~\cite{Mohr:2012tt}.
Previous analyses of electron scattering results using the high
statistics data set taken at the Mainz Microtron (MAMI)
yielded~\cite{Bernauer:2013tpr, Priv_Distler, Arrington:2015yxa} $r_E=0.879(11)$~fm and
$r_M=0.777(19)$~fm, in both cases neglecting uncertainty associated
with two-photon exchange corrections~\cite{Arrington:2011kv}.  A
previous global analysis of world data~\cite{Zhan:2011ji}, excluding
the Mainz data, yielded $r_E = 0.875(10)$~fm and $r_M = 0.867(20)$~fm. 
Similar results were obtained in an independent global analysis which included
constraints on the large-distance behavior of an inferred proton
charge distribution~\cite{Sick:2012zz, Arrington:2015ria}.  So not only
are electron- and muon-based extractions of the charge radius
inconsistent, but there is also a $\sim 3\sigma$ disagreement between
extractions of $r_M$ from different electron scattering data sets.

Here, we address the issue of radius extraction from electron-proton
scattering data. A prominent uncertainty in the extracted radius
arises from the shape of the form factor assumed when extrapolating to
$q^2=0$ where the radius is defined in terms of the form factor
slope. This can be the dominant uncertainty, as happens in 
particular in the A1 Collaboration's extraction of the charge radius
from Mainz data~\cite{Bernauer:2013tpr}.  In
Ref.~\cite{Hill:2010yb}, one of us investigated the implications of
analyticity for the form factors of electromagnetic lepton-nucleon
scattering.  Reference~\cite{Hill:2010yb} considered a representative
data set consisting of extracted electric form factor values from cross
section data prior to 2007.  In the present paper, we extend this
analysis by fitting directly to cross section data, eliminating
possible systematic uncertainties associated with the reduction from
cross sections to form factors prior to the $q^2\to 0$ extrapolation
that defines the radius observable.  We consider the most recent data,
including separately a ``Mainz data set''~\cite{Bernauer:2013tpr} and
a ``world data set'' (defined below in Sec.~\ref{sec:radexpt},
excluding Mainz data). We extend
the analysis of the electric form factor of the proton to consider
also the magnetic form factor (see also Ref.~\cite{Epstein:2014zua}),
necessary to connect with cross section data.  We discuss the impact
of uncertainties arising from the fitting procedure, theoretical
corrections to the cross sections, and experimental systematic
uncertainties.  We focus here on understanding the implications of
electron-proton scattering data in isolation and do not include
further constraints arising from isospin decomposition in combination
with electron-neutron, $\pi\pi\to N\bar{N}$, or other
data~\cite{Belushkin:2006qa,Hill:2010yb,Epstein:2014zua}.  As we will
see, several critical issues in the electron scattering data demand
attention before the inclusion of such further ingredients. 

The remainder of the paper is structured as follows.
Section~\ref{sec:notation} introduces notations and conventions.
Section~\ref{sec:z} discusses the constraints of the form factor shape
arising from analyticity and perturbative scaling.
Section~\ref{sec:rad}
reviews the status of radiative corrections and defines the default
models used in the remainder of the paper.
Section~\ref{sec:defaultfit} analyzes the Mainz data set, employing
exactly the same analysis strategy as detailed in
Ref.~\cite{Bernauer:2013tpr}, with the exception that the bounded $z$
expansion is used in place of polynomial or spline functions to
represent the form factors.  Section~\ref{sec:systematics} studies a
range of possible systematic effects.  Section~\ref{sec:results}
provides updated extractions of the charge and magnetic radii and
uncertainties, for both the Mainz data and the world data.
Section~\ref{sec:summary} presents a summary and conclusions. 
Supplemental Material \cite{Supplement} includes the data used for fits in 
Sections~\ref{sec:defaultfit},~\ref{sec:systematics}, and \ref{sec:results}.

\section{Conventions and notation \label{sec:notation} }

The Dirac and Pauli form factors of the proton, $F_1$ and $F_2$,
respectively, are defined by
\begin{align}
  \langle p(p')|J^\mu_{\rm em}|p(p)\rangle = \bar u(p')
  \Gamma^{(p)\mu}(p^\prime,p) u(p)\,,
\end{align}
where
\begin{align}\label{eq:protonvertex}
\Gamma^{(p)\mu}(p^\prime,p) &= \gamma^\mu F_1(q^2) + {i \over  2m_p} \sigma^{\mu\nu} q_\nu F_2(q^2) \,,
\end{align}
with $q^\mu = p^{\prime\mu}-p^\mu$.  The Sachs electric and magnetic
form factors are related to the Dirac--Pauli basis by
\begin{align}
\label{eq:sachs}
G_E(q^2) &= F_1(q^2) + \frac{q^2}{4m_p^2} F_2(q^2) \,,
\nl
G_M(q^2) &=
F_1(q^2) + F_2(q^2) \,,
\end{align}
where $G_E(0) = 1$, $G_M(0) = \mu_p \approx
2.793$~\cite{Nakamura:2010}.  The electric and magnetic radii, $r_E$ and $r_M$, are
defined as the slopes of the Sachs form factors at $q^2 = 0$, i.e.,
\begin{align}
{ G_{E,M} (q^2) \over G_{E,M}(0) }= 1 + \frac{1}{6} r_{E,M}^2 q^2 + \order(\alpha, q^4)
\,.
\label{eqn:raddef}
\end{align}
In terms of $G_{E}$ and $G_{M}$, the cross section for electron-proton scattering in single photon
exchange approximation is
\begin{align}
\left( \frac{d\sigma}{d \Omega} \right)_{0} = \left(
\frac{d\sigma}{d\Omega} \right)_{\rm Mott} \frac{\epsilon G_E^2 + \tau
  G_M^2}{\epsilon (1+\tau)} \,,
\label{eqn:xs1photon}
\end{align}
where $(d\sigma/d\Omega)_{\rm Mott}$ is the recoil-corrected
relativistic point particle (Mott) result,
\begin{align}
\left( \frac{d\sigma}{d\Omega} \right)_{\rm Mott}
=
{\alpha^2 \over 4 E^2 \sin^4{\theta\over 2}} {E^\prime \over E} \cos^2{\theta\over 2}
\,. 
\end{align}
Here, $Q^2=-q^2$, $E$ is the initial electron energy,
$E^\prime=E/[1+(2E/m_p)\sin^2(\theta/2)]$ and $\theta$ are the energy
and angle with respect to the beam direction of the final state electron, 
and $\epsilon$, $\tau$ are the dimensionless kinematic variables
\be
\begin{split}
\tau = \frac{Q^2}{4m_p^2} \,, \quad
\epsilon = \left[ 1 + 2 (1 + \tau) \tan^2 \frac{\theta}{2} \right]^{-1} \,.
\end{split}
\ee
In fits to the Mainz data below, we employ the beam energy $E$ and
the acceptance-averaged $Q^2$ as independent variables, 
as dictated by the presentation of experimental results in this data set. 
In fits to world data excluding Mainz data, we employ $E$ and $\theta$.

\section{Form factor shape \label{sec:z} }

When performing statistical analyses that constrain the form factors
and derived quantities such as the radius, it is important that the
class of allowed functions be large enough to contain the true form
factors but sufficiently constrained for meaningful values to
be obtained, i.e., without arbitrarily large errors, and such that
overfitting to statistical noise does not bias parameter extractions.
We summarize here our knowledge about the analytic structure of the
form factors, introduce notation for the $z$ expansion, and explain an
important property of the $z$ expansion with regard to convexity and
stability of fits involving many parameters.

\subsection{Analyticity and $z$ expansion}

Let us recall the analytic structure of the form factors $F_1(q^2)$,
$F_2(q^2)$, or equivalently $G_E(q^2)$, $G_M(q^2)$. The form
factors may be extended to functions of the complex variable $t=q^2$,
analytic outside of a cut at timelike values of $t$, beginning at
the two-pion production threshold, $t\ge 4m_\pi^2$.%
\footnote{Here and throughout, $m_\pi=140\,{\rm MeV}$ denotes the
 charged pion mass. Accounting for isospin violation would imply a
 smaller threshold at $4m_{\pi^0}^2$. A conservative approach to
 accounting for this effect would be to lower the threshold; we have
 verified that the difference is inconsequential to the fits. }
In the restricted kinematic region accessed in a given experimental data set, 
$-Q_{\rm max}^2 \le t \le 0$, the finite distance to singularities
implies the existence of a small expansion parameter, $|z|_{\rm max}<1$.
To see this, we perform a conformal mapping of the domain of analyticity onto the unit circle,%
\footnote{For a discussion and further references, see Ref.~\cite{Hill:2006ub}.}
\be\label{eq:z}
z(t,t_{\rm cut},t_0) = \frac{\sqrt{t_{\rm cut} - t} - \sqrt{t_{\rm cut} - t_0}}{\sqrt{t_{\rm cut} - t} + \sqrt{t_{\rm cut} - t_0} } \,,
\ee
where $t_{\rm cut} = 4m_\pi^2$ and $t_0$ is a free parameter
representing the point mapping onto $z=0$. By the ``optimal''
choice $t_0^{\rm opt}(Q^2_{\rm max}) = t_{\rm cut} \left( 1 - \sqrt{1+ Q^2_{\rm
 max}/t_{\rm cut}} \right)$, the maximum value of $|z|$ is
minimized: $|z| \le |z|_{\rm max}^{\rm opt} = [(1+Q_{\rm max}^2/t_{\rm
 cut})^\frac14 - 1]/[(1+Q_{\rm max}^2/t_{\rm cut})^\frac14 + 1]$.
For example, with $Q^2_{\rm max} =$ 0.05, 0.5, and $1\,{\rm GeV}^2$, 
we have $|z|_{\rm max}^{\text{opt}} \approx$ 0.06, 0.25, and $0.32$, respectively.%
\footnote{For $t_0=0$, these numbers become approximately twice as large,
  $|z|_{\rm max} \approx $ 0.12, 0.46, and 0.58.
}
Expanding the form factors as
\be
  G_E(q^2) = \sum_{k=0}^{k_{\rm max}} a_k \, z^k \,, \quad
  G_M(q^2) = \sum_{k=0}^{k_{\rm max}} b_k \, z^k \,,
\label{eqn:zFF}
\ee
we find that higher-order terms are suppressed by powers of
this small parameter.

\subsection{Coefficient bounds and large-$k$ scaling} \label{sec:bounds}

The identity~\cite{Bhattacharya:2011ah},
\be\label{eq:sumrule0}
\sum_{k=0}^\infty a_k^2 = {1\over\pi} \int_{t_{\rm cut}}^{\infty} {dt\over t-t_0}
\sqrt{t_{\rm cut}-t_0\over t-t_{\rm cut}} |G|^2 < \infty \,,
\ee
ensures that the coefficients multiplying $z^k$ are not only bounded in size 
but must decrease at large $k$.
This guarantees that a finite number of parameters is necessary to
describe the form factor with a given precision throughout the kinematic region of interest.
In the fits performed in this paper, we focus on the class of form factors
(\ref{eqn:zFF}), with a uniform bound on $|a_k/a_0|$ and $|b_k/b_0|$.
A study of form factor models, explicit spectral functions, and scattering data
motivates the conservative bound of $|a_k/a_0|_{\rm max}=|b_k/b_0|_{\rm max}=5$ for either $t_0=0$ or
$t_0=t_0^{\rm opt}$ when limiting the fit to $Q^2_{\rm max} \approx 1$~GeV$^2$~\cite{Bhattacharya:2011ah, Epstein:2014zua}.
We adopt this bound for our default fits but study also the case of modified bounds.

In fact, a stronger statement can be made regarding the large-$k$ behavior of the expansion coefficients.
Since at large spacelike values of momentum transfer, $Q^2\to\infty$,
the Sachs form factors are known to fall as $1/Q^4$ up to logarithms~\cite{Lepage:1980fj},
we have that $ Q^k G(-Q^2) \to 0$, $k=0,\dots,3$. From Eq.~(\ref{eqn:zFF}) this implies
\be
\frac{d^n}{dz^n} G_E \bigg|_{z=1} = 0 \,, \quad n=0,1,2,3,
\ee
or, equivalently, the series of four sum rules,
\be\label{eqn:sumrules}
\sum_{k=n}^\infty k(k-1)\cdots (k-n+1) a_{k} = 0 \,, \quad n=0,1,2,3.
\ee
Absolute convergence of these series corresponds to $a_k = {\cal O}(1/k^4)$.%
\footnote{Absolute convergence may
  be verified by inspecting the analog of Eq.~(\ref{eq:sumrule0}) applied to $|d^nG/dz^n|^2$ in place of $|G|^2$.}

\subsection{Convexity and $\chi^2$ minimization}

An important feature of Taylor expanded amplitudes facilitates
efficient and stable numerical fits.%
\footnote{While this observation has emerged from a particular
 example of fits to electron scattering data, the argument applies to
 general quantum mechanical observables represented as squares of
 Taylor-expanded amplitudes.} Consider a $\chi^2$ function of
schematic form
\be\label{eqn:chi2example}
\chi^2 = \sum_i \frac{( A_i^2 - M_i)^2}{E_i^2} \,, \quad A_i = \sum_n a_n x_i^n \,,
\ee
where the sum is over data points labeled by index $i$, $M$ represents
a measurement, $E$ is the error on the measurement, and $A$ is the
theoretical amplitude expressed in terms of a kinematic variable $x$
[e.g., $x_i=z(q_i^2)$ in the present application]. The Taylor
expansion coefficients $a_n$ are fit parameters to be determined by
minimizing $\chi^2$.

If $\chi^2(\{a_n\})$ is a convex function of its arguments, $a_n$,
then any local minimum is necessarily a global minimum, 
and the relevant optimization problem is amenable to efficient numerical algorithms.
In general, determining convexity of a multivariate quartic polynomial is NP hard~\cite{NP}.
We notice, however, that the matrix of second derivatives is
\begin{align}\label{eqn:Asum}
\frac{\partial^2 \chi^2}{\partial a_n \partial a_m}
&= 4 \sum_i x_i^{n+m} \left( \frac{3 A_i^2 - M_i}{E_i^2} \right) \,.
\end{align}
Each term in the sum over $i$ is seen to be a positive semidefinite
matrix provided that amplitudes satisfy $3A_i^2 > M_i$. 
Each contribution in Eq.~(\ref{eqn:chi2example}) is thus convex throughout 
the parameter regime where this physical condition is satisfied. 
Since a linear combination of convex polynomials with positive coefficients is convex, 
the sum of terms in Eq.~(\ref{eqn:chi2example}) is also convex in this regime. 
It is thus straightforward to build up solutions to the numerical $\chi^2$
minimization problem over a large number of parameters by successively
increasing the number of parameters and data points, using the previous
solution $\{a_n\}$ as the initial condition. The convexity condition
ensures that this procedure does not yield a solution that is a
local but not a global minimum.

The preceding arguments strictly apply when a single
form factor dominates, which is, for example, the case for $G_E$ in
Eq.~(\ref{eqn:xs1photon}) at low $Q^2$. In the general case in which
$G_E$ and $G_M$ are fit simultaneously, the $\chi^2$ function takes a
more general form involving the sum of probabilities, $A_i^2 \to
A_i^2 +B_i^2$, for which the simple convexity theorem following from
Eq.~(\ref{eqn:Asum}) no longer applies. It may be interesting to pursue
more general ``physical convexity'' theorems involving multiple
probability sums and correlated errors.

\subsection{Advantages over other parametrizations}

We remark that several parametrizations of the proton form factors in common use rely on
somewhat arbitrary expansions. A simple Taylor expansion in $q^2$~\cite{Arrington:2003qk} is
only guaranteed to converge below the pion production threshold $q^2\le 4m_\pi^2 \approx
0.08\,{\rm GeV}^2$. Convergence of a sequence of Pad\'{e} approximants, implemented either
directly as a ratio of polynomials~\cite{Kelly:2004hm, Arrington:2007ux} or as a continued
fraction~\cite{Sick:2003gm, Arrington:2006hm}, requires positivity of the spectral
function in the dispersive representations of the form factors, a
property which is not satisfied.%
\footnote{ That it cannot be satisfied is readily seen from the
 asymptotic behavior $Q^{-2}$ for the form factor represented by such
 a spectral function. }

While these functions may be able to provide a sufficiently precise representation of the
form factors with enough fit parameters, the parameters tend to be highly correlated.
Without any way to bound the parameters, these correlations can lead to a large uncertainty
on any given parameter (such as the radius) that grows as the number of parameters increases.
Because of this, it may be difficult or impossible to include enough parameters to properly
reproduce the data while at the same time achieving a meaningful limit on the extracted radius.
The correlation between different parameters may also lead to the situation in which overfitting 
the noise in data at high $Q^2$ biases the extracted radius.  
This concern applies especially for the magnetic form factor for which the data at low $Q^2$ have
larger uncertainties than the higher $Q^2$ data.

\section{Radiative corrections \label{sec:rad}}

In this section, we provide a brief summary of one-loop radiative corrections and Sudakov resummation
in electron-proton scattering.
We extract the radius according to Eq.~(\ref{eqn:raddef}), 
using data to which corrections have been applied to extract a Born cross section.
To understand the impact of the different corrections applied to various
data sets, we begin in Secs.~\ref{sec:1photon}--\ref{sec:sudakov}
with a brief overview of notation and results for
one-photon exchange, two-photon exchange, real photon emission, and Sudakov resummation as they
impact cross section measurements.
In Sec.~\ref{sec:radexpt}, we return to a discussion of experimental implementations
for the data sets employed in the remainder of this paper. 

\subsection{One-photon exchange\label{sec:1photon}}

The (on-shell, renormalized) scattering amplitude for the
electron-proton scattering process $e(k)p(p) \to e(k^\prime) p(p^\prime)$ 
involving one exchanged photon may
be written 
\begin{multline}\label{eq:1photon}
{\cal M}_1 = - { 4\pi \alpha \over q^2}{1\over 1 - \hat{\Pi}(q^2) }
\bar{u}^{(e)}(k^\prime) \Gamma^{(e)\mu}(k^\prime,k) u^{(e)}(k)
\\
\times 
\bar{u}^{(p)}(p^\prime) \Gamma^{(p)}_\mu(p^\prime,p) u^{(p)}(p) \,,
\end{multline}
and includes radiative corrections involving the proton and electron vertices
(and wave function renormalization) and vacuum polarization.    
Here, $\alpha=7.297 \times 10^{-3}$ is the fine structure constant.
The proton vertex function $\Gamma^{(p)}(p^\prime,p)$ is expressed, as in Eq.~(\ref{eq:protonvertex}),
in terms of the IR divergent on-shell form factors discussed below.  
The electron vertex function $\Gamma^{(e)}(k^\prime,k)$ is similarly expressed
in terms of on-shell form factors 
normalized as $F_1^{(e)}(0) \equiv 1$, $F_2^{(e)}(0) \equiv a_e \approx \alpha/(2\pi)$.
The photon propagator correction $\hat{\Pi}(q^2)$ accounts for contributions
of both leptonic and hadronic vacuum polarization.   

The on-shell form factors appearing in Eq.~(\ref{eq:1photon})
are necessarily infrared divergent at nonzero momentum
transfer, as deduced by the cancellation with bremsstrahlung emission.
In terms of a photon mass, let us introduce conventional ``Born''
form factors which are finite including first-order radiative
effects in the $\lambda\to 0$ limit.   We employ the tilde
notation $\tilde{F}_i$ to denote the on-shell form factor with the
corresponding Born form factor $F_i$:
\begin{align}\label{eq:redFF}
  \tilde{F}_i(q^2)
  \equiv \bigg\{
    1 - {\alpha\over 2\pi} \big[ K(p,p^\prime)-K(p,p) \big] 
    \bigg\}  F_i(q^2)
    \,.
\end{align}
Here $K(p_1,p_2)$ denotes the
integral~\cite{Tsai:1961zz,Maximon:2000hm}
\begin{multline}\label{eq:Kfunc}
  K(p_1,p_2) = {2 p_1\cdot p_2 \over -i\pi^2}\int d^4L
  {1\over L^2-\lambda^2+i0}
   \\
  \times{1\over L^2+2L\cdot p_1 + i0}{1\over L^2 + 2L\cdot p_2 + i0} \,,
\end{multline}
and is readily evaluated in analytic form.  The electric and magnetic
radii are now defined
as the slopes of the Born form factors with respect to $q^2$,
\be\label{eq:sachsborn}
{G^\prime_{E,M}(0)\over G_{E,M}(0)} = \frac16 r_{E,M}^2 \,. 
\ee
Infrared divergences are absorbed into the extracted prefactors in Eq.~(\ref{eq:redFF})
and will cancel upon including the effects of real photon emission.
The electron form factors may be calculated analytically in QED, with 
infrared divergences similarly cancelling against real emission. 
For completeness, let us note that for the
IR divergent on-shell Sachs form factors we have 
\begin{align}\label{eq:sachsrad}
  {\tilde{G}_{E,M}^\prime(0)\over \tilde{G}_{E,M}(0)} &\equiv \frac16 r_{E,M}^2 + {\alpha \over 3\pi m_p^2} \left( \log{m_p\over \lambda} + \frac14 \right) \,.
\end{align}

Leptonic vacuum polarization contributions to $\hat{\Pi}(q^2)$ are
readily computed analytically, and the hadronic contributions are constrained
by $e^+ e^- \to {\rm hadrons}$ data.%
\footnote{Alternatively, one could model the
  hadronic contributions by quark loop diagrams~\cite{Ent:2001hm}; however,
  strong interaction corrections are not controlled at small $Q^2 \lesssim \Lambda_{\rm QCD}^2$.}
For the purposes of determining the radii, we may simply absorb the
hadronic contribution $\hat{\Pi}_{\rm had}(q^2)$ into an alternate definition of the
reduced form factors:
\begin{align}\label{eq:hadvp}
F_i(q^2) \to [1-\hat{\Pi}_{\rm had}(q^2) ]^{-1} F_i(q^2) \,.
\end{align}

Several remarks are in order. First, we note that Eq.~(\ref{eq:redFF}) is not
a {\it calculation} of proton-vertex radiative corrections but rather a
{\it definition} of Born form factors in the presence of radiative corrections. 
The definitions of the radii following from 
Eqs.~(\ref{eq:redFF}) and (\ref{eq:sachsborn}) differ slightly from the definition of
Maximon and Tjon~\cite{Maximon:2000hm} which includes an additional contribution
(there denoted $\delta_{el}^{(1)}$)
involving a sticking-in-form-factors (SIFF) ansatz for the proton vertex.
However, in many analyses (including, in particular, Ref.~\cite{Bernauer:2013tpr}),
the additional contributions beyond those in Eq.~(\ref{eq:redFF}) are anyways
ignored.  The convention (\ref{eq:redFF}) does not require the specification of
a form factor model and is closely aligned with standard treatments of electron scattering.
Let us further remark that this
convention differs slightly from a convention commonly used in atomic physics
applications~\cite{Hill:2011wy}. However, the difference,
represented by the term $\alpha/(12\pi m_p^2)$ in Eq.~(\ref{eq:sachsrad}), corresponds
to a relative shift of $\sim 2\times 10^{-5}$ in $r_E$, well below current experimental sensitivities
in either electron scattering or muonic hydrogen.

Second, we remark that if hadronic vacuum polarization is not removed explicitly
before fitting then the resulting proton form factors
should be interpreted with the alternate definition (\ref{eq:hadvp}).
With this definition, the fitted radius now corresponds to
\be\label{eq:rhad}
[r_{E,M}^2]^{\rm fit} = r_{E,M}^2 + 6 \hat{\Pi}_{\rm had}^\prime(0) \,. 
\ee
A dispersive analysis of $e^+e^- \to {\rm hadrons}$ data yields~\cite{Jegerlehner:1996ab,Friar:1998wu},
\be
\hat{\Pi}^{\prime}_{\rm had}(0) = -9.31(20) \times 10^{-3} \,{\rm GeV}^{-2} \,.
\ee
This correction leads to a small shift, $\sim 0.001\,{\rm fm}$, in $r_E$,
and a more careful error analysis does not appear to be
warranted at the current level of precision.
We note that Ref.~\cite{Bernauer:2013tpr} did not account for hadronic vacuum polarization
explicitly and hence implicitly employed the alternate definition (\ref{eq:hadvp}).
Experiments in the world data set used explicit models to account for
hadronic vacuum polarization or included uncertainties to account for the neglect of this correction. 
Hence, the extracted radii should differ slightly from the Mainz
value according to the replacement (\ref{eq:rhad}), but the effect is well below the
current experimental precision.%
\footnote{
An explicit correction for hadronic vacuum polarization is typically
applied in atomic physics analyses.
This is the case in particular for the CREMA analysis of muonic hydrogen~\cite{Pohl:2010zza}.
To avoid a double counting, the shift (\ref{eq:rhad})
should therefore in principle be applied to electron scattering extractions
that absorb this correction into the definition of the radius, before input or comparison
to atomic physics extractions.}
The treatment in Eq.~(\ref{eq:hadvp}) efficiently accounts for
the effects of hadronic vacuum polarization on the radii in terms of the single number
$\hat{\Pi}^\prime_{\rm had}(0)$.
When interpreting form factors at finite momentum transfer, care must be taken to account for the
$q^2$ dependence of hadronic vacuum polarization.

Finally, we note that the analytic structures of the functions
$K(p,p^\prime)$ in Eq.~(\ref{eq:redFF}) and of
$\hat{\Pi}_{\rm had}(q^2)$ in Eq.~(\ref{eq:hadvp})
do not upset the assumptions going into
the $z$ expansion, since these functions are analytic outside of a cut
at timelike $q^2 \ge 4 m_p^2$ for $K(p,p^\prime)$
and at $q^2 \ge 4 m_\pi^2$ for $\hat{\Pi}_{\rm had}(q^2)$.

\subsection{Two-photon exchange\label{sec:2photon}}

The two-photon exchange (TPE) contribution may be written
\begin{align}\label{eq:2photon}
  {\cal M}_2 &= {\alpha\over 2\pi} \bigg[ -K(p,-k) - K(p^\prime,-k^\prime) + K(p,k^\prime)
    \nl
    &\quad + K(p^\prime,k) \bigg]
  {\cal M}_1 + \hat{\cal M}_2^{\rm MoTs}
  \nl
  &= {\alpha \over \pi} \bigg[
    -{E\over \sqrt{E^2-m_e^2}}\log\left( {E+\sqrt{E^2-m_e^2} \over m_e} \right)
    \nl
    &\quad
    +{E^\prime \over \sqrt{E^{\prime 2} - m_e^2}}\log\left( {E^\prime + \sqrt{ E^{\prime 2} - m_e^2} \over m_e} \right)
    \bigg]
  \nl
  &\quad 
  \times \log{Q^2\over \lambda^2} {\cal M}_1 + \hat{\cal M}_2^{\rm MaTj}
  \,. 
\end{align}
The $K(p_1,p_2)$ functions are defined above in Eq.~(\ref{eq:Kfunc}).
As indicated in Eq.~(\ref{eq:2photon}),
two conventions exist in the literature for isolating an IR finite
TPE contribution: $\hat{\cal M}_2^{\rm MoTs}$~\cite{Tsai:1961zz} (Mo-Tsai)
and $\hat{\cal M}_2^{\rm MaTj}$~\cite{Maximon:2000hm} (Maximon-Tjon). 
As long as the full correction ${\cal M}_2$ is applied to the data, 
the results are independent of the convention used to separate IR divergent and IR finite contributions.
Our hadronic model for the finite contribution is based on the Maximon-Tjon convention and so yields the complete 
${\cal M}_2$ when applied to the Mainz data, which uses the same convention. 
For the world data, the Mo-Tsai convention is used, and so our calculation of $\hat{{\cal M}}_2^{\rm MaTj}$ yields 
a total ${\cal M}_2$ contribution that differs by $-0.4$\% to $0.1$\% at the cross section level
compared to the consistent combination. 
This small error is accounted for in the radiative correction uncertainties quoted for these measurements, 
and we have verified that such differences have an insignificant impact on the extracted radii.

\begin{table}[t]
  \caption{\label{tab:tpeblucoefs}
    Expansion coefficients for
    Eq.~(\ref{eqn:SIFFBlun}) in the SIFF TPE prescription of Ref.~\cite{Blunden:2005ew}.
    Note that $n_3$ is determined by $F_{1,2}(0) = \sum_j n_j/d_j$.}
  \begin{ruledtabular}
    \begin{tabular}{ccc}
      & $F_1$ & $F_2$ \\
      \hline
      $n_1$ & 0.38676 & 1.01650 \\
      $n_2$ & 0.53222 & -19.0246 \\
      $d_1$ & 3.29899 & 0.40886 \\
      $d_2$ & 0.45614 & 2.94311 \\
      $d_3$ & 3.32682 & 3.12550
    \end{tabular}
  \end{ruledtabular}
\end{table}

As a default, we employ the SIFF ansatz to estimate the
TPE correction~\cite{Blunden:2003sp, Blunden:2005ew, Carlson:2007sp,Arrington:2011dn}.
We have computed ${\cal M}_2$ using two form factor models.
The first uses dipole $F_1$, $F_2$ form factors,
\be
F_{1,2}(q^2) \to F_{1,2}(0) \bigg( 1 - \frac{q^2}{\Lambda^2} \bigg)^{-2} \,,
\label{eqn:SIFFdip}
\ee
with a value $\Lambda^2={0.71\,{\rm GeV}^2}$. The second model represents
$F_1$, $F_2$ as a sum of monopoles,
\be
  F_{1,2}(q^2) \to \sum_{j=1}^{N} \frac{n_j}{d_j - q^2}
  \,.
\label{eqn:SIFFBlun}
\ee
To compare with previous results in the
literature~\cite{Blunden:2005ew}, we consider in particular the case
$N=3$ with parameter values $n_j, d_j$ given in
Table~\ref{tab:tpeblucoefs}.
We compare these models for ${\cal M}_2$
to results with
vanishing finite TPE correction in the Maximon-Tjon convention,
\be\label{eq:MaTjzero}
\hat{\cal M}_2^{\rm MaTj}({\rm no\,\,TPE}) = 0 \,,
\ee
and to results setting $\hat{\cal M}_2^{\rm MaTj}/{\cal M}_1$ equal to the complete, ``Feshbach''~\cite{McKinley:1948zz},
result for ${\cal M}_2/{\cal M}_1$ in the $m_p \to\infty$ limit%
\footnote{An imaginary part in Eq.~(\ref{eq:Fesh}) is ignored since it affects the
  cross section only at relative order $\alpha^2$.  For definiteness, we have expressed
  the result in terms of $(E,\theta)$ instead of the variables ($E,Q^2$) before taking the
  $m_p\to\infty$ limit, to match the expression used in Ref.~\cite{Bernauer:2013tpr}.  
}:
\be\label{eq:Fesh}
\hat{\cal M}_2^{\rm MaTj}({\rm Feshbach}) =
\bigg( 1 + {\pi \alpha}{\sin{\theta\over 2} \over 1 +  \sin{\theta\over 2} } \bigg) 
    {\cal M}_1 \,. 
\ee

\subsection{Soft bremsstrahlung\label{sec:brem}}

The soft bremsstrahlung contribution to the cross section is
\begin{multline}\label{eq:brem}
d\sigma_{\rm brem.} = -{\alpha\over 4\pi^2} d\sigma_0
\int{d^3\ell \over \sqrt{\bm{\ell}^2+\lambda^2} }\bigg|_{|\bm{\ell}| \le (E/E^\prime)\Delta E}
\\
\left( {k^\prime \over k^\prime \cdot \ell} - {k\over k\cdot \ell} - {p^\prime \over p^\prime\cdot \ell}
+ {p \over p\cdot \ell} \right)^2 \,,
\end{multline}
where
$\Delta E$ is the
accepted energy cut interval for the final state electron in the lab frame.  This integral
may be evaluated analytically~\cite{Maximon:2000hm}. After cancellation of infrared divergences, 
the differential cross section including first-order real and virtual radiative effects
may be written
\begin{align}\label{eq:radxs}
d\sigma = (d\sigma)_{0} (1 + \delta) \,. 
\end{align}
Here, $(d\sigma)_0$ is the cross section (\ref{eqn:xs1photon}), expressed
in terms of Born form factors, and $\delta$ is a finite correction
depending on kinematic variables that accounts for 
vertex, vacuum polarization, and TPE radiative corrections.  

\subsection{Large log resummation\label{sec:sudakov}}

The preceding subsections (Secs.~\ref{sec:1photon}--\ref{sec:brem})
summarize a complete treatment of first-order radiative corrections. 
The hadronic input, apart from the form factors to be determined, 
consists of a TPE model for $\hat{\cal M}_2$ in Eq.~(\ref{eq:2photon}) and
the number $\hat{\Pi}^\prime_{\rm had}(0)$ (the latter impacts the radius at a level below
current uncertainties).
However, we wish to describe scattering data with momentum transfers as large
as $Q^2 \sim 1\,{\rm GeV}^2$. In this regime, large logarithms from
electron radiative corrections cause a poor convergence,
or even breakdown, of the naive perturbation theory, since
\be
{\alpha\over \pi} \log^2{Q^2 \over m_e^2} \bigg|_{Q^2 \sim 1\,{\rm GeV}^2} \approx 0.5 \,.
\ee
Thus, first-order radiative corrections are insufficient for percent-level accuracy.  

When $Q\sim E\sim E^\prime$ and $m_e \sim \Delta E$,  
the leading series of logarithms $\alpha^n \log^{2n}(Q^2/m_e^2)$ are resummed by making
in (\ref{eq:radxs}) the replacement,  
\be\label{eq:resum}
1+\delta \to \exp( \delta ) \,.  
\ee
Two-loop corrections without logarithmic enhancement are below the relevant experimental precision.
For definiteness, in our analysis of the Mainz data,
we employ the prescription used in Ref.~\cite{Bernauer:2013tpr},
exponentiating all first-order corrections in (\ref{eq:resum})
except the finite TPE contributions.  
We return to a discussion of deficiencies in this treatment in Sec.~\ref{sec:largelog}. 

\subsection{Summary of experimental implementations \label{sec:radexpt}}

Apart from TPE and, in some cases, large log resummation, radiative corrections have
already been applied to all of the cross sections
we include in our fit, as part of the original analysis of the experiments.
We will examine the impact of different TPE prescriptions, with final results based
on the SIFF sum of monopoles TPE correction as in Eq.~(\ref{eqn:SIFFBlun}) and Table~\ref{tab:tpeblucoefs}.
Possible deficiencies in radiative corrections are treated at the same level as experimental
systematic errors.

Consider first the Mainz data set. 
The A1 Collaboration's data analysis applied radiative corrections based on the prescription of
Refs.~\cite{Maximon:2000hm, Vanderhaeghen:2000ws}, as detailed in Ref.~\cite{Bernauer:2013tpr}.
This includes TPE corrections using the Feshbach prescription~\cite{McKinley:1948zz} and the large
log resummation given in Eq.~(\ref{eq:resum}) above (excluding the finite TPE contribution from the exponentiation).
In the analysis of correlated systematic uncertainties,
the cutoff on the bremsstrahlung tail was varied, yielding rms cross section variations
well below 0.1\%.  These variations in the cross sections were
used to determine the impact of radiative correction uncertainties on the radius.
No uncertainty was included in the cross sections for the TPE contribution.

Consider now the world data set.
The world data come from many different experiments, and the details of the radiative 
corrections vary. However, they are all based on the general formalism
of Mo and Tsai~\cite{Tsai:1961zz, Mo:1968cg, Tsai:1971qi}, with improvements and modifications 
added in later works, e.g., Refs.~\cite{Walker:1993vj, Ent:2001hm}. Our compilation of world data
comes from Ref.~\cite{Arrington:2003ck}, along with additional low $Q^2$ and more recent cross
section~\cite{Frerejacque:1965ic, Ganichot:1972mb, Qattan:2004ht, Christy:2004rc} and recoil
polarization or polarized target measurements~\cite{Milbrath:1997de, Gayou:2001qt, Pospischil:2001pp, Strauch:2002wu,
MacLachlan:2006vw, Jones:2006kf, Crawford:2006rz, Puckett:2010ac, Zhan:2011ji}, with
earlier polarization transfer results~\cite{Jones:1999rz, Gayou:2001qd, Ron:2007vr} replaced by the
results of final, updated analyses~\cite{Punjabi:2005wq, Puckett:2011xg, Ron:2011rd}.
Further details of radiative corrections, in particular for earlier experiments,
are presented in Ref.~\cite{Arrington:2003df}. Our compilation includes the corrections applied to earlier
measurements discussed in that work; furthermore, we include additional vacuum polarization terms; exclude 
small angle data, $\theta<20^\circ$, from Ref.~\cite{Walker:1993vj}; and include separate normalization factors for
data taken with different detectors or under very different conditions~\cite{Goitein:1970pz, Bartel:1973rf, Andivahis:1994rq}.

The original publications of the experiments comprising the world data set
did not apply TPE corrections, and different prescriptions were used to
approximate Eq.~(\ref{eq:resum}).
For the most part, these experiments quoted normalization and uncorrelated uncertainties
of 0.5--1\% each to account for uncertainties in the radiative corrections applied, dominated by uncertainty
associated with TPE corrections.%
\footnote{
  While this turned out to be smaller than the size of TPE corrections in recent
calculations~\cite{Carlson:2007sp, Arrington:2011dn}, it appears to be a significant overestimate
of the residual uncertainty at lower $Q^2$ values, based on the consistency between low-$Q^2$
estimates of the corrections~\cite{Arrington:2012dq}.
  }
In this case, we will apply TPE corrections
similar to those applied in the previous global analysis~\cite{Arrington:2007ux}, for which the errors
assigned in previous experiments were taken to be sufficient to account for uncertainties after applying
a hadronic calculation of the TPE corrections. Note that one experiment~\cite{Simon:1980hu} did not include
uncertainties associated with these corrections and so had much smaller total uncertainties than other
experiments. Following Ref.~\cite{Arrington:2007ux}, we thus include an additional systematic uncertainty
to the data of Ref.~\cite{Simon:1980hu}: 
we increase the normalization uncertainty by 1\% (to a total of 1.5\%)
and add 0.5\% in quadrature to the point-to-point uncertainty.

In Sec.~\ref{sec:results}, we will include constraints on 
the form factor ratio $G_E/G_M$ from polarization measurements.
In the kinematic range considered, with $Q^2\le 1\,{\rm GeV}^2$,
the TPE correction, estimated from a simple hadronic
model~\cite{Blunden:2005ew}, is small compared to experimental errors.%
\footnote{
  The hadronic model for TPE corrections~\cite{Blunden:2005ew} predicts a
  correction not larger than $\sim 0.5\%$ over the full $\epsilon$ range for $Q^2\le 1\,{\rm GeV}^2$.
  Furthermore, the 41 data points with $Q^2\le 1\,{\rm GeV}^2$
  are concentrated at large epsilon, $\epsilon \approx 0.7 \pm 0.2$,
  where the TPE correction model predicts a correction $\lesssim 0.2\%$. 
}
Following Ref.~\cite{Arrington:2007ux}, this model-dependent correction is thus
omitted from the fits.  We will find that the polarization data
points do not have a strong influence on the radius fits, and thus do not pursue
a more detailed treatment of radiative corrections to these data points. 

\section{Updated fit of the Mainz data set \label{sec:defaultfit}}

In this section, we extract the charge and magnetic radii from the Mainz data set, retaining the original
treatment of statistical and systematic uncertainties and correction factors from
Ref.~\cite{Bernauer:2013tpr} but incorporating
our knowledge of the structure of the form factors as presented in Sec.~\ref{sec:z}. We first reproduce
the Mainz polynomial and inverse polynomial fits and then provide an updated extraction using
the bounded $z$ expansion. To highlight differences in the theoretical treatment of the form factors,
we fit to the full data set (1422 points) and apply the Feshbach correction as the only TPE correction,
as was done in the primary radius extraction from Ref.~\cite{Bernauer:2013tpr}.
We then discuss the impact of moving from polynomial fits to 
fits using the bounded $z$ expansion and comment on
other attempts to extract the radii from the Mainz data.

Note that, because of the way the data and uncertainties are parametrized
for the Mainz data, the uncertainties from such a fit represent only
part of the total uncertainty. Meaningful error estimates
require the examination of correlated effects arising, e.g., from experimental
systematic errors and radiative corrections.
In this section, we focus on how the improved form factor parametrization modifies the extracted
radii and fit uncertainties. 
Section~\ref{sec:systematics} will include an examination of the corrections applied to the Mainz
data and the treatment of systematic uncertainties presented in their analysis.

We take the cross section data as provided in the Supplemental Material of Ref.~\cite{Bernauer:2013tpr},
which includes the Feshbach correction for TPE, and scaling of the statistical uncertainties to account
for unidentified systematic errors, as discussed in Sec.~\ref{sec:systematics}. We allow the 
normalization parameters to float freely in the fit, in accordance with Ref.~\cite{Bernauer:2013tpr}.
In addition to examining the full Mainz data set, we also provide results obtained by restricting to momentum transfers
below a given $Q^2_{\rm max}$. We use the $\chi^2$ function
\be
 \chi^2_\sigma = \sum_{i=1}^{N_\sigma} \frac{(\sigma_i - \sigma_{i,\text{fit}}/\eta_{i,\text{fit}})^2}{\delta \sigma_i^2} \,.
\label{eqn:chi2_mainz}
\ee
Here, $N_\sigma$ is the number of cross section points for a specified kinematic cut $Q^2_{\text{max}}$,
$\sigma$ is the measured cross section (after accounting for radiative corrections),
$\delta \sigma$ is the (point-to-point, uncorrelated) uncertainty, $\sigma_{\text{fit}}$ is the cross section
calculated using the chosen form factor model, and $\eta_{\text{fit}}$ is a product of normalization
parameters for a given run (i.e., for data taken at a given choice of angle and energy).
There are 31 normalization parameters in the complete data set.

In our default fits, we enforce bounds on the form factor parameters by a $\chi^2$ penalty,
\be\label{eq:chib}
\chi_{\text{b}}^2 = \sum_{i=1}^{k_\text{max}} \left(
\frac{a_{i, \text{fit}}^2}{|a_k|_{\text{max}}^2} + \frac{b_{i, \text{fit}}^2}{|b_k|_{\text{max}}^2}
\right)
\,,
\ee
where $a_{i, \text{fit}}$ and $b_{i, \text{fit}}$ are the fit
values of the coefficients for $G_E^p$ and $G_M^p$,
respectively, and $|a_k|_{\text{max}}$ and $|b_k|_{\text{max}}$
are (Gaussian) bounds on the coefficients. 
For the polynomial and unbounded $z$ expansion fits,
$|a_k|_{\rm max}$ and $|b_k|_{\rm max}$ are taken to be very large, acting simply as numerical
regulators in the fits (they are taken large enough such that fit results represent the infinite bound limit).
For the bounded $z$ expansion, Eq.~(\ref{eq:chib}) 
enforces a Gaussian, vs sharp cutoff, statistical prior on the form factor parameter space,%
\footnote{For a related discussion, see Ref.~\cite{Lepage:2001ym}. See also Ref.~\cite{Schindler:2008fh}.}
typically
taken to be $|a_k/a_0|_{\rm max} = |b_k/b_0|_{\rm max} = 5$.
A more detailed discussion of the dependence of fit results on form factor
priors is postponed to Sec.~\ref{sec:consistchecks}.

\subsection{Polynomial and inverse polynomial fits}

\begin{table}[htb]
\caption{\label{tab:results_poly}
  Results for fits using polynomials of degree $10$ 
  and inverse polynomials of degree $7$
  for the full ($N_\s = 1422$) A1 MAMI data set. The reduced $\chi^2$ is calculated taking
  $N_\text{dof} = N_\s - 2k_\max - N_{\rm norm}$ with $N_{\rm norm}=31$.
}
\begin{ruledtabular}
\begin{tabular}{lcllc}
Fit type & \multicolumn{1}{c}{$r_E$ [fm]} & \multicolumn{1}{c}{$r_M$ [fm]} & $\chi^2$ & $\chi^2_\text{red}$
\\ \hline
   poly 10    & 0.886 & 0.794 &~1561.6~& 1.14
   \\
   inv poly 7 & 0.886 & 0.768 &~1569.1~& 1.14
\end{tabular}
\end{ruledtabular}
\end{table}

The radius central values, minimum $\chi^2$, and reduced $\chi^2$ are displayed
in Table~\ref{tab:results_poly} for fits with form factors represented as polynomials in
$q^2$ of degree 10, or as inverse polynomials of degree 7.
These results are very close, but not identical, to the corresponding results in Table~IV and Fig.~20 of
Ref.~\cite{Bernauer:2013tpr}.
We have compared our results to the output from the example fitting code provided as part of the
Supplemental Material for Ref.~\cite{Bernauer:2013tpr}, finding agreement with the results of this code.
For example, in the case of the polynomial of degree 10, the results of the example fitting code agree with our
results in Table~\ref{tab:results_poly}, both having a minimum $\chi^2$ of 1561.6,
lower than the value 1563 quoted in Table~IV of Ref.~\cite{Bernauer:2013tpr}.%
\footnote{More precisely, the fitting code returned a $\chi^2$ of 1561.60 and $r_M=0.797\,{\rm fm}$.
  Evaluating our $\chi^2$ function with the corresponding parameters yielded an identical 1561.60.
  Using the same initialization conditions as the example fitting code,
  our minimization code independently returned a minimum $\chi^2$ of 1561.58 and $r_M=0.794\,{\rm fm}$,
  as displayed in Table~\ref{tab:results_poly}.
}

\subsection{Bounded $z$ expansion fits}

Let us proceed to consider the implications of the bounded $z$ expansion.
Here we retain the identical data set as employed in Table~\ref{tab:results_poly}.
For the default fit, we take $t_0=0$, $k_{\rm max}=12$,
and a Gaussian bound of $|a_k|_{\rm max} = |b_k|_{\rm max}/\mu_p = 5$. The value $k_{\rm max}=12$
is large enough that the result does not change if $k_{\rm max}$ is increased further.

\begin{figure}[tb]
\begin{center}
\includegraphics[width=0.49\textwidth]{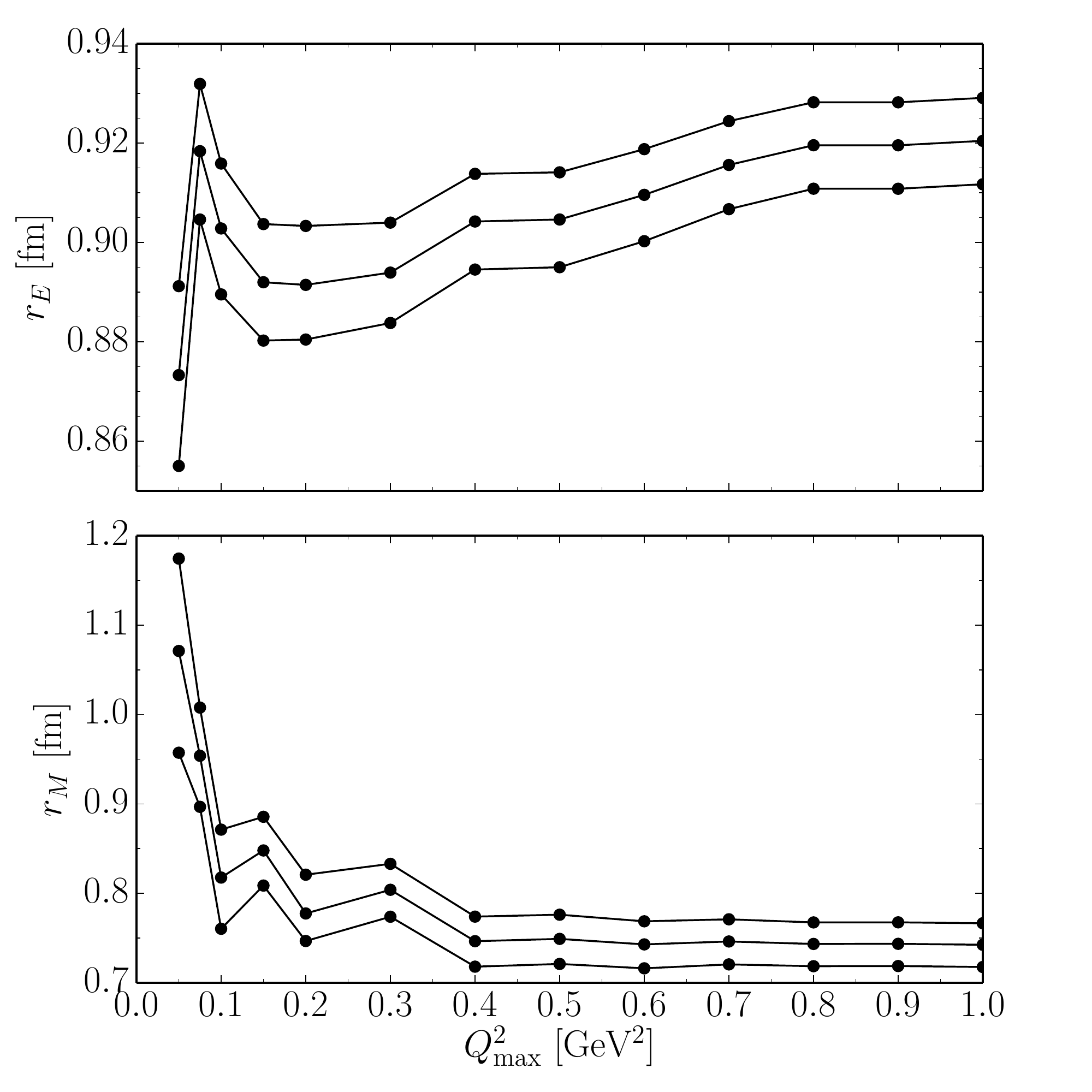}
\caption{Extracted electric (top panel) and magnetic (bottom panel) radii as functions of
the kinematic cut $Q^2_{\rm max}$ on momentum transfer for the 1422 point A1 MAMI data set, 
using the $z$ expansion with $t_0=0$, Gaussian priors with
$|a_k|_{\text{max}} = |b_k|_{\text{max}}/\mu_p = 5$, $k_{\text{max}} = 12$. 
One-$\sigma$ error bands are statistical only.
\label{fig:radvQ2max_mainz}
}
\end{center}
\end{figure}

The results for this fit are displayed in Fig.~\ref{fig:radvQ2max_mainz} as a function of $Q^2_{\rm max}$. 
The extracted radii and $\chi^2$ values are provided for three $Q^2_{\rm max}$ values in Table~\ref{tab:results_mainz}.
The quoted uncertainty includes only the statistical-type uncertainties, i.e.,
counting statistics and uncorrelated systematic uncertainties that are represented by the
rescaling of the statistical errors in the A1 data set. 
The uncertainty is obtained by varying the radius around the best-fit value,
refitting the data while allowing all data set normalizations to float, to map
out the $\chi^2$ contour as a function of radius. The contours are typically symmetric
and very nearly parabolic, and in the tables, we quote the 
average of the change in radius that yields $\Delta\chi^2=1$ on the high and low 
sides of the central value. Note that the primary A1 analysis of the Mainz data, identical except
for the choice of the fitting function, yielded~\cite{Bernauer:2013tpr} $r_E=0.879(5)_{\rm stat}$~fm
and $r_M=0.777(13)_{\rm stat}$~fm,
including only statistical uncertainties for comparison with our bounded $z$ expansion
results in Table~\ref{tab:results_mainz}.

\begin{table}[tb]
  \caption{Results from the fits in Fig.~\ref{fig:radvQ2max_mainz} for three values of
    $Q^2_{\rm max}$.
    $N_\sigma$ is the number of cross section points with $Q^2$ below $Q^2_{\rm max}$, and
    $N_{\rm norm}$ is the number of normalization parameters appearing in the data subset.
\label{tab:results_mainz}
}
\begin{ruledtabular}
\begin{tabular}{lllrrc}
$Q^2_{\rm max}$ $({\rm GeV}^2)$ & $r_E$ (fm) & $r_M$ (fm) & $\chi^2_{\rm min}$ & $N_\sigma$ & \!\!$N_{\text{norm}}$ \\
\hline
0.05 &~$0.873(18)$~&~$1.071(114)$ &  479.4~&  483~& 13 \\
 0.5 &~$0.905(10)$ &~$0.749 (28)$ &~1404.7~&~1285~& 29 \\
 1 &~$0.920(9)$  &~$0.743 (25)$ &~1605.5~&~1422~& 31
\end{tabular}
\end{ruledtabular}
\end{table}

\subsection{Discussion}

Let us remark on three aspects of the fits summarized in Table~\ref{tab:results_mainz}.
First, we remark that the bounded $z$ expansion fit to the entire 1422 point data set
($Q^2_{\rm max}=1\,{\rm GeV}^2$) yields an electric radius significantly larger than the
Mainz A1 extraction~\cite{Bernauer:2013tpr}. Having analyzed identical data sets, this difference
arises solely from requiring the form factors to lie within the class allowed by the bounded $z$
expansion. The difference, 0.041~fm, is large compared to the Mainz estimated systematic
uncertainty. The magnetic radii exhibit a smaller difference, with our
result 0.034~fm below the Mainz extracted value. 

Second, the extracted radii have significant dependence on $Q^2_{\rm max}$. For example,
$r_E=0.873(18)_{\rm stat}$~fm with
$Q^2_{\rm max}=0.05$~GeV$^2$ vs $r_E=0.920(9)_{\rm stat}$~fm with $Q^2_{\rm max}=1$~GeV$^2$.
The difference, $0.047$~fm, is again large compared to the quoted uncertainties.
Furthermore, there is a non-negligible variation of the $r_E$ central value as $Q^2_{\rm max}$
is increased above 0.5~GeV$^2$, even though the region below 0.5~GeV$^2$ includes
more than 90\% of the data points, and (as illustrated below in Fig.~\ref{fig:statQ2}) 
the data above 0.5~GeV$^2$ do not significantly impact the radius uncertainty.
In fits with unbounded parameters, it is not surprising that the extracted radius is sensitive to higher-$Q^2$ data
because the radius may change to provide a better fit to fluctuations in the data that
are accommodated by arbitrarily large parameter values.
This behavior is unexpected in fits with bounded parameters. 
Thus, it is surprising that the small amount of higher-$Q^2$ data
has such a significant impact on the extracted radius.
The dependence on $Q^2_\max$ suggests a possible tension between the lower- and higher-$Q^2$ data.

Third, taking at face value the complete 1422 point data set and error assignments,
the resulting electric radius is $r_E=0.920(9)_{\rm stat}(6)_{\rm other}$~fm,
where for the moment we simply take the A1 evaluation of other contributions to the uncertainty.%
\footnote{The error $(6)_{\rm other}$ results from the quadrature sum for the errors
  $(4)_{\rm syst}(2)_{\rm model}(4)_{\rm group}$ presented in Ref.~\cite{Bernauer:2013tpr}.  These errors
  were added in quadrature in Ref.~\cite{Bernauer:2013tpr}, but it has more recently been
  advocated~\cite{Priv_Distler} to
  add the final error linearly to the quadrature sum of the first two, resulting in $\sim (8.5)_{\rm other}$. 
}
This result is 7$\sigma$ above the muonic hydrogen value,
$r_E=0.84087(30)$~fm~\cite{Antognini:1900ns}.  It is also in tension
with the results extracted from hydrogen spectroscopy, $r_E =
0.8758(77)$~fm~\cite{Mohr:2012tt}, and with a previous global
analysis~\cite{Zhan:2011ji} of world electron-proton scattering data which yielded
$r_E=0.875(10)$. The magnetic radius value of $r_M=0.743(25)_{\rm
  stat}(10)_{\rm other}$ is almost 4$\sigma$ from the value
$r_M=0.867(20)$~fm from the global analysis in Ref.~\cite{Zhan:2011ji}.
However, we note that recent global analyses~\cite{Zhan:2011ji,
  Sick:2012zz}  use different representations of the form factors
compared to the bounded $z$ expansion used in
Table~\ref{tab:results_mainz}.   In Sec.~\ref{sec:global} below,
we will perform our own analysis of the world data  for a
consistent comparison with the analysis of the Mainz data.

Simply replacing the fit functions employed in Ref.~\cite{Bernauer:2013tpr}
with the $z$ expansion does not resolve the discrepancy with muonic hydrogen results.
In fact, the result is a larger difference with muonic hydrogen, as well as a
tension with previous extractions from world electron-proton scattering data.
In addition, the results show an unexpected dependence on the $Q^2$ range of
data included in the fit.  In the following Sec.~\ref{sec:systematics}, we consider in detail
a range of sources of the systematic error before presenting best values for the radii.

\subsection{Further tests related to the $z$ expansion}

\subsubsection{Dependence on $k_{\rm max}$ \label{sec:kmaxdepend}}

\begin{figure}[htb]
\begin{center}
\includegraphics[width=0.99\linewidth]{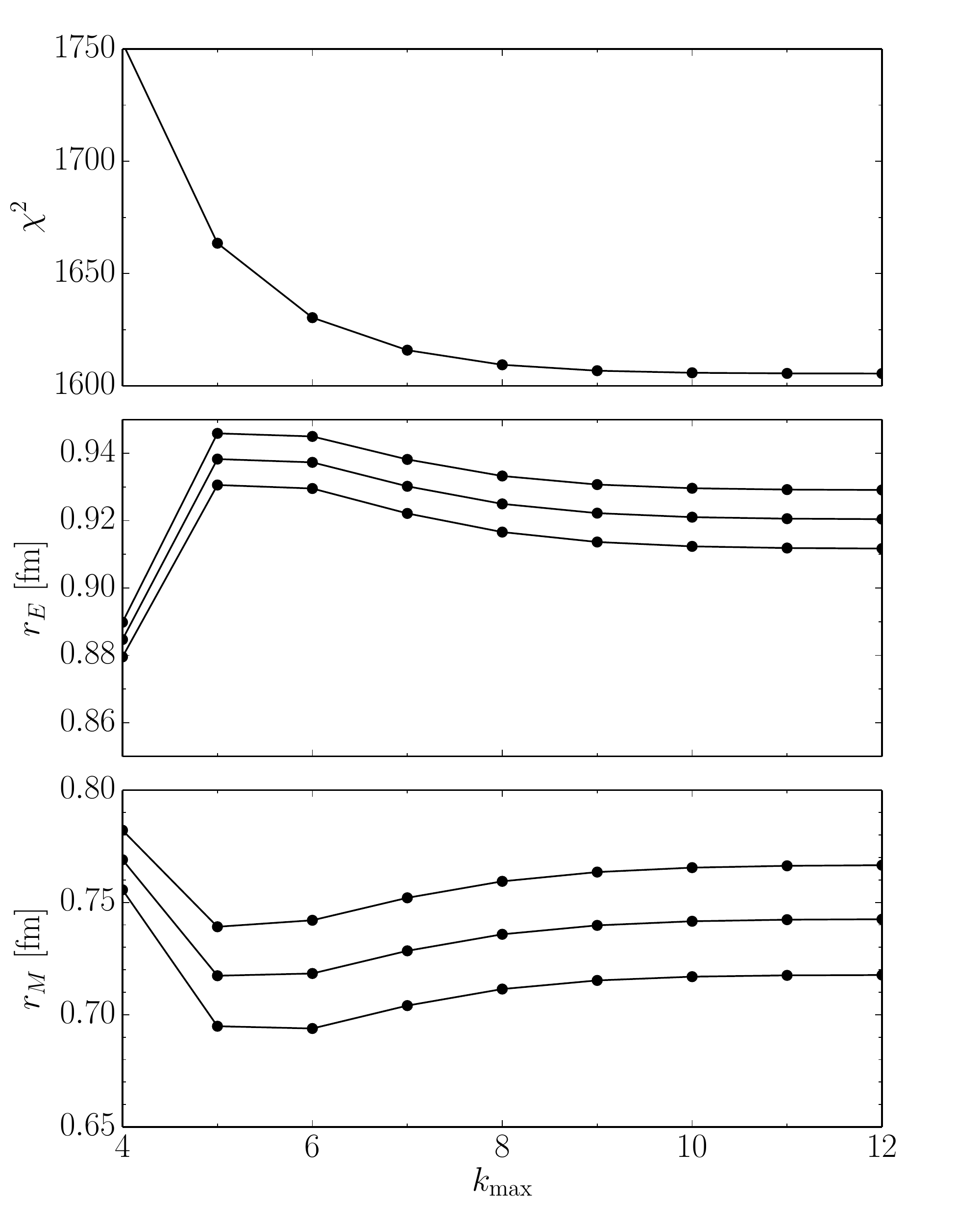}
\caption{Total $\chi^2$ (top panel) and extracted electric (middle panel) and magnetic (bottom panel) radii as
  functions of $k_{\rm max}$ for the 1422 point A1 MAMI data set,
  using the $z$ expansion with $t_0=0$, Gaussian priors with $|a_k|_{\text{max}} = |b_k|_{\text{max}}/\mu_p = 5$.
  One-$\sigma$ error bands are statistical only. 
\label{fig:radvkmax_mainz}
}
\end{center}
\end{figure}

In the bounded $z$ expansion, we may estimate the maximum power of $z$
which can impact the data at a given level when the expansion coefficients
$a_n$ are order unity. Setting the upper limit of the contribution at the
level of $\sim 0.5$\% implies $k_{\rm max} \approx 10$ should be sufficient
for $t_0=0$, $Q^2_{\rm max}=1$~GeV$^2$.
Figure~\ref{fig:radvkmax_mainz} shows the $\chi^2$ values
and radii extracted as a function of $k_{\rm max}$ for the bounded $z$
expansion fit to the full Mainz data set.  The rightmost points at
$k_{\rm max}=12$ correspond to the rightmost points in
Fig.~\ref{fig:radvQ2max_mainz} and to the final row of
Table~\ref{tab:results_mainz}.  In accordance with our power counting
estimate, the minimum $\chi^2$ value and extracted radii have
stabilized by $k_{\rm max}=10$.  For definiteness, we choose
$k_{\rm max}=12$ for all of our bounded $z$ expansion fits. 
While this is significantly more parameters than required for fits to
smaller values of $Q^2_{\rm max}$, the bounds on the fit parameters prevent
the problem of radius instability
due to overfitting of noise in the higher-$Q^2$ data~\cite{Arrington:2015yxa}.

\subsubsection{Unbounded $z$ expansion fits}

\begin{figure}[htb]
\begin{center}
 \includegraphics[width=0.99\linewidth]{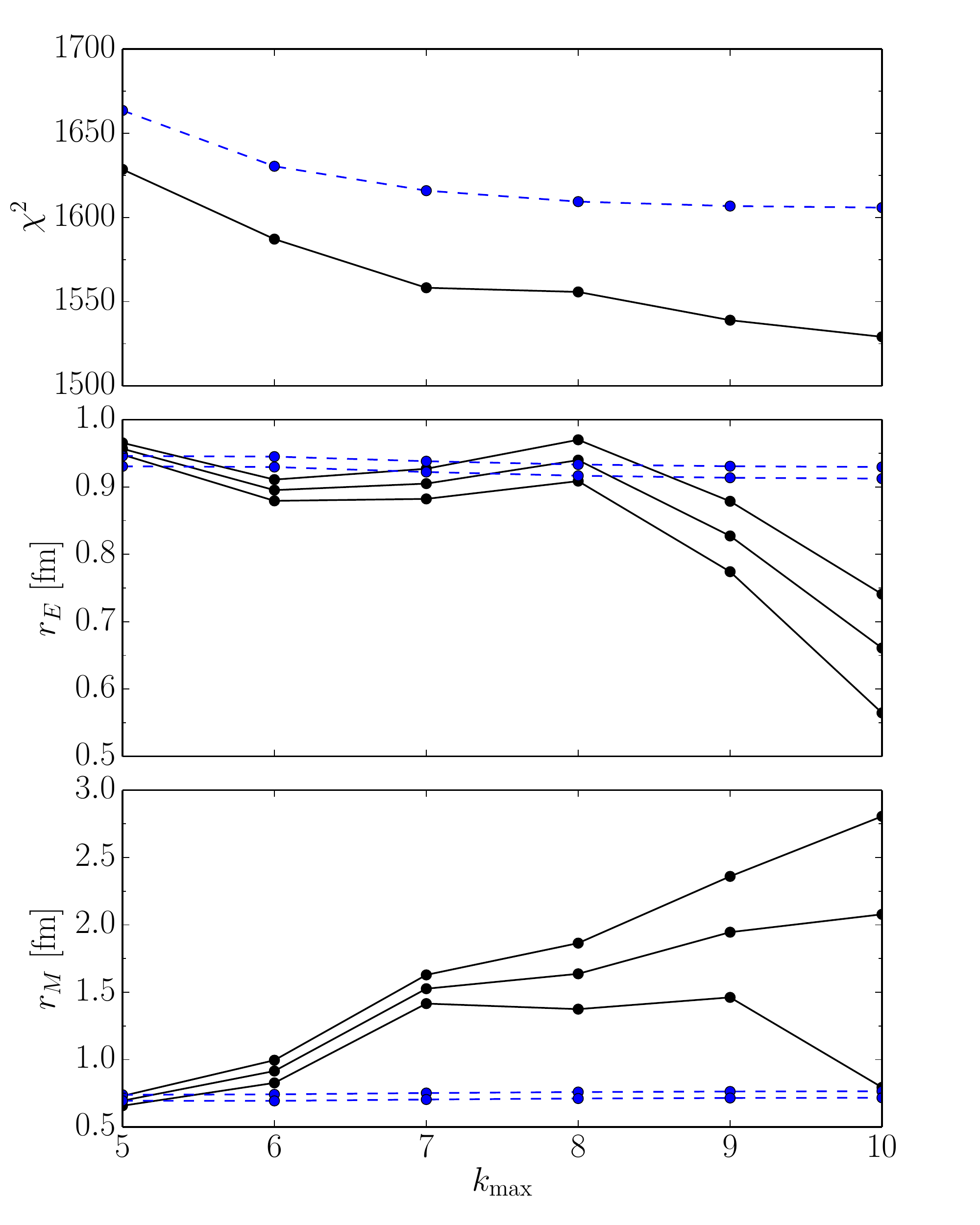}
 \caption{$\chi^2$ (top panel, solid black line) and extracted electric (middle panel, solid black lines)
   and magnetic (bottom panel, solid black lines) radii with $1\sigma$ statistical error bands as functions of
   $k_{\text{max}}$ for the unbounded $z$ expansion fit with $t_0=0$
   to the 1422 point A1 MAMI data set (with floating normalization).
   The dashed blue lines show  the $\chi^2$ and $1\sigma$ error bands from
   the bounded fit in Fig.~\ref{fig:radvkmax_mainz} for comparison.
\label{fig:radvkmax_unbounded_mainz}
}
\end{center}
\end{figure}

The bounded $z$ expansion (the formal $k_{\rm max}\to\infty$ limit
with bounded coefficients) is a particularly well-motivated
implementation of form factor priors.  A different but common choice of priors
corresponds to setting $a_{k} = 0$ for all coefficients beyond a given
order $k>k_{\rm max}$, with the remaining coefficients unconstrained, and
$-\infty <  a_{k} < \infty$ for $k\le k_{\rm max}$.  We perform some
illustrative fits with this modified choice of priors in order to
separate the impact of applying bounds from the impact of changing
from polynomial or inverse polynomial functions to the $z$ expansion.
We again fit the 1422 point A1 data set using the same rescaled errors
and Feshbach TPE correction as in Fig.~\ref{fig:radvQ2max_mainz}, but
now set $|a_k|_{\rm max}, |b_k|_{\rm max} \to \infty$ in
Eq.~(\ref{eq:chib}).

In the limit of large $k_{\rm max}$, the true form factors are guaranteed
to lie in the space of curves described by the unbounded $z$ expansion.
However, many badly behaved form factors (in particular, form factors
in conflict with predictions of QCD, as discussed in Sec.~\ref{sec:z})
also lie in this space of curves, and fits without
constraints on the coefficients lose predictive power at large $k_{\rm max}$.

Figure~\ref{fig:radvkmax_unbounded_mainz} shows results for unbounded
fits with floating normalizations.  The minimum $\chi^2$ value
continues to decrease significantly as  parameters are added through
$k_{\rm max}=10$.
Quantitatively reliable radius estimates are difficult
to obtain from such a fit; for small $k_{\rm max}$, omitted terms in the form factor
expansion can introduce a potentially large, but difficult to quantify, bias
in the fitted radii~\cite{Kraus:2014qua}, while for large $k_{\rm max}$, the
uncertainties grow rapidly.

\subsubsection{Fixed-normalization fits}

\begin{figure}[htb]
\begin{center}
  \includegraphics[width=0.99\linewidth]{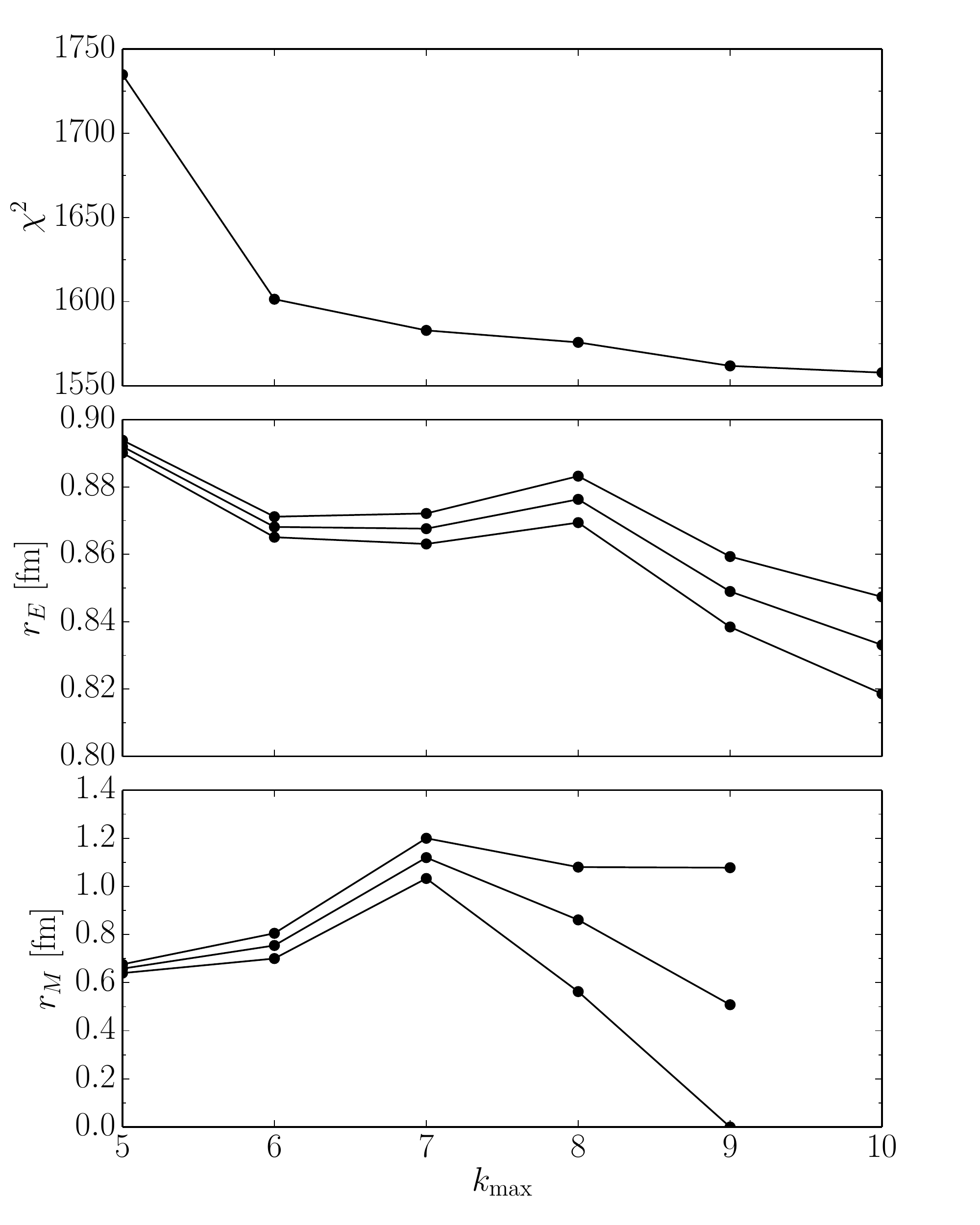}
  \caption{$\chi^2$ (top panel) and extracted electric (middle panel) and magnetic (bottom panel) radii
    with $1\sigma$ statistical error bands as functions of $k_{\text{max}}$ for the unbounded $z$ expansion fit 
    with $t_0=0$ to the 1422 point A1 MAMI data set with fixed normalization parameters.
  }
\label{fig:radvkmax_fixnorm_mainz}
\end{center}
\end{figure}

As noted earlier, the manner in which uncorrelated systematic
uncertainties are treated in the Mainz data set is only complete
when the normalization parameters are allowed to vary and when
correlated systematic uncertainties are estimated
separately~\cite{Bernauer:2013tpr}.  Thus, fits which fix the
normalization of the data based on the default Mainz fit will
underestimate uncertainties and potentially yield different values
for the radii if the fit is performed with a different functional form
than that used to determine the normalization parameters.
Nonetheless, such fits have been performed~\cite{Lorenz:2014vha, Carlson:2015jba}, 
and so we provide a comparison of unbounded fits with and without floating normalization factors. 

The result for fixed normalization factors is displayed in
Fig.~\ref{fig:radvkmax_fixnorm_mainz}.  Comparing to fits with
floating normalizations, one can see that the uncertainties are
significantly smaller in the case of fixed normalizations, with the
fit uncertainties on $r_E$ underestimated by a factor $\gtrsim 5$.%
\footnote{In detail, for $k_{\rm max} = 9 (10)$, the radius
  uncertainty is $\delta r_E = 0.011 (0.014)$ for fixed
  normalizations, compared to $\delta r_E = 0.053 (0.096)$ for
  floating normalizations.}

Even ignoring the artificially small uncertainties that arise when
neglecting the normalization uncertainty of the data sets, it is not
clear that there is any value of $k_{\rm max}$ for which the fit provides a sufficiently precise description 
of the data while still providing meaningful uncertainties on the charge radius. 
For the magnetic radius, the results are even
less clear, with only an upper limit on the radius possible for
$k_{\rm max} \geq 9$.  We note that for large $k_{\rm max}$ the
$r_E$ central values for the fits displayed in Figs.~\ref{fig:radvkmax_unbounded_mainz}
and~\ref{fig:radvkmax_fixnorm_mainz} require very large coefficient
values, in violent conflict with order unity predictions of QCD.%
\footnote{For example, at $k_{\rm max}=9$, requiring the central
  value for $r_E$  in Fig.~\ref{fig:radvkmax_fixnorm_mainz} to lie
  within the $1\sigma$ envelope of a bounded $z$ expansion requires
  coefficient bound $\gtrsim 10^4$.   }

\section{Systematic studies for the Mainz data set \label{sec:systematics}}

Taking the data and error prescriptions of Ref.~\cite{Bernauer:2013tpr}
at face value, we have found radius extractions in tension with each other
for fits of different functions to the same data set (Table~\ref{tab:results_poly}
compared to the last line of Table~\ref{tab:results_mainz}) and for fits of the
same function to subsets of the same data set (Fig.~\ref{fig:radvQ2max_mainz}
and Table~\ref{tab:results_mainz}).

The first observation indicates the strong dependence of the extracted radius on the
specification of physical form factors, as implemented by the bounded $z$ expansion.
We focus solely on this class of form factors in the following.
Since the systematic error analysis of Ref.~\cite{Bernauer:2013tpr} relied on
rescaling statistical errors based on fits to a particular class of form factor models,
we also revisit this analysis using the bounded $z$ expansion.

The second observation, that fits to data subsets are in tension with fits to the
entire data set, indicates the possibility of an underestimated systematic
error. We investigate below a range of correlated errors and their potential impact on
radius extractions.

The analysis of the Mainz data by the A1 Collaboration decomposed the
uncertainties into several different contributions.
Let us briefly review this decomposition. 
The only uncertainty applied directly to the quoted cross sections is called
the statistical uncertainty. This is a combination of counting
statistics and systematic uncertainties which are taken to be
uncorrelated between different data points and normally distributed 
and is thus treated in  the same way as counting statistics. We refer to
the systematic uncertainties that are independent for different data
points as the uncorrelated systematic uncertainties. The A1
Collaboration includes these uncorrelated systematic uncertainties by
introducing  a rescaling factor on the counting statistics, with a
procedure to extract these scaling factors which we summarize below in
Sec.~\ref{sec:rescale}.

Normalization uncertainties for each data subset in the experiment are
accounted for in the extraction of the radius by allowing the 31
normalization factors corresponding to different configurations to
float freely when fitting the form factors. The A1 analysis of
Ref.~\cite{Bernauer:2013tpr} suggests an uncertainty of 3.5--5\% on the
normalization factors, but no constraints were included in the fits. 
Because the cross sections are quoted after the determination of
the normalization factors in their fit, any information on the initial
normalizations is lost, and it is no longer possible for us to make use
of even the limited precision with which these normalization factors
were constrained.

Finally, the A1 Collaboration gives a procedure for estimating the
impact of correlated systematic uncertainties on their data. These are
corrections which potentially have a strong kinematic dependence, and
their impact will not necessarily decrease when one includes a large
number of measurements. Such corrections must be treated
independently from the statistical and uncorrelated systematic errors.
The Mainz treatment of these uncertainties and our examination of
other possible correlated effects are included in Sec.~\ref{sec:corrsys}. 

In the remainder of this section, we introduce the following three modifications to
the analysis. First, in Sec.~\ref{sec:TPE}, 
we study the impact of different TPE correction models on
radius extractions, choosing the SIFF sum of monopoles ansatz as the default in the remaining fits.
Second, after identifying in Sec.~\ref{sec:rescale}
potential shortcomings in the rescaling of statistical errors,
in Sec.~\ref{sec:rebin},
we rebin data taken at identical kinematic settings in order to
incorporate in Sec.~\ref{sec:constsyst}
uncorrelated systematic errors which do not scale with statistics.
Lastly, in Sec.~\ref{sec:corrsys},
we consider a range of correlated systematic errors consistent
with the experimental precision achieved in Ref.~\cite{Bernauer:2013tpr}.

\subsection{TPE model dependence\label{sec:TPE}}

\begin{figure}[htb]
\begin{center}
\includegraphics[width=0.99\linewidth]{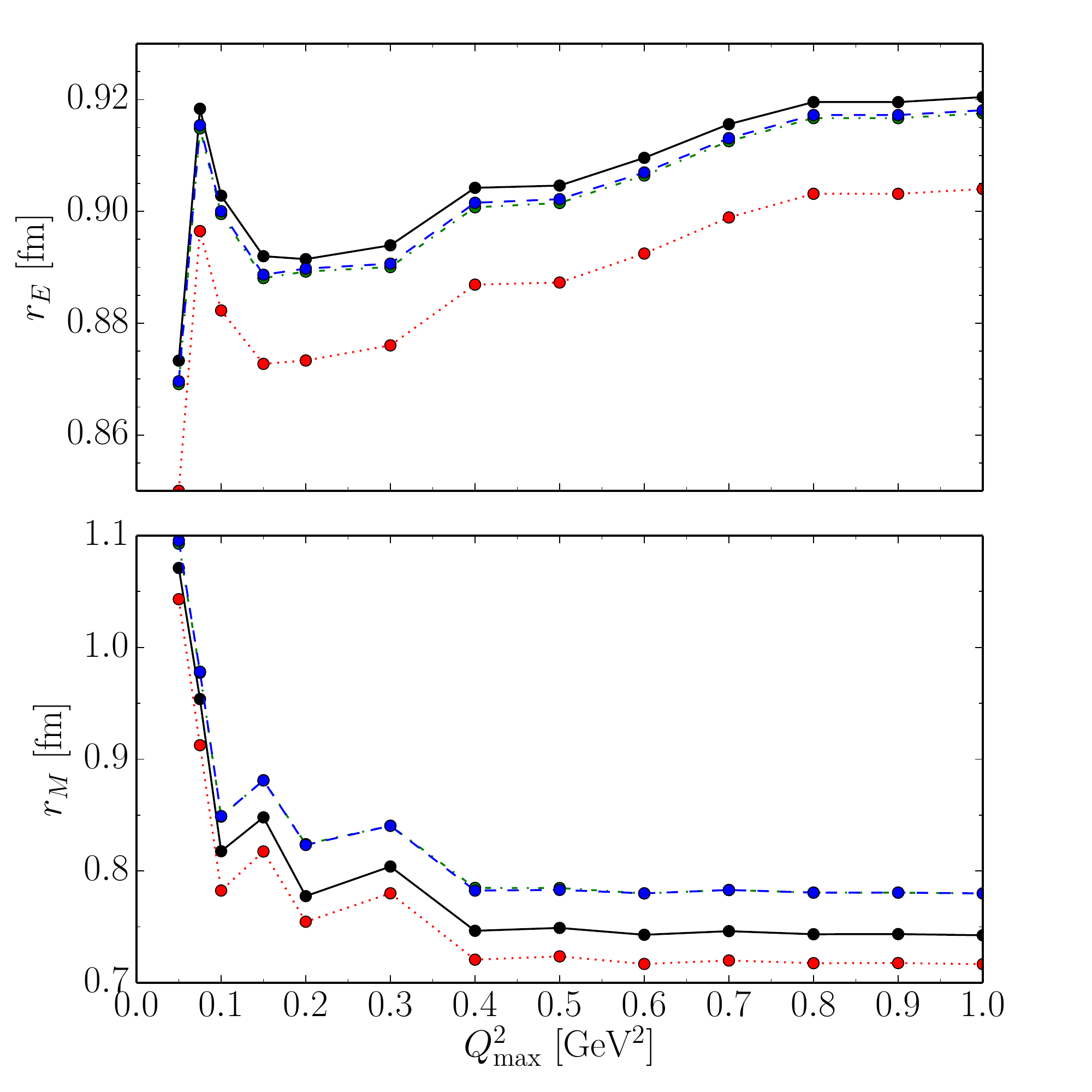}
\caption{
\label{fig:radcorr_mainz}
Extracted electric (top panel) and magnetic (bottom panel) radii as functions of
the kinematic cut $Q^2_{\rm max}$ on momentum transfer for several TPE models,
as discussed in the text: no correction (red, dotted),
Feshbach correction (black, solid),
SIFF dipole (green, dot-dashed), and
SIFF sum of monopoles (blue, dashed).
There is a negligible difference between the SIFF choices of the dipole and the sum of monopoles.
Fits are to the 1422 point A1 MAMI data set, 
using the $z$ expansion with $t_0=0$, Gaussian priors with
$|a_k|_{\text{max}} = |b_k|_{\text{max}}/\mu_p = 5$, $k_{\text{max}} = 12$. 
}
\end{center}
\end{figure}

The variation of the extracted radii with different TPE models is illustrated
in Fig.~\ref{fig:radcorr_mainz}, where the extracted $r_E$ and $r_M$ central values are plotted
vs $Q^2_{\rm max}$. The black curve is identical to the central curve in Fig.~\ref{fig:radvQ2max_mainz}.
The remaining results are obtained by repeating the fit of Sec.~\ref{sec:defaultfit} after removing the
Feshbach TPE correction, Eq.~(\ref{eq:Fesh}),
and then applying the SIFF TPE result [using dipole form factors, Eq.~(\ref{eqn:SIFFdip}),
or those from Blunden \textit{et al.}, Eq.~(\ref{eqn:SIFFBlun})]
or applying no finite TPE correction [in the Maximon-Tjon convention, Eq.~(\ref{eq:MaTjzero})].
As the plot illustrates, expressed as a difference relative to the Feshbach correction, the
results have mild $Q^2_{\rm max}$ dependence. 
Numerical values for $Q^2_{\rm max}=0.05,\, 0.5,\, 1$~GeV$^2$ are given in Table~\ref{tab:TPE}.

\begin{table}[tb]
\caption{Change in the extracted charge and magnetic radii for three different TPE
  corrections, relative to the Feshbach correction applied in the Mainz analysis.
  Results are for the fit with $Q^2_{\rm max}=0.05,\, 0.5,\, 1\,{\rm GeV}^2$ in Fig.~\ref{fig:radcorr_mainz}.  
\label{tab:TPE}
}
\begin{ruledtabular}
\begin{tabular}{ccrr}
$Q^2_{\rm max}$ $({\rm GeV}^2)$ & Model & $\Delta r_E$ (fm) & $\Delta r_M$ (fm) \\
\hline
0.05 & Feshbach    &    $\equiv 0$   &    $\equiv 0$   \\
& SIFF dipole & $-0.004$ & $+0.022$ \\
& SIFF Blunden& $-0.004$ & $+0.025$ \\
& No TPE      & $-0.023$ & $-0.028$ \\
\hline
0.5 & Feshbach    &    $\equiv 0$   &    $\equiv 0$   \\
& SIFF dipole & $-0.003$ & $+0.036$ \\
& SIFF Blunden& $-0.002$ & $+0.034$ \\
& No TPE      & $-0.017$ & $-0.026$ \\
\hline
1 & Feshbach    &    $\equiv 0$   &    $\equiv 0$   \\
& SIFF dipole & $-0.003$ & $+0.038$ \\
& SIFF Blunden& $-0.002$ & $+0.037$ \\
& No TPE      & $-0.016$ & $-0.026$ \\
\end{tabular}
\end{ruledtabular}
\end{table}

The Feshbach correction is the exact result in the formal limit of infinite proton mass 
and is independent of the proton structure.  The exact result for arbitrary kinematics for a pointlike
proton~\cite{Arrington:2011dn} yields a correction that grows with
$Q^2$, approximately doubling the correction between $Q^2=0$ and
1~GeV$^2$.  However, calculations using either
hadronic~\cite{Blunden:2005ew} or partonic~\cite{Carlson:2007sp}
models to account for proton structure indicate that the correction
does not grow with increasing $Q^2$ but instead becomes smaller and
then changes sign. This is the behavior required to explain the
difference between the Rosenbluth and polarization measurements of $\mu_p
G_E/G_M$ for the proton~\cite{Arrington:2003qk} and has been recently
confirmed for $Q^2 \approx 1$--$1.5$~GeV$^2$ by comparisons of positron and
electron scattering from the proton~\cite{Adikaram:2014ykv, Rachek:2014fam}.

There is a significant difference in the charge radius between the
case of no TPE corrections and either the Feshbach or SIFF
corrections.  However, there is a relatively small difference between
Feshbach and SIFF, suggesting that the infinite proton mass limit
provides a significant part of the correction for $r_E$.  For the
magnetic radius, there is a large difference between all three
approaches.
For both the charge and magnetic radii, there is  little sensitivity
to the choice of form factors included in the SIFF calculation.  We
collect in Table~\ref{tab:TPE} the deviations of the extracted radius
using different models in place of the Feshbach correction.  In all
subsequent fits we employ the SIFF ansatz, using for definiteness the
sum of monopoles in Table~\ref{tab:tpeblucoefs} as our default TPE
model. The uncertainty associated with TPE corrections will be 
incorporated into the evaluation of correlated systematic uncertainties
in Sec.~\ref{sec:corrsys}.

\subsection{Uncorrelated systematic uncertainties}

\subsubsection{Summary of the Mainz A1 approach \label{sec:rescale}}

To estimate the uncorrelated systematic uncertainties, the A1
Collaboration performed a fit to the entire 1422 point data set using a
default form factor model (an eight-parameter cubic spline model for each
of $G_E$ and $G_M$).  The data were then grouped according to the beam
energy and the spectrometer used in the measurement.  For each data
group, the uncorrelated systematic uncertainties were taken from
examination of the distribution of the differences between measured and
fit cross sections, scaled by the uncertainty from counting
statistics. (If the counting statistics fully represented the
uncorrelated uncertainties, then this should be a Gaussian
distribution with width one.) This distribution was fit with a
Gaussian, the width of which was then taken as the scaling factor applied to
the counting statistics to determine the combined statistical and
systematic uncorrelated uncertainties. The scaling factors obtained
in this way vary from 1.070 to 2.283, as given in the Supplemental
Material of Ref.~\cite{Bernauer:2013tpr}. 

This rescaling procedure is meant to yield
a reduced $\chi^2$ close to unity when the data are compared to the
original fit.  However, because the Gaussian fit may underestimate the
impact of outliers and the scaling of the uncertainties changes the
relative weighting of the different data sets, the fit to the data set
with updated uncertainties yields a reduced $\chi^2$ somewhat larger
than unity: $\chi^2_{\rm red} \approx 1.15$ for the entire
data set. This suggests that the quoted systematics are somewhat underestimated.

The rescaled statistical errors represent the
minimum additional uncertainty necessary to account for random
scatter around a global fit to the data. 
Any correlated effects will not be included in the extracted uncertainties.
For example, the A1 analysis of Ref.~\cite{Bernauer:2013tpr} does not include any uncertainty associated
with the error in the measurement of the beam energy or offsets in the spectrometer
angles. Such kinematic offsets would yield correlated errors in the cross
sections which would not be captured by this procedure.
  
The A1 rescaling procedure yields systematic uncertainties which
depend on the form factor model used in the fit. We performed a
similar analysis using our bounded $z$ expansion and using the
$\chi^2$ value for each data subset relative to the fit as the square
of the scaling factor. This procedure yielded similar scaling factors,
larger by 6\% on average compared to the A1 procedure (thus yielding a
value of $\chi^2_{\rm red}$ closer to unity), with a typical scatter
around the small average offset of approximately 10\%. Thus, the form factor model
dependence of this rescaling procedure is small, though not
negligible, and related more to the difference in our procedure than
to the change in the fitting function.

The rescaling procedure accounts for undetermined systematic errors that are
assumed to be uncorrelated and to scale with the statistical counting errors. 
However, it is not clear that all the uncorrelated systematics should scale with the statistical uncertainties. 
If one assumes that the statistical and uncorrelated systematic
uncertainties add in quadrature, then one can compare the original (unscaled) and rescaled statistical 
uncertainties to extract the effective uncorrelated systematic uncertainty used in the A1 procedure.  
This inferred uncertainty can be extremely small, as low as 0.05\%, but varies with
kinematics and with the counting statistics, attaining values up to 2\%.
Because the full set of 1422 data points includes many instances of repeated
measurements at identical kinematics, this procedure implies even
greater reduction in the systematic uncertainty associated with each
independent kinematic point, with values as low as 0.02\%,
even though it is experimentally difficult to constrain uncertainties at that level.

To address these concerns, we present a modified treatment of the
uncorrelated systematic uncertainties where a fixed uncorrelated
systematic error is added to all points to account for unknown drifts
or corrections.

\subsubsection{Rebinning studies \label{sec:rebin}}

As noted above, the 1422 data points include many repeated
measurements at the same conditions.  One would expect that many
potential systematic errors would be identical for these points,
e.g. time-dependent efficiencies, rate-dependent corrections, 
or uncertainties in the beam energy or spectrometer angle settings.
Adding a fixed systematic to every one of the 1422
data points would underestimate the systematic uncertainty for data
points with many repeated measurements. Therefore, we begin by
combining data points taken with identical conditions, reducing the
data set from 1422 data points to 658 independent cross section
measurements.

We group (i.e., rebin) all data taken at identical kinematic settings,
using only the  uncertainties from counting statistics (i.e. removing
the A1 scaling factor).  We tested our assumption that the points
within the groups of repeated measurements  were consistent within
statistics by looking at the $\chi^2$ values and confidence levels for
every set of the rebinned data. There were 407 settings with multiple runs
taken under identical conditions, and the confidence-level
distribution for these sets is consistent with a uniform distribution
between 0 and 100\% except for a handful of outliers below 1\% confidence
level, indicating a nonstatistical scatter of the points being
combined. Most of these outliers involved scatter at the 0.1--0.2\%
level, which is contained within the systematic uncertainty we will
add to achieve $\chi^2_{\rm red}$ near unity. 
One setting---$E_{\rm beam}=315$~MeV, $\theta =30.01^\circ$, spectrometer C---had
a single measurement that deviated by $\sim$1.5\% from the two other
measurements at that setting, while the statistical uncertainties were
approximately 0.15\%. We excluded this set of points, yielding a
total of 657 independent cross section measurements when fitting the rebinned data.

We remark that normalization parameter 14 appears for only one point
in the rebinned 657 point data set (and two points in the original 1422
point data set).  Since the normalization parameters have to be allowed
to float freely in the fit, this data point has no impact, but for
definiteness, it is retained. Note that with the Mainz procedure of
applying scaling factors to the counting statistics the rebinning has
no impact on the fit, and the only change at this point would be due
to the exclusion of the one point after rebinning.  However, when
applying a more conventional constant uncorrelated systematic
uncertainty to all points,  the uncertainty is best applied to the
rebinned data points.

\subsubsection{Uncorrelated systematics for rebinned data \label{sec:constsyst}}

\begin{table}[htb]
\caption{Number of data points, reduced $\chi^2$, and confidence
  level for each combination of spectrometer (A, B, or C) and beam
  energy (in MeV) of the rebinned A1 MAMI data set.  
  Columns 4 and 5 give the results after the inclusion of a uniform 0.25\% uncorrelated systematic; 
  columns 6 and 7 give the results after the inclusion of the final 0.3\%--0.4\% uncorrelated systematic. 
  See the text for details.
  \label{tab:rebin}
}
\begin{ruledtabular}
\begin{tabular}{ccrldld}
  Spec. & Beam & $N_\sigma$ & $\chi^2_{\rm red}$ & \multicolumn{1}{c}{CL (\%)} & $\chi^2_{\rm red}$ & \multicolumn{1}{c}{CL (\%)}\\
  \hline
A & 180 &   29  &       0.59 &         96.1  & 0.46 & 99.4\\   
  & 315 &   23  &       0.54 &         96.4  & 0.44 & 99.1\\
  & 450 &   25  &       1.52 &          4.8  & 1.00 & 46.7\\
  & 585 &   28  &       1.54 &          3.4  & 1.03 & 42.8\\
  & 720 &   29  &       1.05 &         39.9  & 0.87 & 66.4 \\
  & 855 &   21  &       0.92 &         56.8  & 0.77 & 76.0\\
\hline
B & 180 &  61 &         0.85 &         79.8  & 0.65 & 98.3 \\  
  & 315 &  46 &         1.05 &         38.5 & 0.76 & 88.5 \\
  & 450 &  68 &         0.90 &         71.7 & 0.67 & 98.2 \\
  & 585 &  60 &         0.61 &         99.2  & 0.50 & 99.96 \\
  & 720 &  57 &         1.29  &          6.9 & 0.97 & 53.7 \\
  & 855 &  66 &         1.88 &          0.002 & 1.15 & 19.6\\
\hline
C & 180 &  24 &         0.88 &         63.3 & 0.68 & 88.0\\
  & 315 &  24 &         1.16 &         27.2 & 0.78 & 76.8\\
  & 450 &  25 &         1.53 &          4.3 & 1.08 & 35.9 \\
  & 585 &  18 &         0.83 &         66.3 & 0.65 & 86.4 \\
  & 720 &  32 &         1.11 &         30.2 & 0.90 & 62.3\\
  & 855 &  21 &         0.79 &         73.7 & 0.62 & 90.5\\
\end{tabular}
\end{ruledtabular}
\end{table}

With the rebinned data set in hand, we proceed to investigate the
inclusion of an uncorrelated systematic error that does
not scale with statistics. We add a fixed systematic uncertainty to
every data point and perform the bounded $z$ expansion fit with
$Q^2_{\rm max}=1\,{\rm GeV}^2$ and our default form factor scheme,
$t_0=0$, $k_{\rm max}=12$, and Gaussian bound $|a_k|_{\rm max}=|b_k|_{\rm max}/\mu_p = 5$. 
We varied the systematic uncertainty until we found a reduced $\chi^2$ value close to unity. 
This required a systematic uncertainty of approximately 0.3\%. 
We then examined the $\chi^2$ contribution from each of the 18
energy-spectrometer combinations to see if any of them had anomalously
large or small $\chi^2_{\rm red}$ values. While the spread of
$\chi^2_{\rm red}$ values was significant, many data subsets have a
relatively small number of points, and the only subset which was an
extreme outlier was the data from spectrometer B at $E_{\rm beam}=855$~MeV. 
We chose to increase the systematic uncertainty on
this data subset to 0.4\%, while keeping 0.3\% for all other data subsets.
The reduced $\chi^2$ and confidence levels for each data sub set
are displayed in Table~\ref{tab:rebin}. The total $\chi^2$ is 520.4 for 657
points, which might suggest that 0.3\% is a slight overestimate of the
uncorrelated systematic, but it is a small effect, with a 0.25\%
correction yielding a reduced $\chi^2$ above 1 by a similar amount.

Table~\ref{tab:results_rebin} shows the radius fit results for the rebinned Mainz data with
the statistical scaling factors from the original analysis replaced by the
constant 0.3\% systematic uncertainty (0.4\% for spectrometer B at 855 MeV beam energy).

\begin{table}[tbh]
\caption{Results for fitting of the 657 point rebinned A1 MAMI data set with 0.3\%--0.4\% uncorrelated systematic
uncertainties at three values of $Q^2_{\rm max}$ using the $z$ expansion with
 $t_0=0$, Gaussian priors with $|a_k|_{\text{max}} = |b_k|_{\text{max}}/\mu_p = 5$,
 $k_{\text{max}} = 12$. $N_\sigma$ is the number of cross section points with $Q^2<Q^2_{\rm max}$, and
$N_{\rm norm}$ is the number of normalization parameters appearing in the data subset.
\label{tab:results_rebin}
}
\begin{ruledtabular}
\begin{tabular}{cllrrc}
$Q^2_{\rm max}$ $({\rm GeV}^2)$ & $r_E$ (fm) & $r_M$ (fm) & $\chi^2_{\rm min}$ & $N_\sigma$ & \!\!$N_{\text{norm}}$ \\
\hline
0.05 & ~$0.856(27)$~ & ~$1.11(14)$~ &  ~110.5~ & ~176~ & 13 \\
0.5  & ~$0.895(14)$~ & ~$0.777(34)$~ & ~442.0~ & ~568~ & 29 \\
1  & ~$0.908(13)$~ & ~$0.767(33)$~ & ~520.4~ & ~657~ & 31
\end{tabular}
\end{ruledtabular}
\end{table}

This procedure introduces enough uncertainty to account for random
scatter of the points around the best-fit curve. However, any errors
that are correlated between multiple points will bias the fit, and
will not be fully reflected in this procedure, making the resulting
uncertainty estimate more of a lower limit. While the impact of
correlated uncertainties will be examined separately, these rely on
specific models for kinematic dependences of any additional
errors. The inclusion of an even larger uncorrelated uncertainty would
allow the data to account for a range of correlated errors, but the
reduced $\chi^2$ would end up significantly smaller than unity.  For illustration,
Table~\ref{tab:results_rebin_largersys} shows the results where we
apply a 0.5\% uncorrelated systematic uncertainty to every data point,
instead of the 0.3\%--0.4\% uncertainties in the previous fit.

\begin{table}[tbh]
  \caption{
Same as Table~\ref{tab:results_rebin}, but with 0.5\% uncorrelated systematic uncertainty. 
\label{tab:results_rebin_largersys}
}
\begin{ruledtabular}
  \begin{tabular}{cllrrc}
$Q^2_{\rm max}$ $({\rm GeV}^2)$ & $r_E$ (fm) & $r_M$ (fm) & $\chi^2_{\rm min}$ & $N_\sigma$ & \!\!$N_{\text{norm}}$ \\
\hline
0.05 &~$0.861(35)$~&~$1.05(18)$~&~48.7~&~176~& 13 \\
0.5  &~$0.891(18)$~&~$0.768(43)$~&~211.5~&~568~& 29 \\
1  &~$0.901(17)$~&~$0.758(42)$~&~250.3~&~657~& 31
\end{tabular}
\end{ruledtabular}
\end{table}

\subsection{Correlated systematic uncertainties \label{sec:corrsys}}

We now consider systematic errors that do not scale with statistical
errors but that are also correlated across data points. We begin by
examining the procedure of Ref.~\cite{Bernauer:2013tpr}. We then examine
modified evaluations of the radius uncertainty associated with the correlated
systematic uncertainties.

\subsubsection{Summary of the Mainz A1 approach}\label{sec:corrsysmainz}

In the A1 MAMI data set, each cross section  is accompanied by two
factors to account for systematic uncertainties. The first is due to
the bremsstrahlung energy cut and is estimated by varying the cut. 
The second is meant to account for efficiency changes, normalization drifts, 
variations in spectrometer acceptance, and background misestimations. 
This second class of systematics is investigated by applying a 
kinematic-dependent correction to the data. 
The complete data set is refit after multiplying or dividing the 
individual cross section ratios by the corresponding
factor for either the energy cut or correlated systematic error, and
the largest difference in radius obtained from multiplying or dividing is taken as the
uncertainty.  The total systematic uncertainty is then obtained by
summing in quadrature:
\be
\Delta r_{\text{syst}} = \sqrt{ (\Delta
  r_{\text{Ecut}})^2 + (\Delta r_{\text{corr}})^2 } \,.
\ee

\begin{table}[tb]
\caption{Results for changes in the radii under increases (upper value for each $Q^2_{\rm max}$)
  or decreases (lower value) in the energy loss cut.
  Fits are for the 657 point rebinned A1 MAMI data set with 0.3\%--0.4\% uncorrelated systematic
  uncertainties at three values of $Q^2_{\rm max}$ using the $z$ expansion with
  $t_0=0$, Gaussian priors with $|a_k|_{\text{max}} = |b_k|_{\text{max}}/\mu_p = 5$,
  $k_{\text{max}} = 12$.
\label{tab:results_rebin_accept}
}
\begin{ruledtabular}
\begin{tabular}{crr}
$Q^2_{\rm max}$ $({\rm GeV}^2)$ & $\Delta r_E$ (fm) & $ \Delta r_M$ (fm) \\
\hline
0.05
& $-0.001$ & $+0.023$ \\
& $-0.005$ &  $0.000$ \\
\hline
0.5
& $+0.003$ & $+0.003$ \\
& $-0.003$ & $+0.003$ \\
\hline
1
& $+0.003$ & $+0.009$ \\
& $-0.002$ & $ 0.000$ 
\end{tabular}
\end{ruledtabular}
\end{table}

The stated cross section uncertainties associated with the variation in the energy cut are small, with
a rms variation of 0.08\%. These mainly introduce an additional scatter into the
cross sections but have little impact on the radius central values.
For the entire data set, this translates to an uncertainty in $r_E$ of $0.003\,{\rm fm}$ 
and in $r_M$ of $0.009\,{\rm fm}$. Explicit results are given in Table~\ref{tab:results_rebin_accept}. 

In the A1 analysis, the kinematic-dependent correlated systematic is assumed to depend linearly on
the scattering angle [cf. Eq.~(\ref{eq:syst}) below, with $x=\theta$],
with a variation of approximately 0.2\% between the minimum and maximum angles for each
energy-spectrometer combination, except the 855 MeV data with spectrometer C (covering large
angles), for which the variation is approximately 0.5\%.%
\footnote{These values are deduced from the appropriate column of the tabulated data set
in the Supplemental Material of Ref.~\cite{Bernauer:2013tpr}.}
We perform a more comprehensive study of correlated systematics below.
  
\subsubsection{Sensitivity to size or kinematic dependence}

The correlated systematics mentioned above could represent either
experimental or theoretical uncertainties.
For example, they could be associated
with radiative corrections (beyond the energy cutoff variation),
background subtraction~\cite{Sick:2012zz}, potential offsets in the absolute
beam energy or angle calibration, etc. The impact of such uncertainties
on the cross sections is difficult to constrain below the 0.5\% level, but because of the floating
normalizations of the different datas ets, these correlated systematic
uncertainties only need to account for the variation within a specific
normalization subset.

While some sources of correlated corrections may be well approximated
by a correction that is linear in the scattering angle over a single
energy-spectrometer setting, this is not the only possible kinematic
dependence, and effects may be relevant over larger or smaller subsets
of data, or may be more important for one spectrometer. Thus, we
examine the impact of different prescriptions for applying the
correlated systematics.  We take a 0.5\% variation in the systematic
correction, but vary the functional form used to go from the minimum
to the maximum kinematic settings within data subsets, and we vary how the full
experiment is broken down. For the latter, we examine three cases:
0.5\% variation over the range of angles for each spectrometer-energy
combination (as done in the A1 analysis, with 18 separate angular
ranges), 0.5\% variation over the full kinematic range for each
spectrometer (with three separate ranges), and 0.5\% variation for each
of the 34 normalization subsets.

We examine eight different approaches to varying the kinematic dependence
of the systematic correction over a given data subset.  We multiply
and divide the cross sections and uncertainties by the factor 
\begin{align}\label{eq:syst}
1 + \delta_{\rm corr} = 1 + a { x - x_{\rm min} \over x_{\rm max} -
  x_{\rm min} } \,, 
\end{align}
where $a=0.005$ and $x$ is a kinematic variable.  We take the variable
$x$ to be proportional or inversely proportional to $\theta$, $Q^2$,
or $E^\prime$,  or to be proportional to $\varepsilon$ or
$1/\sin^4({\theta/2})$.  Note that for a given energy the correction goes
from zero at one extreme of the angular range for the data subset to
0.5\% at the other extreme; these different corrections only
modify the interpolation to intermediate angles.
These illustrative functional forms can be motivated from specific sources,
including kinematic offsets, rate-dependent effects, or simplified models of radiative corrections. 
However, the exact magnitude and precise functional form
cannot be fully determined without further input.

\begin{table}[tb]
  \caption{Results for changes in the radii under multiplication (top sign)
    or division (bottom sign) by a linear perturbation
  as in Eq.~(\ref{eq:syst}) for each beam energy/spectrometer combination,
  with $x=Q^2$, $1/Q^2$, $\theta$, or $1/\theta$.  Fits are for the 657 point rebinned A1 MAMI
  dataset with 0.3\%--0.4\% uncorrelated systematic
  uncertainties at three values of $Q^2_{\rm max}$ using the $z$ expansion with
  $t_0=0$, Gaussian priors with $|a_k|_{\text{max}} = |b_k|_{\text{max}}/\mu_p = 5$,
  $k_{\text{max}} = 12$.
\label{tab:Q2syst}
}
\begin{ruledtabular}
\begin{tabular}{ccrr}
$x$ & $Q^2_{\rm max}$ [GeV$^2$] &  $\Delta r_E$ [fm] & $ \Delta r_M$ [fm] \\
\hline
$Q^2$ &  
0.05 
& $\mp 0.017$ & $\pm 0.021$ \\
&0.5
& $\mp 0.016$ & $\mp 0.022$ \\
&1
& $\mp 0.015$ & $\mp 0.026$ \\
\hline
$1/Q^2$ &  
0.05 
& $\pm 0.041$ & $\mp 0.046$ \\
&0.5
& $\pm 0.025$ & $\pm 0.016$ \\
&1
& $\pm 0.023$ & $\pm 0.021$  \\
\hline
$\theta$ &
0.05
& $\mp 0.022$ & $\pm 0.027$ \\
&0.5
& $\mp 0.018$ & $\mp 0.021$ \\
&1
& $\mp 0.017$ & $\mp 0.025$ \\
\hline
$1/\theta$ &
0.05
& $\pm 0.036$ & $\mp 0.039$ \\
& 0.5
& $\pm 0.024$ & $\pm 0.018$ \\
& 1
& $\pm 0.021$ & $\pm 0.022$ 
\end{tabular}
\end{ruledtabular}
\end{table}

Taking the correction to be linear in the scattering angle, $x=\theta$,
and applied to each of the 18 energy-spectrometer combinations,
we find an uncertainty in the radii from fits
to the entire data set of $\Delta r_E=0.017$~fm, $\Delta r_M=0.025$~fm.
These are roughly 2.5 times larger than the values quoted in the Mainz
analysis, due mainly to the increase from their $\sim$0.2\% to our 0.5\%.
Other functional forms give similar results, with the largest effect
coming from scaling the uncertainties with $1/Q^2$.  The cases
$x=Q^2$, $1/Q^2$, $\theta$, and $1/\theta$ are given in Table~\ref{tab:Q2syst}.
We take the case $x=\theta$ to represent a reasonable average of the
functional forms tested.

\subsubsection{Impact of applying systematic corrections to different data subsets}\label{sec:corrsystests2}

Applying the 0.5\% correction over the full kinematic range for each spectrometer,
rather than over the range corresponding to a single beam energy, yielded somewhat
smaller uncertainties for $r_E$ and somewhat larger uncertainties for $r_M$. 
There was also a wider spread in the uncertainties arising from different
functional forms in Eq.~(\ref{eq:syst}), as expected for the interpolation over a wider
kinematic range. Applying the 0.5\% variation only over the angular
range for each normalization subset yielded uncertainties that were
typically 20\%--30\% larger for $r_E$ compared to the default approach, with
smaller increases for the uncertainty on $r_M$.  We note that similar 
studies using the original 1422 point data set showed much larger increases when
applying the correction to the different normalization subsets.

For simplicity, we have taken the systematic scaling factor, $a$ in
Eq.~(\ref{eq:syst}), identical in sign (i.e., always multiplying or always
dividing by $1+\delta_{\rm corr}$) and magnitude for each data subset.
However, many systematic effects could differ for the different
spectrometers, and the combined effect might be enhanced or suppressed
by the assumption of identical corrections. When applied individually to
each spectrometer, the charge radius uncertainty tends to be dominated by
the corrections applied to spectrometer B. For the magnetic radius, there 
tends to be a significant cancellation between the corrections from the
three spectrometers, and the result of shifting all of the spectrometers
identically (used in Ref.~\cite{Bernauer:2013tpr} and shown in
Table~\ref{tab:Q2syst}) is much smaller than the result of evaluating the
corrections independently for each spectrometer. Because it is not clear
how much the spectrometer corrections may be related, we do not enhance
the uncertainty in $r_M$. We simply note
that the uncertainty on $r_M$ shown in Table~\ref{tab:Q2syst}
could be a significant underestimate if the cancellation in these
tests does not reflect the true nature of any systematic corrections.

To further investigate the impact of applying different correlated systematic 
shifts to different data subsets, consider a fit with distinct parameters $a$ in Eq.~(\ref{eq:syst}) 
for different data subsets. These are allowed to vary as part of the fit, 
which then allows for subpercent kinematic variations in each data subset. 
This could be done with separate parameters for each spectrometer, for each of the 18 energy-spectrometer combinations 
or for each of the 34 different normalization subsets. 
For definiteness, we consider a fit with an independent normalization and slope parameter $a$ for each of the 34 normalization subsets.%
\footnote{Note that in the Mainz analysis and our other fits there are 31 normalization parameters 
which appear in 34 different combinations. For this test, we allow all 34 normalization factors to vary independently.}
This seems most consistent with the breakdown of uncertainties into normalization, correlated systematics, and uncorrelated systematics.
For $Q^2_{\text{max}} = 0.5$ GeV${}^2$, we find $r_E = 0.891(18)$ fm and $r_M = 0.792(49)$ fm, 
compared to $r_E=0.895(20)$ fm and $r_M = 0.776(38)$ fm from Table \ref{tab:results_mainzrebin} below, 
which includes both statistical and uncorrelated systematic uncertainties.
The changes in the extracted radii are consistent with the previously assigned uncertainties associated with the correlated systematics. 
The uncertainties in this fit are somewhat smaller for the charge radius and larger for the magnetic radius, 
in line with the expectation based on applying the corrections separately to each spectrometer.  
This may be a more realistic estimate of the uncertainties and could potentially allow for a combined analysis of 
Mainz and world data by including all of the Mainz systematic uncertainties explicitly in the fit. 
However, most likely neither the Mainz assumption that the corrections
are totally correlated between settings nor the assumption here that they are totally uncorrelated is entirely
realistic. The analysis presented here is included only as an independent 
estimate of the impact of allowing the correlated systematic correction to differ for different kinematic settings.

\subsubsection{Final evaluation of the correlated systematics}\label{sec:corrsysfinal}

It is difficult to determine an optimal approach for evaluating the
impact of unknown systematic errors or corrections.
The analysis strategy for the Mainz data set involves a
breakdown of the uncertainties into uncorrelated, correlated,
and normalization contributions and seems most consistent with 
applying the correlated uncertainty to each normalization subset. 
As noted above, this tends to increase the
uncertainty on $r_E$ in comparison to applying the correlated uncertainty
to each spectrometer or to each spectrometer-energy combination.
Similarly, applying corrections independently to the three spectrometers
tends to decrease the uncertainty on $r_E$ and increase the uncertainty on $r_M$.

We choose to evaluate the correlated systematic error by making simple, minimal
changes to the A1 procedure. We evaluate the impact of a linear angle-dependent
correction ($x=\theta$), applied to each beam-spectrometer combination, 
but choose a fixed 0.4\% variation. The 0.4\% variation ($a=0.004$)
is approximately twice the typical value considered in the A1 analysis,
which seems a reasonable estimate to account for additional
systematic effects such as TPE~\cite{Arrington:2011kv} and background
subtraction~\cite{Sick:2012zz}.
At $Q^2_{\rm max}=1\,{\rm GeV}^2$,
this choice yields uncertainties of 0.014~fm and 0.020~fm for $r_E$ and $r_M$,
  respectively, four-fifths of the uncertainties shown for
$x=\theta$ in Table~\ref{tab:Q2syst}.

These uncertainties are significant, but not sufficient to explain
the discrepancy with muonic hydrogen. Obtaining 
larger shifts due to such corrections would require 
either a systematic shift above the 0.4\% assumed here,
a correction applied over smaller data subsets, a more extreme
functional form for such corrections than considered here, 
or a conspiracy between shifts applied to different
spectrometer-beam combinations.

\section{Radius results from Mainz and world data \label{sec:results}}

Having completed our systematics studies, we proceed to perform a
final fit to the Mainz data and compare with a fit to other world
data using the same theoretical framework.  We close this section with
several consistency checks on the fits, including a discussion of form
factor priors, radiative corrections beyond TPE, and the verification
of the fit consistency between different spectrometer-energy subsets of
the data.

\subsection{Best fit radii from Mainz data \label{sec:newfit}}

\begin{figure}[htb]
  \includegraphics[width=0.49\textwidth]{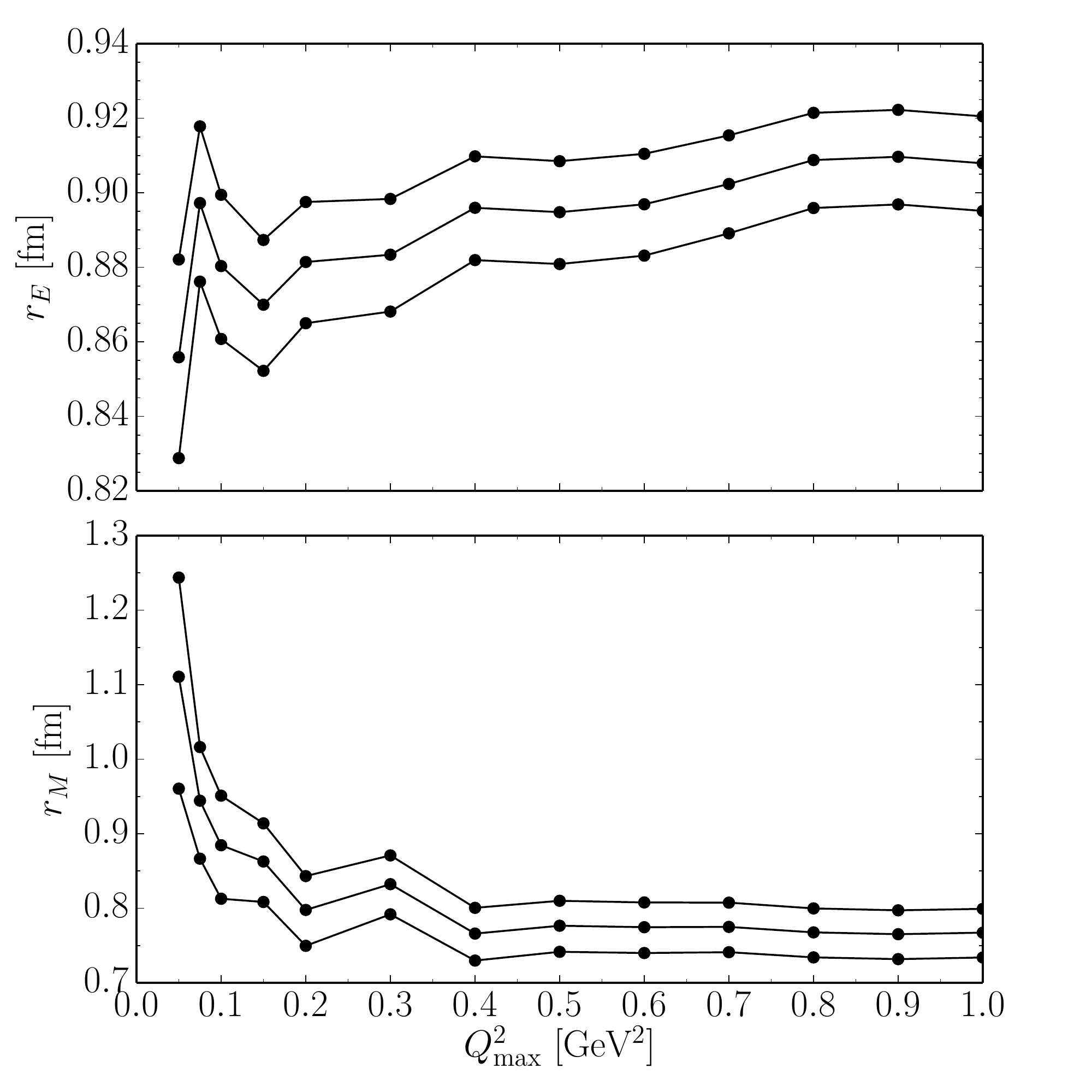}
\caption{Extracted electric (top panel) and magnetic (bottom panel) radii as functions of the
  kinematic cut $Q^2_{\rm max}$ on the momentum transfer for the rebinned 657 point A1 MAMI data set
  with 0.3\%--0.4\% uncorrelated systematic uncertainties, using the $z$ expansion with
  $t_0=0$, Gaussian priors with $|a_k|_{\text{max}} = |b_k|_{\text{max}}/\mu_p = 5$,
  $k_{\text{max}} = 12$. 
  Error bands are statistical and uncorrelated systematic only.
\label{fig:radvQ2max_mainz_rebin}
}
\end{figure}

Let us summarize our final fit to the Mainz data set.  We use the 657
point rebinned data set of Sec.~\ref{sec:rebin}, with the SIFF sum of
monopoles TPE correction from Table~\ref{tab:tpeblucoefs} in place of
the Feshbach correction applied in the  Mainz analysis and with the
A1 statistical rescaling replaced by a fixed 0.3\%--0.4\% uncorrelated
systematic as in Sec.~\ref{sec:constsyst}. We employ our default
form factor scheme,  $t_0=0$, $k_{\rm max}=12$, and Gaussian bound
$|a_k|_{\rm max}=|b_k|_{\rm max}/\mu_p = 5$.  The results are shown in
Fig.~\ref{fig:radvQ2max_mainz_rebin} and
Table~\ref{tab:results_mainzrebin}. The ``statistical'' uncertainty
accounts for both counting statistics and the uncorrelated systematic
uncertainties.  The energy cut correction is taken from
Table~\ref{tab:results_rebin_accept}.  The correlated systematic
uncertainty is obtained from the $x=\theta$ entry of
Table~\ref{tab:Q2syst}, rescaled to $0.4\%$, as described above in
Sec.~\ref{sec:corrsysfinal}. 

\begin{table}[tb]
  \caption{
    Final radius results from the fits to the rebinned 657 point A1 MAMI
    data set with 0.3\%--0.4\% uncorrelated systematic uncertainties
    in Fig.~\ref{fig:radvQ2max_mainz_rebin}, for three values
    of $Q^2_{\rm max}$. The uncertainties labeled ``stat'' include both the
    statistical and uncorrelated systematic uncertainties, while those
    labeled ``$\Delta {\rm E}$" and ``cor" account for the energy cut dependence
    and the correlated systematic uncertainties, respectively.
\label{tab:results_mainzrebin}
}
  \begin{ruledtabular}
  \begin{tabular}{ccc}
$Q^2_{\rm max}$ & $r_E$ & $r_M$  \\
$[{\rm GeV}^2]$ & [fm] & [fm] \\
  \hline
0.05 &$0.856(27)_{\rm stat}(5)_{\Delta \rm E} (18)_{\rm cor}$~&$1.11(14)_{\rm stat}(2)_{\Delta \rm E} (2)_{\rm cor}$~~~ \\
0.5  &$0.895(14)_{\rm stat}(3)_{\Delta \rm E} (14)_{\rm cor}$~&$0.776(34)_{\rm stat}(3)_{\Delta \rm E} (17)_{\rm cor}$   \\
1    &$0.908(13)_{\rm stat}(3)_{\Delta \rm E} (14)_{\rm cor}$~&$0.766(33)_{\rm stat}(9)_{\Delta \rm E} (20)_{\rm cor}$  
\end{tabular}
\end{ruledtabular}
  \end{table}

\subsection{Best fit radii from world data \label{sec:global}}

Now that we have a procedure for analyzing the Mainz data, we perform a similar fit to the global set
of world data, excluding the Mainz data set. We perform this separate analysis in part to obtain
independent results as a check on consistency between the Mainz data set and the world data set.
In addition, it is not clear that there is a reliable way to perform a combined analysis of the Mainz data
with other experiments, given the very different manner in which uncertainties from the Mainz experiment
were presented~\cite{Arrington:2015yxa}. The inclusion of the correlated systematic correction coefficients
$a$ in the fit, as discussed in Sec.~\ref{sec:corrsystests2}, yields a fit to the Mainz data where all
uncertainties are accounted for in the fit and would allow a combined analysis with the world data.
However, this approach allows the correlated systematic corrections to be different for each subset, 
which may not be significantly better than the assumption in the Mainz analysis that these corrections are identical for different subsets. 
We thus present separate fits to the Mainz and world data sets 
so that the comparison can be made without worrying about how to consistently treat Mainz and world uncertainties.

For the analysis of world data, we take the $\chi^2$ function 
\be\label{eqn:chi2_world} 
\chi^2_{\text{w}} = \chi_\sigma^2 + \chi_{\text{b}}^2 + \chi_{\text{n}}^2 \,. 
\ee
Here, $\chi^2_\sigma$ and $\chi^2_{\rm b}$ are identical to those used for the Mainz
analysis, Eqs.~(\ref{eqn:chi2_mainz}) and~(\ref{eq:chib}).
Because these experiments provide a normalization uncertainty, 
we follow previous analyses and include $\chi_{\text{n}}^2$ for
the floating normalization parameters assigned to each experiment, where
\be
\chi_{\text{n}}^2 = \sum_{i=1}^{N_{\text{exp}}} \frac{(1 - \eta_{i,\text{fit}})^2}{d\eta_i^2} \,.
\ee
Below $Q^2=1\,{\rm GeV}^2$, there are $N_{\text{exp}} = 23$ independent experiments, and $d\eta_i$ ranges from 1.5\% to 4.6\%.
For the fit to the combined world and polarization data sets, we include an additional term
for the recoil polarization and polarized target measurements of $\mu_p G_E/G_M$,
\be
\begin{split}
\chi^2_{\text{w+p}} = \chi^2_{\text{w}} + \sum_{i=1}^{N_{\text{rat}}} \frac{(R_i - R_{i,\text{fit}})^2}{dR_i^2} \,,
\label{eqn:chi2_worldpol}
\end{split}
\ee
where $R_i = \mu_p G_E(q^2_i)/G_M(q^2_i)$.

\begin{table}[tb]
\begin{ruledtabular}
  \caption{
    Final radius results from the fits 
    to the world cross section data in Fig.~\ref{fig:radvQ2max_world} (first line) and
    to the combined world cross section and polarization data in Fig.~\ref{fig:radvQ2max_worldpol}
    (second line).  There is no polarization data below $Q^2=0.05\,{\rm GeV}^2$.   
    \label{tab:results_world}
}
\begin{tabular}{ccccrrr}
$Q^2_{\rm max}$ & $r_E$ & $r_M$ & $\chi^2$ & $N_\sigma$ & $N_{\text{rat}}$ & $N_{\text{exp}}$ \\
$({\rm GeV}^2)$ & (fm) & (fm) \\
  \hline
  0.05& $0.846(42)$ & $1.04(11)$ & 52.9 & 111 & 0 & 8 \\
\hline
0.5 & $0.910(25)$ & $0.919(38)$ & 163.4 & 269 & 0 & 15 \\
    & $0.927(24)$ & $0.899(38)$ & 234.5 & 269 & 30 & 15 \\
\hline
1 & $0.916(24)$ & $0.914(34)$ & 260.9 & 363 & 0  & 23 \\
    & $0.919(23)$ & $0.913(34)$ & 366.0 & 363 & 41 & 23
\end{tabular}
\end{ruledtabular}
\end{table}

\begin{figure}[htb]
\begin{center}
  \includegraphics[width=0.99\linewidth]{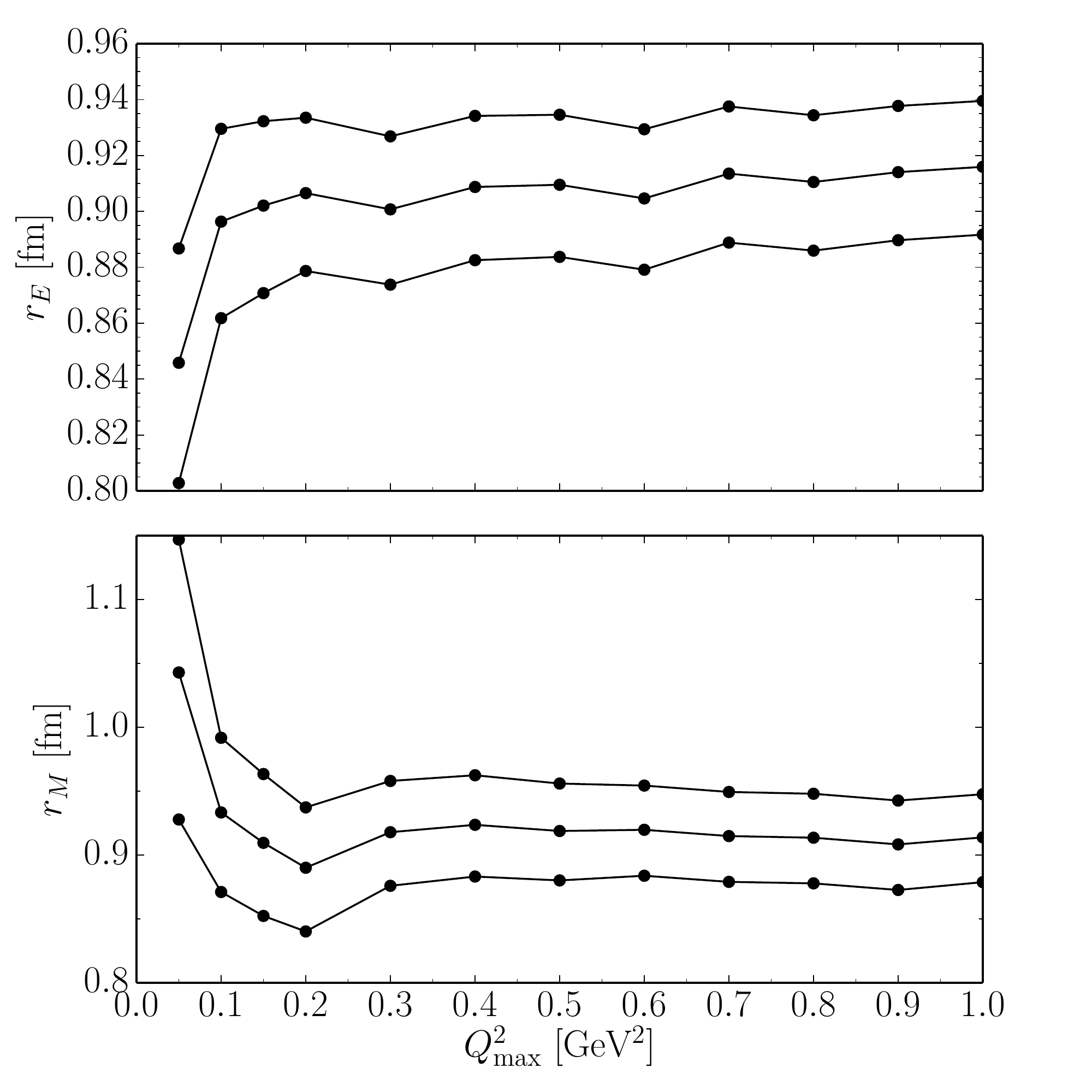}
\caption{Extracted electric (top panel) and magnetic (bottom panel) radii as functions of the
  kinematic cut $Q^2_{\rm max}$ on momentum transfer for the world cross section data set,
  using the $z$ expansion with
  $t_0=0$, Gaussian priors with $|a_k|_{\text{max}} = |b_k|_{\text{max}}/\mu_p = 5$, $k_{\text{max}} = 12$.
  Error bands include statistical and systematic uncertainties.
\label{fig:radvQ2max_world}
}
\end{center}
\end{figure}

\begin{figure}[htb]
\begin{center}
  \includegraphics[width=0.99\linewidth]{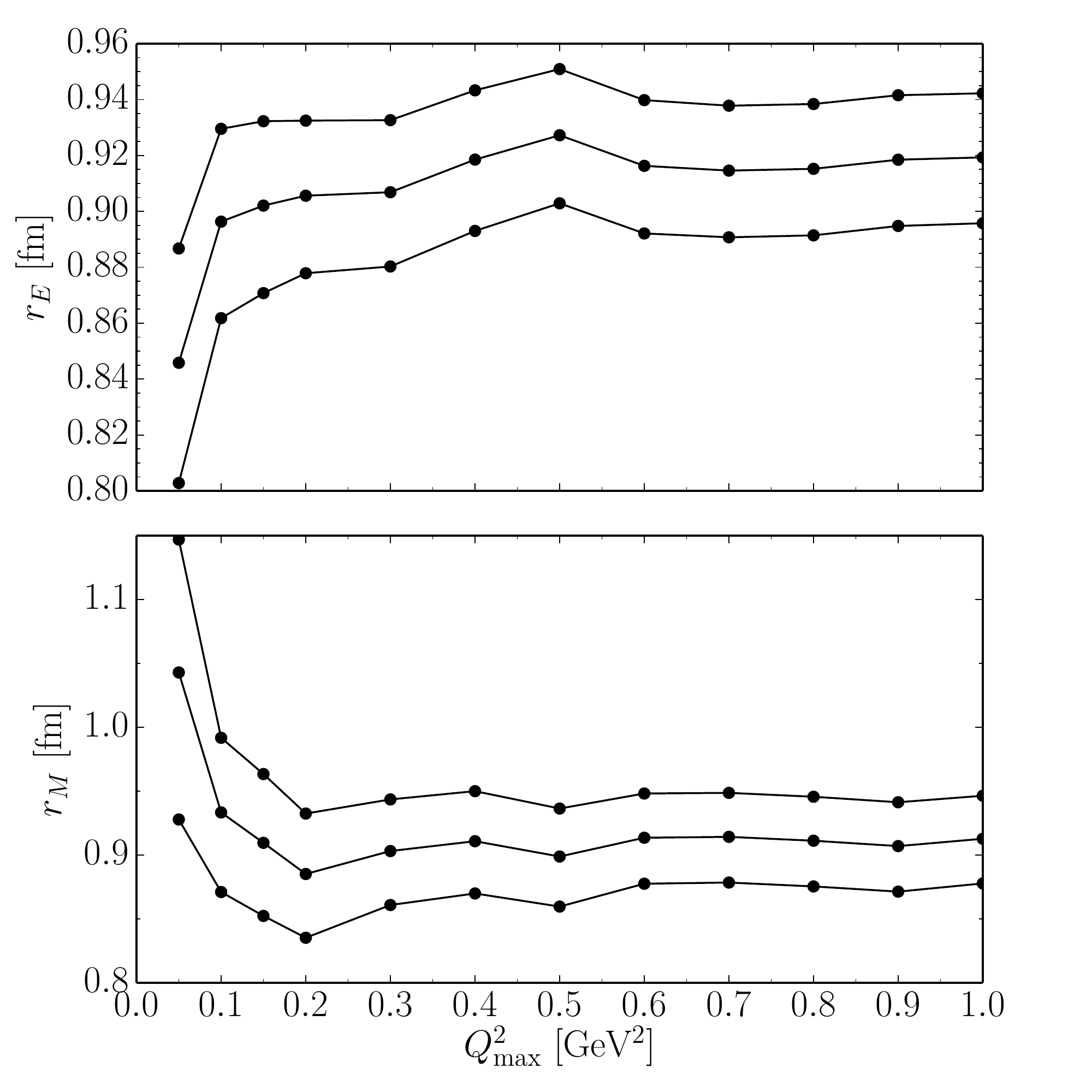}
  \caption{Same as Fig.~\ref{fig:radvQ2max_world} but with both the world polarization data in addition
    to the world cross section data. 
\label{fig:radvQ2max_worldpol}
}
\end{center}
\end{figure}

The TPE model for the cross section data in these analyses is the SIFF prescription
with the form factor as a sum of monopoles from Table~\ref{tab:tpeblucoefs}.%
\footnote{The use of different conventions in the world data to isolate the IR finite TPE contribution, 
as detailed after Eq. (\ref{eq:2photon}), changes $r_E$ and $r_M$ by less than $0.003$ fm.}
As in the fit to the A1 MAMI data set, we find little difference in the $r_E$ results
using either a dipole form factor or the Feshbach correction 
but significant differences in the $r_M$ results between approaches.  
For $Q^2_{\text{max}} = 1$~GeV$^2$, there are $N_\sigma = 363$ cross
section data points, $N_{\rm rat}=41$ polarization data points, and
$N_{\rm exp}=23$ normalization parameters.  Results for fits using the $z$ expansion
with our default $t_0=0$, $k_{\rm max}=12$, and Gaussian bounds
$|a_k|_{\text{max}} = |b_k|_{\text{max}}/\mu_p =5$,
are displayed in Figs.~\ref{fig:radvQ2max_world}
and~\ref{fig:radvQ2max_worldpol}.  Table~\ref{tab:results_world}
contains the radii values and error budget for particular values of
$Q^2_{\rm max}$.  The results in Table~\ref{tab:results_world}
indicate that the inclusion of the polarization data does not
significantly change the extracted radii. The results for the electric
radius are in agreement with the fit to the Mainz data in
Table~\ref{tab:results_mainzrebin}, while magnetic radius values
disagree by $2.7 \sigma$ if the uncertainties are added in
quadrature.

\subsection{Consistency checks\label{sec:consistchecks}}

We have derived best-fit values for $r_E$ and $r_M$ from the Mainz
data set and from a world data set excluding the Mainz data. We
observe a significant dependence of the Mainz radius on the $Q^2$
range included in the fit as well as a disagreement between the
Mainz and world data extractions of $r_M$. Here, we
describe several consistency checks on the fits we have performed.
We also consider the possibility of a common systematic not
specific to a particular experiment, and reexamine subleading radiative corrections which become enhanced at large $Q^2$. 

\subsubsection{Priors \label{sec:priors}}

Let us revisit the dependence on the class of form factors over which the
fit is performed, defined in the bounded $z$ expansion by the choice of 
$k_{\rm max}$, $t_0$, and coefficient bound. 

As discussed above in Sec.~\ref{sec:kmaxdepend},
we have taken $k_{\rm max}$ large enough such that the fit results are independent
of the precise value of $k_{\rm max}$, removing this choice from the discussion
of prior dependence. 

With our imposition of coefficient bounds,
the chosen form factor class depends on $t_0$.%
\footnote{
Modifications (multiplication by suitable analytic function $\phi$)
to the $z$ expansion can ensure that the form factor class is
rigorously independent of $t_0$ in the large $k_{\rm max}$ limit
if the bound is placed on $\sum_k a_k^2$. See, e.g., Ref.~\cite{Becher:2005bg}.
}
We have redone selected fits with different scheme choices, e.g.,
$t_0 = t_0^{\rm opt}$ defined after Eq.~(\ref{eq:z}), finding negligible
dependence on $t_0$ in the large $k_{\rm max}$ limit.

Regarding Gaussian vs sharp priors, 
we have employed Gaussian priors for numerical ease but have checked that our results
are not significantly changed if sharp priors are used.
Central values for both $r_E$ and $r_M$ differ by a negligible amount between
the two priors, and the difference in radius errors is small. 

We note that enforcing a bound on the radius parameters could in principle bias the
radius fits.  For example, at $t_0=0$ (or default choice), the squared radii
are proportional to fit parameters $a_1$ and $b_1$, and a bound on these parameters
would tend to bias fits toward smaller radii.  We have checked that fitting
with the bounds on $a_1$ and $b_1$ removed has a negligible impact.

Finally, consider the choice of the numerical value for the bound. 
We choose a Gaussian bound of 5, i.e., $|a_k/a_0|_{\text{max}} = |b_k/b_0|_{\rm max}= 5$,
for our fits, based on the sum rules and studies discussed in Sec.~\ref{sec:bounds}.
Our implementation of the bounds is meant to be very conservative,
especially at large $k$, where the coefficients must fall as $1/k^4$.
We examined the impact of tightening the constraint for larger values of $k$, taking
a bound of 5 for $k=1,\dots,4$ and a bound of 20/$k$ for larger $k$ values. 
This yields small changes in the central radius values ($< 0.3\,\sigma$), with only slightly smaller uncertainties.
Because tighter high-$k$ constraints have a minimal impact, for simplicity,
we use a fixed bound of $|a_k/a_0|_{\text{max}} = |b_k/b_0|_{\rm max}= 5$.
More aggressive priors or $k_{\rm max}$ truncations could be invoked to
reduce the statistical/fit uncertainty on the radius at the expense of introducing
model-dependent truncation errors.  This may allow for the possibility
of reduced uncertainties in the extracted radii, if one can verify that the reduction
in uncertainty coming from tighter bounds or truncations is not replaced with a bias
that yields a larger net uncertainty.  In this work, we are focused on minimizing any such
biases and so do not attempt to further constrain the fits.

We also fit the data and obtained statistical errors for larger bounds, 
$|a_k/a_0|_{\text{max}} = |b_k/b_0|_{\rm max}= 10$.
The change in the extracted radii is very small for fits with large $k_{\rm max}$
(in particular for our default $k_{\rm max}=12$). No additional 
uncertainty is applied as typical changes in the fit were significantly smaller
than the statistical or the correlated systematic uncertainties.
Fits in which $k_{\rm max}$ was not sufficiently large to give fully converged results
showed larger changes, but are not included in the final results presented here.

\subsubsection{Data set exclusions}

\begin{table}[htb]
\caption{
  Change in extracted $r_E$ and $r_M$ when each data subset is excluded.
  Fits are for rebinned A1 MAMI data set with 0.3\%--0.4\% uncorrelated systematic
  uncertainties at $Q^2_{\rm max}=0.5\,{\rm GeV}^2$ (568 data points), using the $z$ expansion with
  $t_0=0$, Gaussian priors with $|a_k|_{\text{max}} = |b_k|_{\text{max}}/\mu_p = 5$,
  $k_{\text{max}} = 12$.
\label{tab:exclusion5}
}
\begin{ruledtabular}
\begin{tabular}{cccrr }
  Spec. & Beam & $N_\sigma$ & $\Delta r_E$ $({\rm fm})$ & $\Delta r_M$ $({\rm fm})$ \\ 
  \hline
A & 180 & 539 &   $-0.008$ & $-0.031$ \\
  & 315 & 545 &  $+0.001$ & $-0.008$ \\
  & 450 & 543 &  $-0.004$ & $+0.008$ \\
  & 585 & 540 &  $0.000$ & $-0.009$ \\
  & 720 & 552 & $-0.003$ & $-0.002$ \\
  & 855 & 561 &  $0.000$ & $0.000$ \\
\hline
B & 180 & 507 & $-0.001$ & $+0.034$ \\
  & 315 & 522 & $+0.001$ & $+0.003$ \\
  & 450 & 500 & $+0.003$ & $-0.017$ \\
  & 585 & 508 & $+0.005$ & $+0.005$ \\
  & 720 & 511 & $-0.002$ & $-0.006$ \\
  & 855 & 502 & $+0.005$ & $+0.019$ \\
\hline
C & 180 & 544 & $-0.002$ & $+0.030$ \\
  & 315 & 544 & $0.000$ & $-0.023$ \\
  & 450 & 543 & $-0.006$ & $+0.032$ \\
  & 585 & 561 & $-0.001$ & $+0.001$ \\
  & 720 & 566 &  $0.000$ & $0.000$ \\
  & 855 & 568 & --- & ---
\end{tabular}
\end{ruledtabular}
\end{table}

\begin{table}[tb]
  \caption{
  Same as Table~\ref{tab:exclusion5}, but for  $Q^2_{\rm max}=1\,{\rm GeV}^2$ (657 data points). 
    \label{tab:exclusion1}
  }
  \begin{ruledtabular}
    \begin{tabular}{cccrr }
      Spec. & Beam & $N_\sigma$ & $\Delta r_E$ $({\rm fm})$ & $\Delta r_M$ $({\rm fm})$ \\ 
      \hline
    A & 180 & 628 & $-0.008$ & $-0.035$ \\
      & 315 & 634 & $-0.001$ & $-0.007$ \\
      & 450 & 632 & $-0.004$ & $+0.012$ \\
      & 585 & 629 & $-0.002$ & $-0.015$ \\
      & 720 & 628 & $-0.004$ & $-0.003$ \\
      & 855 & 636 & $-0.001$ & $-0.004$ \\
      \hline
    B & 180 & 596 & $0.000$ & $+0.041$ \\
      & 315 & 611 & $0.000$ & $+0.004$ \\
      & 450 & 589 & $+0.004$ & $-0.016$ \\
      & 585 & 597 & $+0.005$ & $+0.006$ \\
      & 720 & 600 & $-0.004$ & $-0.007$ \\
      & 855 & 591 & $+0.007$ & $+0.020$ \\
      \hline
    C & 180 & 633 & $-0.003$ & $+0.036$ \\
      & 315 & 633 & $+0.001$ & $-0.017$ \\
      & 450 & 632 & $-0.006$ & $+0.021$ \\
      & 585 & 639 & $+0.001$ & $-0.000$ \\
      & 720 & 625 & $-0.002$ & $-0.005$ \\
      & 855 & 636 & $+0.001$ & $+0.001$ 
    \end{tabular}
  \end{ruledtabular}
\end{table}

To verify that the fits to the Mainz data set are not biased by one particular subset of the data,
we redo our best fits for $r_E$ and $r_M$ 18 times, excluding in each fit
a particular energy/spectrometer combination.
The results are displayed in Tables~\ref{tab:exclusion5} and~\ref{tab:exclusion1}
for $Q^2_{\rm max}=0.5$ and $1\,{\rm GeV}^2$, respectively. 
For the electric radius, the impact of each subset exclusion is typically less than half
of the total statistical error (taken from Table~\ref{tab:results_mainzrebin}).
Several subset exclusions impact the magnetic radius at a level comparable to or greater
than the total statistical error. For the 180 MeV data, excluding any one of the
three spectrometers gives a 0.030--0.041~fm shift in $r_M$. This is much larger than the
estimated systematic uncertainty and much larger than one might expect based on an
exclusion of 4\%--11\% of the data points.
While this suggests the need for a larger uncertainty in the quoted value of $r_M$,
it is hard to quantify what uncertainty would be appropriate as we are comparing
highly correlated fits.  As such, we simply note this as another potential issue,
similar to the anomalous $Q^2_{\rm max}$ dependence observed in the extraction of the
charge radius.

\subsubsection{Subleading radiative corrections \label{sec:largelog}}

We have used standard prescriptions for the electron vertex
and bremsstrahlung radiative corrections. As noted above, in the $Q^2 \sim {\rm GeV}^2$
regime, it is critical to resum the leading $\alpha\log^2(Q^2/m_e^2)$ terms.
However, numerically enhanced subleading logarithms can also have a significant
impact, as illustrated by the following considerations.
This exercise is presented both as an illustration of how a correction
would need to deviate from the assumptions of Sec.~\ref{sec:corrsysfinal}
in order to reconcile muonic hydrogen with electron scattering measurements
of the charge radius, and to point out
the potential impact of a class of naively subleading but numerically
enhanced  radiative corrections. 

Recall the explicit form for the sum of the one-loop electron vertex and real
bremsstrahlung radiative corrections,
\begin{align}
\delta &= {\alpha \over \pi} \bigg\{
\bigg[ \log{Q^2\over m_e^2} - 1 \bigg] \log{ (\eta \Delta E)^2 \over E E^\prime }
+ {13\over 6} \log{Q^2\over m_e^2}
+ \dots 
\bigg\} \,,
\end{align}
where $\eta = E/E^{\prime}$ and the ellipsis denotes terms not containing large logarithms.
In the regime where $\Delta E \sim m_e$ (and $E \sim E^\prime \sim Q$)
the numerically relevant leading log term
\begin{align}
\delta &= {\alpha \over \pi} \bigg\{ -\log^2{Q^2 \over m_e^2} + \dots  \bigg\} \,,
\end{align}
is fixed by infrared divergences whose forms are dictated by soft photon theorems~\cite{Yennie:1961ad}.
Equivalently, an effective theory renormalization analysis between hard ($\sim Q$)
and soft ($\sim m_e$) scales determines the relevant Sudakov form factor.
However, in practice, $\Delta E$ can be large compared to $m_e$, introducing another
scale into the problem, and associated large logarithms not captured by
the naive exponentiation of one-loop corrections.
A complete analysis is outside the scope of the present paper, but to
illustrate the potential impact, let us consider in place of the
ansatz that makes the replacement (\ref{eq:resum}) in Eq.~(\ref{eq:radxs}) 
the following expressions:
\begin{multline}\label{eq:alpha2}
(1+\delta)\to \left[ 1 \pm \left( \delta + {\alpha\over \pi} \log^2{Q^2\over m_e^2} \right)\right]^{\pm 1}
  \\
  \times \exp\left( -{\alpha\over \pi} \log^2{Q^2\over m_e^2} \right) \,.
\end{multline}
These expressions agree with the known corrections through one-loop order and resum the leading
logarithms to all orders in perturbation theory when there is only one large ratio of scales.

\begin{figure}[tb]
\begin{center}
\includegraphics[width=0.99\linewidth]{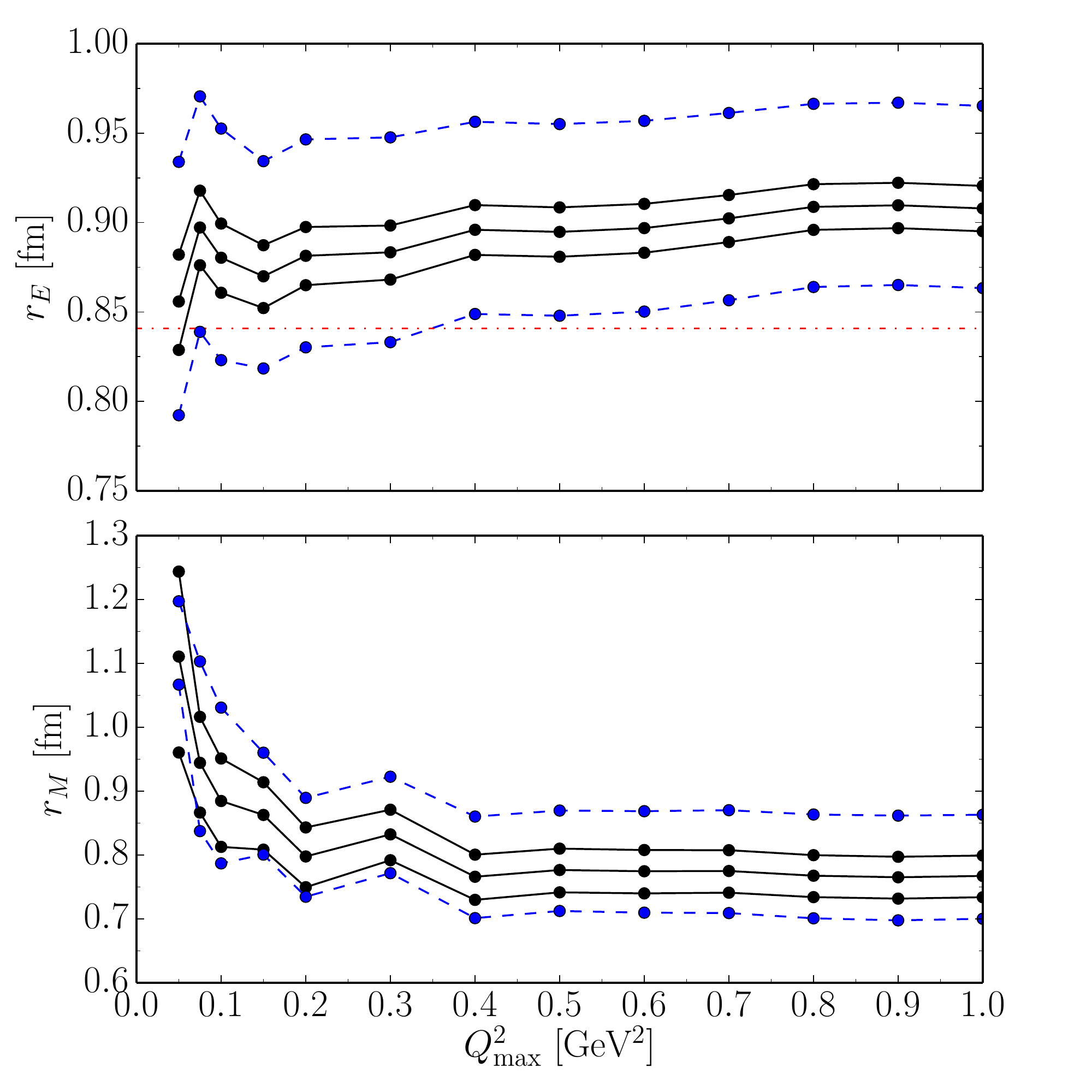}
\caption{
  Illustrative fit  with modified radiative corrections given by Eq.~(\ref{eq:alpha2}) using
  $\Delta E=10\,{\rm MeV}$.  Lower and upper dashed blue lines correspond to the plus sign and
  minus sign in Eq.~(\ref{eq:alpha2}), respectively.  
  Fits are for the 657 point rebinned A1 MAMI data set with 0.3\%--0.4\% uncorrelated systematic
  uncertainties using the $z$ expansion with
  $t_0=0$, Gaussian priors with $|a_k|_{\text{max}} = |b_k|_{\text{max}}/\mu_p = 5$,
  $k_{\text{max}} = 12$.  Black solid lines reproduce the curves in Fig.~\ref{fig:radvQ2max_mainz_rebin}.
  For orientation, the dash-dotted red line indicates the muonic hydrogen value for $r_E$. 
  \label{fig:logs}
}
\end{center}
\end{figure}

Figure~\ref{fig:logs} illustrates the impact of applying the 
correction on the right-hand side of Eq.~(\ref{eq:alpha2}) in place of the ansatz (\ref{eq:resum}). 
For definiteness, the plot takes $\Delta E = 10\,{\rm MeV}$.
As indicated in the figure, the shifts in the radii under this correction
are a factor $\sim 2$--$3$ larger than those allowed 
in Table~\ref{tab:Q2syst}, which considered corrections varying by 0.5\% over
beam-energy/spectrometer combinations.
The variation of the correction (\ref{eq:alpha2})
over beam-energy/spectrometer combinations [i.e., the magnitude of $a$ in Eq.~(\ref{eq:syst})]
ranges between 0.9\% and 2.6\%, with an average 1.5\%.

\subsection{Final radius extractions}

\begin{figure}[htb]
\begin{center}
\hspace{-2mm}\includegraphics[width=0.47\textwidth]{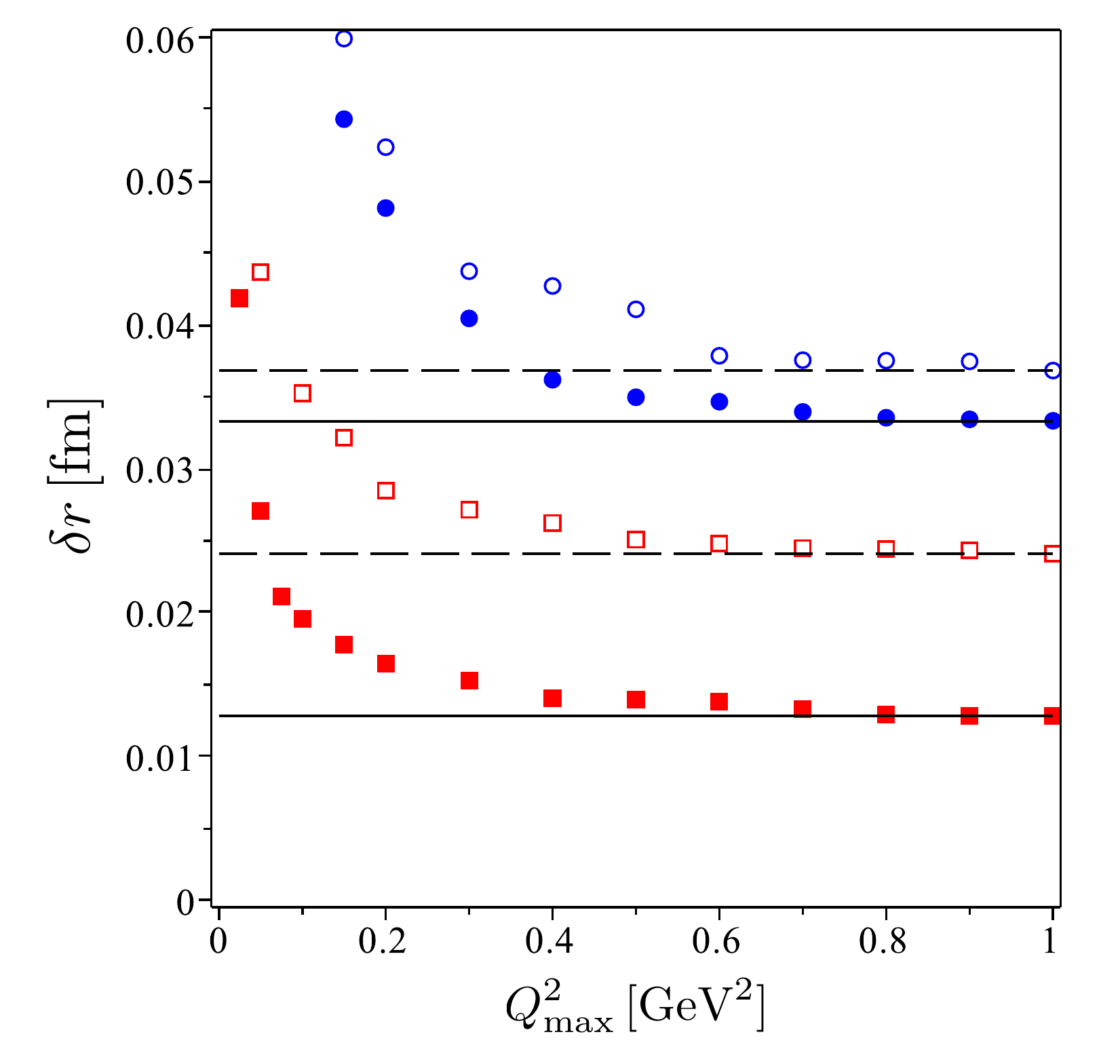}
\caption{Statistical error on $r_E$ (bottom, red squares) and $r_M$ (top, blue circles)
  as a function of $Q^2_{\rm max}$.  Solid symbols are for the 1422 point A1 MAMI data set,
  and open symbols are for the world cross section and polarization data set. 
  Fits use the $z$ expansion with $t_0=0$, Gaussian priors with
  $|a_k|_{\text{max}} = |b_k|_{\text{max}}/\mu_p = 5$, $k_{\text{max}} = 12$. 
  \label{fig:statQ2}
}
\end{center}
\end{figure}

A global analysis combining Mainz and other world data will artificially favor the Mainz data, 
as the uncertainties associated with each cross section measurement include only a small part of the total uncertainty.
Thus, we provide best-fit values separately for our analyses of the Mainz and world data.  
To determine an optimal $Q^2_{\rm max}$, 
Fig.~\ref{fig:statQ2} illustrates the statistical uncertainty on $r_E$ and $r_M$
found using our default fit both to the 1422 point Mainz data set and to the world data set.
For the Mainz data, the uncertainty is minimized by taking
$Q^2_{\rm max}\gtrsim 0.5\,{\rm GeV}^2$, with negligible improvement beyond this point.  
To maximize the statistical power of the data,
while minimizing potential systematic effects in higher $Q^2$ data, we 
take for definiteness the $Q^2_{\rm max}=0.5$~GeV$^2$ results of the previous sections.%
\footnote{
  A similar choice was made in Ref.~\cite{Hill:2010yb} based on radius sensitivity in the world data summarized by extracted
  form factors~\cite{Arrington:2007ux}. 
  Related argumentation for the $Q^2_{\rm max}$ dependence of radius sensitivity, based
  on continued-fraction expansion, is given in Ref.~\cite{Sick:2003gm}. 
  }

We then have for the Mainz data set, from Table~\ref{tab:results_mainzrebin},
\begin{align}\label{eq:rmainz}
  r_E^{\rm Mainz} =0.895(14)(14) \,,
  \quad
  r_M^{\rm Mainz} =0.776(34)(17) \,, 
\end{align}
where the first error comes from counting statistics and uncorrelated systematics 
and the second error comes from variation of the bremsstrahlung energy cut and correlated systematics. 
For the world data set, including cross section and polarization measurements,
we take a slightly higher $Q^2_{\rm max}=0.6\,{\rm GeV}^2$
based on Fig.~\ref{fig:statQ2}.   We then have 
\begin{align}\label{eq:rworld}
  r_E^{\rm world} =0.916(24) \,, 
  \quad
  r_M^{\rm world} =0.914(35) \,. 
\end{align}
These values correspond to the same analysis as presented in Table~\ref{tab:results_world}, but for a
fit with $Q^2_{\rm max}=0.6\,{\rm GeV}^2$.
In contrast to the Mainz data, the world data have combined statistical and systematic uncertainties
at the cross section level and so have only a single combined uncertainty.

The electric charge radius results are consistent with each other 
and between one and two standard deviations higher 
than the atomic physics measurements based on atomic hydrogen which yield
$r_E=0.8758(77)\,{\rm fm}$~\cite{Mohr:2012tt}.  They are well above the muonic hydrogen result
$r_E = 0.84087(39)\,{\rm fm}$~\cite{Antognini:1900ns}.
The magnetic radii differ significantly, indicating an unresolved tension between the Mainz
data set and the world data set.

A simple combination of the results (\ref{eq:rmainz}) and (\ref{eq:rworld}) yields
\begin{align}\label{eq:avg}
  r_E^{\rm avg.} =0.904(15) \,,
  \quad
  r_M^{\rm avg.} =0.851(26) \,. 
\end{align}
While the Mainz and world data sets have comparable total uncertainties,
the high statistics of the Mainz dataset imply that in this case the errors are dominantly systematic. 
It is not entirely clear that a simple average of the Mainz and world results is appropriate~\cite{Arrington:2015yxa, Arrington:2015ria}.
The magnetic radii differ by 2.7$\sigma$,  suggesting an inconsistency between Mainz and world cross sections that is ignored in 
averaging the results. In addition, the simple combination
assumes that the uncertainties for the Mainz and world data
analyses are independent, which may not be the case if there
is a common error, e.g., due to approximations in the radiative correction procedures.

\section{Summary and discussion \label{sec:summary} }

We have performed a comprehensive analysis of electron-proton scattering data
to determine the proton electric and magnetic radii.
Our analysis incorporates constraints of analyticity and perturbative scaling which
enforce model-independent bounds on form factor shape. The bounded $z$ expansion
ensures that the true form factor is guaranteed to lie within the space of considered
curves, while at the same time being sufficiently restrictive to enable meaningful
radius extractions. We focused on the high-statistics Mainz data set, and performed a wide-
ranging study of the impact of potential systematic errors.
We discussed potential flaws in the procedure of rescaling statistical errors and 
addressed these by rebinning data taken at identical kinematic settings and applying
a constant uncorrelated
systematic error that is not assumed to scale with statistics.  We also reevaluated the
correlated systematic uncertainties, increasing the size of these effects to include
contributions neglected in the original analysis, and examining different approaches to
evaluating the impact of such corrections on the radius.

Table~\ref{tab:radii} displays the progression of results
leading up to the final Mainz radius values, as various improvements
are included in the analysis. The data exhibit several issues that
suggest the need for additional uncertainties, but for which it is
difficult to quantify an appropriate correction or uncertainty 
contribution. There is an unusually large variation of $r_E$ and $r_M$
with the $Q^2$ range included in the fit, as illustrated in
Fig.~\ref{fig:radvQ2max_mainz_rebin}. In addition, the exclusion of
individual data sets, in particular at low beam energy, has an
unusually large impact on the extracted radii.
Despite these anomalous features, inclusion of the above improvements 
leads to a proton charge radius that is larger than extracted in the original
A1 analysis~\cite{Bernauer:2013tpr}, and, even with the larger uncertainty, almost
3$\sigma$ above the value $r_E \approx 0.84$~fm inferred from muonic hydrogen.
Based on our examination of the systematic uncertainties, resolving this
discrepancy would require correlated systematic effects well above the
0.5\% level that was considered in our analysis.

\begin{table}[tb]
\caption{Charge and magnetic radii as determined in Ref.~\cite{Bernauer:2013tpr}
  compared to the sequence of fits leading to the final values determined in this paper. 
  For the Mainz data set, the first error is a combination of statistics and uncorrelated
  systematics, and the second error is from correlated systematics. The entry labeled
  ``alternate approach'' is the test from Sec.~\ref{sec:corrsystests2} which evaluates
  the impact of the correlated systematic uncertainties as part of the fit, rather than
  evaluating it separately.}
\label{tab:radii}
\begin{ruledtabular}
\begin{tabular}{lcll}
& Source  & $r_E$ (fm) & $r_M$ (fm) \\  \hline
  \\[-2mm]
A1 spline &  \cite{Bernauer:2013tpr}          &~0.879(5)(10)  &~0.777(13)(14) \\
Bounded $z$ exp.    & Tab.\ref{tab:results_mainz}         &~0.920(9)(-)   &~0.743(25)(-) \\
+Hadronic TPE  &    Tab.\ref{tab:TPE}                  &~0.918(9)(-)   &~0.780(25)(-)   \\
Rebin, 0.3\%--0.4\% syst.  &  Tab.\ref{tab:results_mainzrebin}            &~0.908(13)(-)  &~0.767(33)(-) \\
+0.4\% corr. syst. & Tab.\ref{tab:results_mainzrebin}                &~0.908(13)(14) &~0.767(33)(22) \\
$Q^2_{\rm max}=0.5$~GeV$^2$  &  (\ref{eq:rmainz})              &~0.895(14)(14) &~0.776(34)(17) \\
(Alternate approach)      & \ref{sec:corrsystests2}  &~0.891(18)(-)  &~0.792(49)(-) \\[2mm] \hline \\[-2mm]
New fit to world data     & (\ref{eq:rworld})                &~0.916(24)     &~0.914(35) \\[2mm] \hline \\[-2mm]
Simple avg. & (\ref{eq:avg})           &~0.904(15)     &~0.851(26)\\
\end{tabular}
\end{ruledtabular}
\end{table}

As an independent check of the radius, we
performed the analogous fit to the world data excluding Mainz data.
The systematic error treatment in the world data set differs in that
the systematic errors are included in each cross section data point as
opposed to deduced from a combination of statistics rescaling and
model-dependent correlated systematics analysis.  It is thus not
straightforward to perform a meaningful combined fit, but observables
such as radius values may be compared to verify consistency. 
In this comparison, the Mainz and world $r_E$ values are in good
agreement, and the $r_M$ values differ by $2.7 \sigma$.
This is perhaps not surprising, given the
clear disagreement between the Mainz form factors and world data, 
in particular for $G_M$ at low $Q^2$~\cite{Bernauer:2013tpr}.
Putting aside the discrepancy in magnetic radii, the charge radius puzzle
persists, with the world value for $r_E$ 3$\sigma$ high compared to
muonic hydrogen and the combined Mainz + world average discrepant
at the 4.2$\sigma$ level.
In light of this, it is also important to inquire to what
extent an underestimated systematic effect or theoretical correction
could be common to both data sets.

The theoretical input to the radius extraction consists of specifying the
form factor class and defining the radiative correction model.  We
have examined in detail the uncertainties associated with form factor
shape assumptions.  We find a large impact from fitting to the
physical form factor class defined by the bounded $z$ expansion,
compared to polynomial or inverse polynomial fits.  Somewhat
surprisingly, the central value for the charge radius goes in the
direction of {\it increasing} tension between the electron scattering and
muonic hydrogen.  We have further examined the dependence of radius
values on form factor priors, finding that such a residual dependence is
small compared to other uncertainties. 

The other theoretical input is the radiative correction model,  as
described in Sec.~\ref{sec:rad}. For the most part, the corrections
are known precisely or, for model-dependent terms including hadronic
vacuum polarization or proton vertex corrections, the uncertainties
are estimated to be small compared to the uncertainty in the radius extraction.
Through one-loop order, the only essential model dependence of the radiative
corrections enters from the description of the TPE.  Varying over models in
the literature reveals no large dependence on the applied TPE correction.
The $Q^2 \sim {\rm GeV}^2$ regime demands QED
radiative corrections beyond one-loop order.
In the counting $m_e \sim \Delta E$, leading logarithms
are resummed by a standard ansatz.  Subleading logarithms then enter
at a level expected to be contained within the 0.4\% systematic error
budget. Possible enhancements, either simply numerical or due to the
hierarchy $\Delta E \gg m_e$, could potentially give rise to larger effects.
The Mainz and world data sets differ in their treatment of bremsstrahlung
radiation and approximations based on (\ref{eq:resum}), and in the
uncertainties ascribed to these effects. 
More refined calculations, including a detailed examination
of experimental conditions and the interplay with background modeling
and subtraction, are required in order to fully address this
question~\cite{largelogs}.   

Further constraints may be placed on the proton form factors in combination
with electron-proton scattering data. In particular, the inclusion of 
either electron-neutron scattering data, or both electron-neutron
and pion-nucleon data, allows the threshold $t_{\rm cut}$ appearing
in the definition of the $z$ expansion (\ref{eq:z}) for the isoscalar/isovector
decomposition of the form factors to be raised from
$4 m_\pi^2$ to either $9 m_\pi^2$ or $16 m_\pi^2$.
This yields a smaller $|z|_{\rm max}$ and hence tighter constraints on the
form factors and smaller radius uncertainties.
Isospin violating corrections and systematic uncertainties in the additional
data must be properly accounted for.
These additional constraints by themselves cannot offer a satisfactory
resolution to the proton radius puzzle, since they would 
then be inconsistent with the results of the fit to the electron scattering data alone.
Similar remarks hold for the model spectral function analysis in Ref.~\cite{Lorenz:2014yda}.
We note that it is not feasible to reconstruct accurate spectral functions for form factors, 
${\rm Im} G(t)$, from scattering data, since these have support at $|z|=1$.

Electron scattering data from a polarized target have been taken and
will provide low-$Q^2$ measurements of $\mu_p G_E/G_M$, down to
$Q^2 \approx 0.02$~GeV$^2$, yielding a more sensitive measurement
of the magnetic form factor in this region~\cite{E08-007}. Future
experiments will
obtain low-$Q^2$ ($\sim 10^{-4}-10^{-2}\,{\rm GeV}^2$) proton form
factor measurements less prone to high-$Q^2$
systematics~\cite{Gasparian:2014rna,Mihovilovic:2014aya}.  Muon-proton
scattering promises to provide further insight~\cite{Gilman:2013eiv}.
Independent of scattering measurements, new results are anticipated
from hydrogen spectroscopy that may impact the charge radius
discrepancy, including a new microwave measurement of the $2S$--$2P$ Lamb
shift~\cite{Vutha}, measurement of $2S$--$4P$
transitions~\cite{Beyer:2013daa}, and $1S$--$3S/3D$
transitions~\cite{Arnoult,Peters}.   Next-generation lattice QCD
simulations may provide another
handle~\cite{Yamazaki:2009zq,Bratt:2010jn,Alexandrou:2013joa,Bhattacharya:2013ehc,Green:2014xba};
in particular, resolving the
$\sim 8\%$ discrepancy between $r_E \approx 0.84\,{\rm fm}$ of muonic hydrogen
and $r_E \approx 0.91\,{\rm fm}$ of the simple fit to Mainz electron scattering data
is a viable present-day target.  Such first-principles calculations would be
independent of radiative corrections involving the electron,
thus avoiding the reliance on hadronic models for TPE, and detector-dependent modeling of radiative tails. 

Regardless of its resolution, the proton radius puzzle has important
implications across particle, nuclear, and atomic physics. For
example, understanding and controlling systematic effects, including
radiative corrections, at the percent level will be critical for
measurements at future long baseline neutrino
experiments~\cite{Adams:2013qkq}. Any deficiency in the theoretical
treatment of electron-proton scattering will be exacerbated in neutrino
applications by the presence of additional flux uncertainties and
nuclear corrections. Much more intriguing is the possibility that
updated measurements and a detailed examination of the radiative corrections
will not resolve the discrepancy.  Already much work has been performed to
find explanations in terms of physics beyond the Standard
Model~\cite{Barger:2010aj,TuckerSmith:2010ra,Batell:2011qq,Pohl:2013yb,Izaguirre:2014cza,Pauk:2015oaa,Carlson:2015poa,Carlson:2015jba}.  
Future measurements will provide more stringent tests of the discrepancy
in electron scattering and atomic hydrogen, with plans to directly compare
electron-proton and muon-proton scattering as a test of lepton nonuniversality.

\vspace{5mm}
\noindent
 {\bf Acknowledgments}
We thank Z.~Jiang for collaboration during the early stages of this work and G.~Paz and I.~Sick for discussions. 
Research was supported by a NIST Precision Measurement Grant, the U.S. Department of Energy,
Office of Science, Office of High Energy Physics (DOE Grant No. DE-FG02-13ER41958), and
Office of Nuclear Physics (DOE Grant No. DE-AC02-06CH11357). G.~L. also acknowledges
support by the ICORE Program of Planning and Budgeting Committee and by ISF Grant Nos. 1937/12.


\noindent

\begin{thebibliography}{99}

\bibitem{Mohr:2012tt}
 P.~J.~Mohr, B.~N.~Taylor, and D.~B.~Newell,
 %``CODATA Recommended Values of the Fundamental Physical Constants: 2010,''
 Rev.\ Mod.\ Phys.\ {\bf 84}, 1527 (2012).
% [arXiv:1203.5425 [physics.atom-ph]].
 %%CITATION = ARXIV:1203.5425;%%

 \bibitem{deGouvea:2013onf}
 A.~de Gouvea {\it et al.} [Intensity Frontier Neutrino Working Group Collaboration],
 %``Working Group Report: Neutrinos,''
 arXiv:1310.4340 [hep-ex].
 %%CITATION = ARXIV:1310.4340;%%

\bibitem{Pohl:2010zza}
 %R.~Pohl, A.~Antognini, F.~Nez, F.~D.~Amaro, F.~Biraben, J.~M.~R.~Cardoso, D.~S.~Covita and A.~Dax {\it et al.},
 R.~Pohl {\it et al.},
 %``The size of the proton,''
 Nature {\bf 466}, 213 (2010).
 %%CITATION = NATUA,466,213;%%

\bibitem{Antognini:1900ns}
 %A.~Antognini, F.~Nez, K.~Schuhmann, F.~D.~Amaro, F.~Biraben, J.~M.~R.~Cardoso, D.~S.~Covita and A.~Dax {\it et al.},
 A.~Antognini {\it et al.},
 %``Proton Structure from the Measurement of $2S-2P$ Transition Frequencies of Muonic Hydrogen,''
 Science {\bf 339}, 417 (2013).
 %%CITATION = SCIEA,339,417;%%
 %85 citations counted in INSPIRE as of 21 Oct 2014
  
\bibitem{Bhattacharya:2011ah}
 B.~Bhattacharya, R.~J.~Hill and G.~Paz,
 %``Model independent determination of the axial mass parameter in quasielastic neutrino-nucleon scattering,''
 Phys.\ Rev.\ D {\bf 84}, 073006 (2011).
% [arXiv:1108.0423 [hep-ph]].
 %%CITATION = ARXIV:1108.0423;%%

\bibitem{Day:2012gb} 
  M.~Day and K.~S.~McFarland,
  %``Differences in Quasi-Elastic Cross-Sections of Muon and Electron Neutrinos,''
  Phys.\ Rev.\ D {\bf 86}, 053003 (2012).
%  [arXiv:1206.6745 [hep-ph]].
  %%CITATION = ARXIV:1206.6745;%%

\bibitem{Pohl:2013yb} 
  R.~Pohl, R.~Gilman, G.~A.~Miller and K.~Pachucki,
  %``Muonic hydrogen and the proton radius puzzle,''
  Ann.\ Rev.\ Nucl.\ Part.\ Sci.\  {\bf 63}, 175 (2013).
%  [arXiv:1301.0905 [physics.atom-ph]].
  %%CITATION = ARXIV:1301.0905;%%

\bibitem{Carlson:2015jba} 
  C.~E.~Carlson,
  %``The Proton Radius Puzzle,''
  Prog.\ Part.\ Nucl.\ Phys.\  {\bf 82}, 59 (2015).
%  [arXiv:1502.05314 [hep-ph]].
  %%CITATION = ARXIV:1502.05314;%%
  
\bibitem{Bernauer:2013tpr}
 J.~C.~Bernauer {\it et al.} [A1 Collaboration],
 %``The electric and magnetic form factors of the proton,''
 Phys.\ Rev.\ C {\bf 90}, 015206 (2014).
% [arXiv:1307.6227 [nucl-ex]].
 %%CITATION = ARXIV:1307.6227;%%

\bibitem{Priv_Distler}
  M. Distler, private communication.

\bibitem{Arrington:2015yxa} 
  J.~Arrington,
  %``An examination of proton charge radius extractions from e-p scattering data,''
  J.\ Phys.\ Chem.\ Ref.\ Data {\bf 44}, 031203 (2015)
    arXiv:1506.00873 [nucl-ex].
  %%CITATION = ARXIV:1506.00873;%%

\bibitem{Arrington:2011kv} 
  J.~Arrington,
  %``Comment on `High-Precision Determination of the Electric and Magnetic Form Factors of the Proton',''
  Phys.\ Rev.\ Lett.\  {\bf 107}, 119101 (2011).
%  [arXiv:1108.3058 [nucl-ex]].
  %%CITATION = ARXIV:1108.3058;%%

\bibitem{Zhan:2011ji} 
  %X.~Zhan, K.~Allada, D.~S.~Armstrong, J.~Arrington, W.~Bertozzi, W.~Boeglin, J.-P.~Chen and K.~Chirapatpimol {\it et al.},
  X.~Zhan {\it et al.},
  %``High Precision Measurement of the Proton Elastic Form Factor Ratio $\mu_pG_E/G_M$ at low $Q^2$,''
  Phys.\ Lett.\ B {\bf 705}, 59 (2011).
%  [arXiv:1102.0318 [nucl-ex]].
  %%CITATION = ARXIV:1102.0318;%%

  \bibitem{Sick:2012zz} 
  I.~Sick,
  %``Problems with proton radii,''
  Prog.\ Part.\ Nucl.\ Phys.\  {\bf 67}, 473 (2012).
  %%CITATION = PPNPD,67,473;%%

\bibitem{Arrington:2015ria}
  J.~Arrington and I.~Sick, 
  %``Evaluation of the proton charge radius from e-p scattering'',
  J.\ Phys.\ Chem.\ Ref.\ Data {\bf 44}, 031204 (2015)
  [arXiv:1505.02680 [nucl-ex]].
  %CITATION = ARXIV:1505.02680;%%

\bibitem{Hill:2010yb}
 R.~J.~Hill and G.~Paz,
 %``Model independent extraction of the proton charge radius from electron scattering,''
 Phys.\ Rev.\ D {\bf 82}, 113005 (2010).
% [arXiv:1008.4619 [hep-ph]].
 %%CITATION = ARXIV:1008.4619;%%

 \bibitem{Epstein:2014zua} 
  Z.~Epstein, G.~Paz and J.~Roy,
  %``Model independent extraction of the proton magnetic radius from electron scattering,''
  Phys.\ Rev.\ D {\bf 90}, no. 7, 074027 (2014).
  %  [arXiv:1407.5683 [hep-ph]].
  %%CITATION = ARXIV:1407.5683;%%
  
\bibitem{Belushkin:2006qa} 
  M.~A.~Belushkin, H.-W.~Hammer and U.-G.~Meissner,
  %``Dispersion analysis of the nucleon form-factors including meson continua,''
  Phys.\ Rev.\ C {\bf 75}, 035202 (2007).
%  [hep-ph/0608337].
  %%CITATION = HEP-PH/0608337;%%
  
\bibitem{Supplement}
  See supplemental material at \url{http://link.aps.org/supplemental/10.1103/PhysRevD.92.013013} for the compilations of Mainz and world cross section data and world polarization data used in this analysis.

\bibitem{Nakamura:2010}
 K.~Nakamura {\it et al.} [Particle Data Group],
 %``Review of particle physics,''
 J.\ Phys.\ G {\bf 37}, 075021 (2010).

\bibitem{Hill:2006ub}
For a review and further references, see
 R.~J.~Hill,
 %``The modern description of semileptonic meson form factors,''
%{\it Proceedings of 4th Flavor Physics and CP Violation Conference (FPCP 2006), %Vancouver, British Columbia, Canada, 9-12 Apr 2006, pp 027}
%``The Modern description of semileptonic meson form factors,''
  eConf C {\bf 060409}, 027 (2006)
  [hep-ph/0606023].
 %%CITATION = HEP-PH/0606023;%%
 %%CITATION = ECONF,C060409,027;%%

\bibitem{Lepage:1980fj}
 G.~P.~Lepage and S.~J.~Brodsky,
 %``Exclusive Processes in Perturbative Quantum Chromodynamics,''
 Phys.\ Rev.\ D {\bf 22}, 2157 (1980).
 %%CITATION = PHRVA,D22,2157;%%

\bibitem{NP}
A.~Ahmadi, A.~Olshevsky, P.~A.~Parrilo and J.~N.~Tsitsiklis,
Math.\ Prog., {\bf 137}, nos. 1--2, 453 (2013).
%[arXiv:1012.1908].

\bibitem{Arrington:2003qk} 
  J.~Arrington,
  %``Implications of the discrepancy between proton form-factor measurements,''
  Phys.\ Rev.\ C {\bf 69}, 022201 (2004).
%  [nucl-ex/0309011].
  %%CITATION = NUCL-EX/0309011;%%

\bibitem{Kelly:2004hm}
 J.~J.~Kelly,
 %``Simple parametrization of nucleon form factors,''
 Phys.\ Rev.\ C {\bf 70}, 068202 (2004).
 %%CITATION = PHRVA,C70,068202;%%

\bibitem{Arrington:2007ux}
 J.~Arrington, W.~Melnitchouk and J.~A.~Tjon,
 %``Global analysis of proton elastic form factor data with two-photon exchange corrections,''
 Phys.\ Rev.\ C {\bf 76}, 035205 (2007).
% [arXiv:0707.1861 [nucl-ex]].
 %%CITATION = ARXIV:0707.1861;%%

\bibitem{Sick:2003gm}
 I.~Sick,
 %``On the RMS radius of the proton,''
 Phys.\ Lett.\ B {\bf 576}, 62 (2003).
% [nucl-ex/0310008].
 %%CITATION = NUCL-EX/0310008;%%

\bibitem{Arrington:2006hm} 
  J.~Arrington and I.~Sick,
  %``Precise determination of low-Q nucleon electromagnetic form factors and their impact on parity-violating e-p elastic scattering,''
  Phys.\ Rev.\ C {\bf 76}, 035201 (2007).
%  [nucl-th/0612079].
  %%CITATION = NUCL-TH/0612079;%%

\bibitem{Tsai:1961zz}
 Y.~-S.~Tsai,
 %``Radiative Corrections to Electron-Proton Scattering,''
 Phys.\ Rev.\ {\bf 122}, 1898 (1961).
 %%CITATION = PHRVA,122,1898;%%

\bibitem{Maximon:2000hm}
 L.~C.~Maximon and J.~A.~Tjon,
 %``Radiative corrections to electron proton scattering,''
 Phys.\ Rev.\ C {\bf 62}, 054320 (2000).
% [nucl-th/0002058].
 %%CITATION = NUCL-TH/0002058;%%

\bibitem{Ent:2001hm} 
  R.~Ent, B.~W.~Filippone, N.~C.~R.~Makins, R.~G.~Milner, T.~G.~O'Neill and D.~A.~Wasson,
  %``Radiative corrections for (e, e-prime p) reactions at GeV energies,''
  Phys.\ Rev.\ C {\bf 64}, 054610 (2001).
  %%CITATION = PHRVA,C64,054610;%%

\bibitem{Hill:2011wy}
 R.~J.~Hill and G.~Paz,
 %``Model independent analysis of proton structure for hydrogenic bound states,''
 Phys.\ Rev.\ Lett.\ {\bf 107}, 160402 (2011).
% [arXiv:1103.4617 [hep-ph]].
 %%CITATION = ARXIV:1103.4617;%%

\bibitem{Jegerlehner:1996ab}
 F.~Jegerlehner,
 %``Hadronic vacuum polarization contribution to g-2 of the leptons and alpha (M(Z)),''
 Nucl.\ Phys.\ Proc.\ Suppl.\ {\bf 51C}, 131 (1996).
% [hep-ph/9606484].
 %%CITATION = HEP-PH/9606484;%%

\bibitem{Friar:1998wu}
 J.~L.~Friar, J.~Martorell and D.~W.~L.~Sprung,
 %``Hadronic vacuum polarization and the Lamb shift,''
 Phys.\ Rev.\ A {\bf 59}, 4061 (1999).
% [nucl-th/9812053].
 %%CITATION = NUCL-TH/9812053;%%

\bibitem{Blunden:2005ew}
 P.~G.~Blunden, W.~Melnitchouk and J.~A.~Tjon,
 %``Two-photon exchange in elastic electron nucleon scattering,''
 Phys.\ Rev.\ C {\bf 72}, 034612 (2005).
% [arXiv:nucl-th/0506039].
 %%CITATION = PHRVA,C72,034612;%%

\bibitem{Blunden:2003sp}
 P.~G.~Blunden, W.~Melnitchouk and J.~A.~Tjon,
 %``Two-photon exchange and elastic electron proton scattering,''
 Phys.\ Rev.\ Lett.\ {\bf 91}, 142304 (2003).
% [arXiv:nucl-th/0306076].
 %%CITATION = PRLTA,91,142304;%%

\bibitem{Carlson:2007sp}
 C.~E.~Carlson and M.~Vanderhaeghen,
 %``Two-Photon Physics in Hadronic Processes,''
 Ann.\ Rev.\ Nucl.\ Part.\ Sci.\ {\bf 57}, 171 (2007).
% [hep-ph/0701272].
% %%CITATION = HEP-PH/0701272;%%

\bibitem{Arrington:2011dn} 
  J.~Arrington, P.~G.~Blunden and W.~Melnitchouk,
  %``Review of two-photon exchange in electron scattering,''
  Prog.\ Part.\ Nucl.\ Phys.\  {\bf 66}, 782 (2011).
%  [arXiv:1105.0951 [nucl-th]].
  %%CITATION = ARXIV:1105.0951;%%

\bibitem{McKinley:1948zz}
 W.~A.~McKinley and H.~Feshbach,
 %``The Coulomb Scattering of Relativistic Electrons by Nuclei,''
 Phys.\ Rev.\ {\bf 74}, 1759 (1948).
 %%CITATION = PHRVA,74,1759;%%
 %57 citations counted in INSPIRE as of 03 Nov 2014

\bibitem{Vanderhaeghen:2000ws} 
  M.~Vanderhaeghen, J.~M.~Friedrich, D.~Lhuillier, D.~Marchand, L.~Van Hoorebeke and J.~Van de Wiele,
  %``QED radiative corrections to virtual Compton scattering,''
  Phys.\ Rev.\ C {\bf 62}, 025501 (2000).
%  [hep-ph/0001100].
  %%CITATION = HEP-PH/0001100;%%

\bibitem{Mo:1968cg} 
  L.~W.~Mo and Y.~S.~Tsai,
  %``Radiative Corrections to Elastic and Inelastic e p and mu p Scattering,''
  Rev.\ Mod.\ Phys.\  {\bf 41}, 205 (1969).
  %%CITATION = RMPHA,41,205;%%

\bibitem{Tsai:1971qi} 
  Y.~S.~Tsai,
  %``Radiative Corrections To Electron Scatterings,''
  SLAC-PUB-0848; SLAC-PUB-848.
  %%CITATION = SLAC-PUB-0848, SLAC-PUB-848;%%

\bibitem{Walker:1993vj} 
  %R.~C.~Walker, B.~Filippone, J.~Jourdan, R.~Milner, R.~McKeown, D.~H.~Potterveld, L.~Andivahis and R.~Arnold {\it et al.},
  R.~C.~Walker {\it et al.},
  %``Measurements of the proton elastic form-factors for 1-GeV/c**2 <= Q**2 <= 3-GeV/C**2 at SLAC,''
  Phys.\ Rev.\ D {\bf 49}, 5671 (1994).
  %%CITATION = PHRVA,D49,5671;%%

\bibitem{Arrington:2003ck} 
  J.~Arrington,
  %``Evidence for two photon exchange contributions in electron proton and positron proton elastic scattering,''
  Phys.\ Rev.\ C {\bf 69}, 032201 (2004).
%  [nucl-ex/0311019].
  %%CITATION = NUCL-EX/0311019;%%

\bibitem{Frerejacque:1965ic} 
  D.~Frerejacque, D.~Benaksas and D.~J.~Drickey,
  %``Proton Form-factors From Proton Observation,''
  Phys.\ Rev.\  {\bf 141}, 1308 (1966).
  %%CITATION = PHRVA,141,1308;%%

\bibitem{Ganichot:1972mb} 
  D.~Ganichot, B.~Grossetete and D.~B.~Isabelle,
  %``Backward electron-deuteron scattering below 280 mev,''
  Nucl.\ Phys.\ {\bf A178}, 545 (1972).
  %%CITATION = NUPHA,A178,545;%%

\bibitem{Qattan:2004ht} 
  %I.~A.~Qattan, J.~Arrington, R.~E.~Segel, X.~Zheng, K.~Aniol, O.~K.~Baker, R.~Beams and E.~J.~Brash {\it et al.},
  I.~A.~Qattan {\it et al.},
  %``Precision Rosenbluth measurement of the proton elastic form-factors,''
  Phys.\ Rev.\ Lett.\  {\bf 94}, 142301 (2005).
%  [nucl-ex/0410010].
  %%CITATION = NUCL-EX/0410010;%%

\bibitem{Christy:2004rc} 
  M.~E.~Christy {\it et al.}  [E94110 Collaboration],
  %``Measurements of electron proton elastic cross-sections for 0.4 < Q**2 < 5.5 (GeV/c)**2,''
  Phys.\ Rev.\ C {\bf 70}, 015206 (2004).
%  [nucl-ex/0401030].
  %%CITATION = NUCL-EX/0401030;%%

\bibitem{Milbrath:1997de} 
  B.~D.~Milbrath {\it et al.}  [Bates FPP Collaboration],
  %``A Comparison of polarization observables in electron scattering from the proton and deuteron,''
  Phys.\ Rev.\ Lett.\  {\bf 80}, 452 (1998)
  [Erratum-ibid.\  {\bf 82}, 2221 (1999)].
%  [nucl-ex/9712006].
  %%CITATION = NUCL-EX/9712006;%%

\bibitem{Gayou:2001qt} 
  %O.~Gayou, K.~Wijesooriya, A.~Afanasev, M.~Amarian, K.~Aniol, S.~Becher, K.~Benslama and L.~Bimbot {\it et al.},
  O.~Gayou {\it et al.},
  %``Measurements of the elastic electromagnetic form-factor ratio mu(p) G(Ep) / G(Mp) via polarization transfer,''
  Phys.\ Rev.\ C {\bf 64}, 038202 (2001).
  %%CITATION = PHRVA,C64,038202;%%

\bibitem{Pospischil:2001pp} 
  T.~Pospischil {\it et al.}  [A1 Collaboration],
  %``Measurement of G(E(p))/G(M(p)) via polarization transfer at Q**2 = 0.4-GeV/c**2,''
  Eur.\ Phys.\ J.\ A {\bf 12}, 125 (2001).
  %%CITATION = EPHJA,A12,125;%%

\bibitem{Strauch:2002wu} 
  S.~Strauch {\it et al.}  [Jefferson Lab E93-049 Collaboration],
  %``Polarization transfer in the He-4 (polarized-e, e-prime polarized-p) H-3 reaction up to Q**2 = 2.6-(GeV/c)**2,''
  Phys.\ Rev.\ Lett.\  {\bf 91}, 052301 (2003).
%  [nucl-ex/0211022].
  %%CITATION = NUCL-EX/0211022;%%

\bibitem{MacLachlan:2006vw} 
  %G.~MacLachlan, A.~Aghalarian, A.~Ahmidouch, B.~D.~Anderson, R.~Asaturian, O.~Baker, A.~R.~Baldwin and D.~Barkhuff {\it et al.},
  G.~MacLachlan {\it et al.},
  %``The ratio of proton electromagnetic form factors via recoil polarimetry at Q**2 = 1.13-(GeV/c)**2,''
  Nucl.\ Phys.\ {\bf A764}, 261 (2006).
  %%CITATION = NUPHA,A764,261;%%

\bibitem{Jones:2006kf} 
  M.~K.~Jones {\it et al.}  [Resonance Spin Structure Collaboration],
  %``Proton G(E)/G(M) from beam-target asymmetry,''
  Phys.\ Rev.\ C {\bf 74}, 035201 (2006).
%  [nucl-ex/0606015].
  %%CITATION = NUCL-EX/0606015;%%

\bibitem{Crawford:2006rz} 
  %C.~B.~Crawford, A.~Sindile, T.~Akdogan, R.~Alarcon, W.~Bertozzi, E.~Booth, T.~Botto and J.~Calarco {\it et al.},
  C.~B.~Crawford {\it et al.},
  %``Measurement of the proton electric to magnetic form factor ratio from vector H-1(vector e, e' p),''
  Phys.\ Rev.\ Lett.\  {\bf 98}, 052301 (2007).
%  [nucl-ex/0609007].
  %%CITATION = NUCL-EX/0609007;%%

\bibitem{Puckett:2010ac} 
  %A.~J.~R.~Puckett, E.~J.~Brash, M.~K.~Jones, W.~Luo, M.~Meziane, L.~Pentchev, C.~F.~Perdrisat and V.~Punjabi {\it et al.},
  A.~J.~R.~Puckett {\it et al.},
  %``Recoil Polarization Measurements of the Proton Electromagnetic Form Factor Ratio to Q^2 = 8.5 GeV^2,''
  Phys.\ Rev.\ Lett.\  {\bf 104}, 242301 (2010).
%  [arXiv:1005.3419 [nucl-ex]].
  %%CITATION = ARXIV:1005.3419;%%

\bibitem{Jones:1999rz} 
  M.~K.~Jones {\it et al.}  [Jefferson Lab Hall A Collaboration],
  %``G(E(p)) / G(M(p)) ratio by polarization transfer in polarized e p ---> e polarized p,''
  Phys.\ Rev.\ Lett.\  {\bf 84}, 1398 (2000).
%  [nucl-ex/9910005].
  %%CITATION = NUCL-EX/9910005;%%

  \bibitem{Gayou:2001qd} 
  O.~Gayou {\it et al.}  [Jefferson Lab Hall A Collaboration],
  %``Measurement of G(Ep) / G(Mp) in polarized-e p ---> e polarized-p to Q**2 = 5.6-GeV**2,''
  Phys.\ Rev.\ Lett.\  {\bf 88}, 092301 (2002).
%  [nucl-ex/0111010].
  %%CITATION = NUCL-EX/0111010;%%

\bibitem{Ron:2007vr} 
  %G.~Ron, J.~Glister, B.~Lee, K.~Allada, W.~Armstrong, J.~Arrington, A.~Beck and F.~Benmokhtar {\it et al.},
  G.~Ron {\it et al.},
  %``The Proton Elastic Form Factor Ratio mu(p) G**p(E)/G**p(M) at Low Momentum Transfer,''
  Phys.\ Rev.\ Lett.\  {\bf 99}, 202002 (2007).
%  [arXiv:0706.0128 [nucl-ex]].
  %%CITATION = ARXIV:0706.0128;%%

\bibitem{Punjabi:2005wq} 
  %V.~Punjabi, C.~F.~Perdrisat, K.~A.~Aniol, F.~T.~Baker, J.~Berthot, P.~Y.~Bertin, W.~Bertozzi and A.~Besson {\it et al.},
  V.~Punjabi {\it et al.},
  %``Proton elastic form-factor ratios to Q**2 = 3.5-GeV**2 by polarization transfer,''
  Phys.\ Rev.\ C {\bf 71}, 055202 (2005);
  {\bf 71}, 069902(E) (2005).
%  [nucl-ex/0501018].
  %%CITATION = NUCL-EX/0501018;%%

\bibitem{Puckett:2011xg} 
  %A.~J.~R.~Puckett, E.~J.~Brash, O.~Gayou, M.~K.~Jones, L.~Pentchev, C.~F.~Perdrisat, V.~Punjabi and K.~A.~Aniol {\it et al.},
  A.~J.~R.~Puckett {\it et al.},
  %``Final Analysis of Proton Form Factor Ratio Data at $\mathbf{Q^2 = 4.0}$, 4.8 and 5.6 GeV$\mathbf{^2}$,''
  Phys.\ Rev.\ C {\bf 85}, 045203 (2012).
%  [arXiv:1102.5737 [nucl-ex]].
  %%CITATION = ARXIV:1102.5737;%%

\bibitem{Ron:2011rd}
 G.~Ron {\it et al.} [Jefferson Lab Hall A Collaboration],
 %``Low $Q^2$ measurements of the proton form factor ratio $mu_p G_E / G_M$,''
 Phys.\ Rev.\ C {\bf 84}, 055204 (2011).
% [arXiv:1103.5784 [nucl-ex]].
 %%CITATION = ARXIV:1103.5784;%%

\bibitem{Arrington:2003df} 
  J.~Arrington,
  %``How well do we know the electromagnetic form-factors of the proton?,''
  Phys.\ Rev.\ C {\bf 68}, 034325 (2003).
%  [nucl-ex/0305009].
  %%CITATION = NUCL-EX/0305009;%%

\bibitem{Goitein:1970pz} 
  %M.~Goitein, R.~J.~Budnitz, L.~Carroll, J.~R.~Chen, J.~R.~Dunning, K.~Hanson, D.~C.~Imrie and C.~Mistretta {\it et al.},
  M.~Goitein {\it et al.},
  %``Elastic electron-proton scattering cross-sections measured by a coincidence technique,''
  Phys.\ Rev.\ D {\bf 1}, 2449 (1970).
  %%CITATION = PHRVA,D1,2449;%%

\bibitem{Bartel:1973rf} 
  %W.~Bartel, F.~W.~Busser, W.~r.~Dix, R.~Felst, D.~Harms, H.~Krehbiel, P.~E.~Kuhlmann and J.~McElroy {\it et al.},
  W.~Bartel {\it et al.},
  %``Measurement of proton and neutron electromagnetic form-factors at squared four momentum transfers up to 3-GeV/c$^2$,''
  Nucl.\ Phys.\ {\bf B58}, 429 (1973).
  %%CITATION = NUPHA,B58,429;%%

\bibitem{Andivahis:1994rq} 
  %L.~Andivahis, P.~E.~Bosted, A.~Lung, L.~M.~Stuart, J.~Alster, R.~G.~Arnold, C.~C.~Chang and F.~S.~Dietrich {\it et al.},
  L.~Andivahis {\it et al.},
  %``Measurements of the electric and magnetic form-factors of the proton from Q**2 = 1.75-GeV/c**2 to 8.83-GeV/c**2,''
  Phys.\ Rev.\ D {\bf 50}, 5491 (1994).
  %%CITATION = PHRVA,D50,5491;%%

  \bibitem{Arrington:2012dq} 
  J.~Arrington,
  %``Coulomb corrections in the extraction of the proton radius,''
  J.\ Phys.\ G {\bf 40}, 115003 (2013).
%  [arXiv:1210.2677 [nucl-ex]].
  %%CITATION = ARXIV:1210.2677;%%

\bibitem{Simon:1980hu} 
  G.~G.~Simon, C.~Schmitt, F.~Borkowski and V.~H.~Walther,
  %``Absolute electron Proton Cross-Sections at Low Momentum Transfer Measured with a High Pressure Gas Target System,''
  Nucl.\ Phys.\ {\bf A333}, 381 (1980).
  %%CITATION = NUPHA,A333,381;%%

\bibitem{Lepage:2001ym}
 G.~P.~Lepage, B.~Clark, C.~T.~H.~Davies, K.~Hornbostel, P.~B.~Mackenzie, C.~Morningstar and H.~Trottier,
 %``Constrained curve fitting,''
 Nucl.\ Phys.\ Proc.\ Suppl.\ {\bf 106}, 12 (2002).
% [hep-lat/0110175].
 %%CITATION = HEP-LAT/0110175;%%

\bibitem{Schindler:2008fh} 
  M.~R.~Schindler and D.~R.~Phillips,
  %``Bayesian Methods for Parameter Estimation in Effective Field Theories,''
  Ann.\ Phys.\  {\bf 324}, 682 (2009);
  {\bf 324}, 2051(E) (2009).
%  [arXiv:0808.3643 [hep-ph]].
  %%CITATION = ARXIV:0808.3643;%%

\bibitem{Kraus:2014qua} 
  E.~Kraus, K.~E.~Mesick, A.~White, R.~Gilman and S.~Strauch,
  %``Polynomial fits and the proton radius puzzle,''
  Phys.\ Rev.\ C {\bf 90}, no. 4, 045206 (2014).
%  [arXiv:1405.4735 [nucl-ex]].
  %%CITATION = ARXIV:1405.4735;%%

\bibitem{Lorenz:2014vha}
 I.~T.~Lorenz and U.~G.~Mei§ner,
 %``Reduction of the proton radius discrepancy by 3?,''
 Phys.\ Lett.\ B {\bf 737}, 57 (2014).
% [arXiv:1406.2962 [hep-ph]].
 %%CITATION = ARXIV:1406.2962;%%

\bibitem{Adikaram:2014ykv} 
  D.~Adikaram {\it et al.}  [CLAS Collaboration],
  %``Towards a resolution of the proton form factor problem: new electron and positron scattering data,''
  Phys.\ Rev.\ Lett.\  {\bf 114}, no. 6, 062003 (2015).
%  [arXiv:1411.6908 [nucl-ex]].
  %%CITATION = ARXIV:1411.6908;%%

\bibitem{Rachek:2014fam} 
  %I.~A.~Rachek, J.~Arrington, V.~F.~Dmitriev, V.~V.~Gauzshtein, R.~E.~Gerasimov, A.~V.~Gramolin, R.~J.~Holt and V.~V.~Kaminskiy {\it et al.},
  I.~A.~Rachek {\it et al.},
  %``Measurement of the two-photon exchange contribution to the elastic $e^{\pm}p$ scattering cross sections at the VEPP-3 storage ring,''
  Phys.\ Rev.\ Lett.\  {\bf 114}, no. 6, 062005 (2015).
%  [arXiv:1411.7372 [nucl-ex]].
  %%CITATION = ARXIV:1411.7372;%%

\bibitem{Becher:2005bg}
 T.~Becher and R.~J.~Hill,
 %``Comment on form factor shape and extraction of |V(ub)| from B --> pi l
 %nu,''
 Phys.\ Lett.\ B {\bf 633}, 61 (2006).
% [arXiv:hep-ph/0509090].
 %%CITATION = PHLTA,B633,61;%%

\bibitem{Yennie:1961ad}
 D.~R.~Yennie, S.~C.~Frautschi and H.~Suura,
 %``The infrared divergence phenomena and high-energy processes,''
 Ann. Phys.\ {\bf 13}, 379 (1961).
 %%CITATION = APNYA,13,379;%%

\bibitem{largelogs}
  R.~J.~Hill,
  ``Effective field theory for Sudakov logarithms in lepton-nucleon scattering,'' 
  in preparation. 

\bibitem{Lorenz:2014yda} 
  I.~T.~Lorenz, U.~G.~Meißner, H.-W.~Hammer and Y.-B.~Dong,
  %``Theoretical Constraints and Systematic Effects in the Determination of the Proton Form Factors,''
  Phys.\ Rev.\ D {\bf 91}, no. 1, 014023 (2015).
%  [arXiv:1411.1704 [hep-ph]].
  %%CITATION = ARXIV:1411.1704;%%
  
\bibitem{E08-007}
  J.~Arrington, D.~Day, R.~Gilman, D.~Higinbotham, G.~Ron, and A.~Sarty, Jefferson Lab experiment
  E08-007, https://www.jlab.org/exp\_prog/generated/apphalla.html.

\bibitem{Gasparian:2014rna} 
  A.~Gasparian [PRad at JLab Collaboration],
  %``The PRad experiment and the proton radius puzzle,''
  Eur.\ Phys.\ J.\ Web Conf.\  {\bf 73}, 07006 (2014).
  %%CITATION = 00776,73,07006;%%
  
\bibitem{Mihovilovic:2014aya} 
  M.~Mihovilovic {\it et al.}  [A1 Collaboration],
  %``Initial state radiation experiment at MAMI,''
  Eur.\ Phys.\ J.\ Web Conf.\  {\bf 72}, 00017 (2014).
  %%CITATION = 00776,72,00017;%%
  
  \bibitem{Gilman:2013eiv} 
  R.~Gilman {\it et al.}  [MUSE Collaboration],
  %``Studying the Proton "Radius" Puzzle with \mu p Elastic Scattering,''
  arXiv:1303.2160 [nucl-ex].
  %%CITATION = ARXIV:1303.2160;%%
  
\bibitem{Vutha}
  A.~Vutha et al., ``Progress towards a new microwave measurement of the hydrogen $n$=2 lamb
  shift: a measurement of the proton charge radius", BAPS.2012.DAMOP.D1.138, 2012.
  
\bibitem{Beyer:2013daa} 
  %A.~Beyer, C.~G.~Parthey, N.~Kolachevsky, J.~Alnis, K.~Khabarova, R.~Pohl, E.~Peters and D.~C.~Yost {\it et al.},
  A.~Beyer {\it et al.},
  %``Precision Spectroscopy of Atomic Hydrogen,''
  J.\ Phys.\ Conf.\ Ser.\  {\bf 467}, 012003 (2013).
  %%CITATION = 00462,467,012003;%%

\bibitem{Arnoult}
  O.~Arnoult, F.~Nez, L.~Julien, and F.~Biraben,
  %''Optical frequency measurement of the 1S–3S two-photon transition in hydrogen''
  Eur.\ Phys.\ J.\ {\bf D} 60, 243 (2010).
  
\bibitem{Peters}
  E.~Peters, D.~C.~Yost, A.~Matveev, T.~W.~H\"{a}nsch, and T.~Udem,
  %''Frequency-comb spectroscopy of the hydrogen 1S-3S and 1S-3D transitions''
  Ann. Phys. {\bf 525}, L29 (2013).

\bibitem{Yamazaki:2009zq} 
  T.~Yamazaki, Y.~Aoki, T.~Blum, H.~W.~Lin, S.~Ohta, S.~Sasaki, R.~Tweedie and J.~Zanotti,
  %``Nucleon form factors with 2+1 flavor dynamical domain-wall fermions,''
  Phys.\ Rev.\ D {\bf 79}, 114505 (2009).
%  [arXiv:0904.2039 [hep-lat]].
  %%CITATION = ARXIV:0904.2039;%%

  \bibitem{Bratt:2010jn} 
  J.~D.~Bratt {\it et al.}  [LHPC Collaboration],
  %``Nucleon structure from mixed action calculations using 2+1 flavors of asqtad sea and domain wall valence fermions,''
  Phys.\ Rev.\ D {\bf 82}, 094502 (2010).
%  [arXiv:1001.3620 [hep-lat]].
  %%CITATION = ARXIV:1001.3620;%%
  
\bibitem{Alexandrou:2013joa} 
  C.~Alexandrou, M.~Constantinou, S.~Dinter, V.~Drach, K.~Jansen, C.~Kallidonis and G.~Koutsou,
  %``Nucleon form factors and moments of generalized parton distributions using $N_f=2+1+1$ twisted mass fermions,''
  Phys.\ Rev.\ D {\bf 88}, no. 1, 014509 (2013).
%  [arXiv:1303.5979 [hep-lat]].
  %%CITATION = ARXIV:1303.5979;%%
  
  \bibitem{Bhattacharya:2013ehc} 
  T.~Bhattacharya, S.~D.~Cohen, R.~Gupta, A.~Joseph, H.~W.~Lin and B.~Yoon,
  %``Nucleon Charges and Electromagnetic Form Factors from 2+1+1-Flavor Lattice QCD,''
  Phys.\ Rev.\ D {\bf 89}, no. 9, 094502 (2014).
%  [arXiv:1306.5435 [hep-lat]].
  %%CITATION = ARXIV:1306.5435;%%

\bibitem{Green:2014xba} 
  J.~R.~Green, J.~W.~Negele, A.~V.~Pochinsky, S.~N.~Syritsyn, M.~Engelhardt and S.~Krieg,
  %``Nucleon electromagnetic form factors from lattice QCD using a nearly physical pion mass,''
  Phys.\ Rev.\ D {\bf 90}, 074507 (2014).
%  [arXiv:1404.4029 [hep-lat]].
  %%CITATION = ARXIV:1404.4029;%%

\bibitem{Adams:2013qkq} 
  C.~Adams {\it et al.}  [LBNE Collaboration],
  %``The Long-Baseline Neutrino Experiment: Exploring Fundamental Symmetries of the Universe,''
  arXiv:1307.7335 [hep-ex].
  %%CITATION = ARXIV:1307.7335;%%
  
  \bibitem{Barger:2010aj} 
  V.~Barger, C.~W.~Chiang, W.~Y.~Keung and D.~Marfatia,
  %``Proton size anomaly,''
  Phys.\ Rev.\ Lett.\  {\bf 106}, 153001 (2011).
%  [arXiv:1011.3519 [hep-ph]].
  %%CITATION = ARXIV:1011.3519;%%

  \bibitem{TuckerSmith:2010ra} 
  D.~Tucker-Smith and I.~Yavin,
  %``Muonic hydrogen and MeV forces,''
  Phys.\ Rev.\ D {\bf 83}, 101702 (2011).
%  [arXiv:1011.4922 [hep-ph]].
  %%CITATION = ARXIV:1011.4922;%%
  
\bibitem{Batell:2011qq} 
  B.~Batell, D.~McKeen and M.~Pospelov,
  %``New Parity-Violating Muonic Forces and the Proton Charge Radius,''
  Phys.\ Rev.\ Lett.\  {\bf 107}, 011803 (2011).
%  [arXiv:1103.0721 [hep-ph]].
  %%CITATION = ARXIV:1103.0721;%%

\bibitem{Izaguirre:2014cza} 
  E.~Izaguirre, G.~Krnjaic and M.~Pospelov,
  %``Probing New Physics with Underground Accelerators and Radioactive Sources,''
  Phys.\ Lett.\ B {\bf 740}, 61 (2015).
%  arXiv:1405.4864 [hep-ph].
  %%CITATION = ARXIV:1405.4864;%%

\bibitem{Pauk:2015oaa} 
  V.~Pauk and M.~Vanderhaeghen,
  %``Lepton universality test in the photoproduction of $e^- e^+$ versus $\mu^- \mu^+$ pairs on a proton target,''
  arXiv:1503.01362 [hep-ph].
  %%CITATION = ARXIV:1503.03162;%%
  
\bibitem{Carlson:2015poa} 
  C.~E.~Carlson and M.~Freid,
  %``Extending theories on muon-specific interactions,''
  arXiv:1506.06631 [hep-ph].
  %%CITATION = ARXIV:1506.06631;%%
  
\end{thebibliography}
\end{document}